%% file: PhDThesis.tex
\newcommand{\beq}{\begin{equation}}
\newcommand{\eeq}{\end{equation}}
\newcommand{\beqa}{\begin{eqnarray}}
\newcommand{\eeqa}{\end{eqnarray}}

\documentclass[12pt,a4paper,twoside,openright]{book}

\usepackage{Style/PhDThesis}
\label{•} 
\makeindex

\begin{document}


\fancyhead[LE]{ }
\fancyhead[RO]{ }
\fancyhead[LO,RE]{ }

\include{TitleEnglish}

~\thispagestyle{empty}


\frontmatter


\tableofcontents
\fancyhead[LE]{\thepage~~~~Contents}
\fancyhead[RO]{Contents~~~~~\thepage}
\fancyhead[LO,RE]{}
\newpage
\newpage


\mainmatter


\include{Contents/Abbreviations/Abb}


\include{Contents/Introduction/Intro}


\include{Ch1}


\include{Ch2}


\include{Ch3}


\include{Ch4}


\include{Contents/Conclusions/Conclusions}



\appendix


\include{Contents/Appendices/Appendix_A/AppA}

\include{Contents/Appendices/Appendix_B/AppB}

\include{Contents/Appendices/Appendix_C/AppC}



\renewcommand\listfigurename{List of figures}
\fancyhead[LE]{\thepage~~~~List of figures}
\fancyhead[RO]{List of figures~~~~~\thepage}
\listoffigures

\renewcommand\listtablename{List of tables}
\fancyhead[LE]{\thepage~~~~List of tables}
\listoftables

\fancyhead[RO]{List of tables~~~~~\thepage}




\end{document}

%% file: TitleEnglish.tex

\begin{center}
Departamento de F\'{\i}sica Te\'orica y del Cosmos\\
Universidad de Granada\\
\vspace{-.2cm}
\begin{figure}[h]
\begin{center}
\includegraphics[scale=0.15]{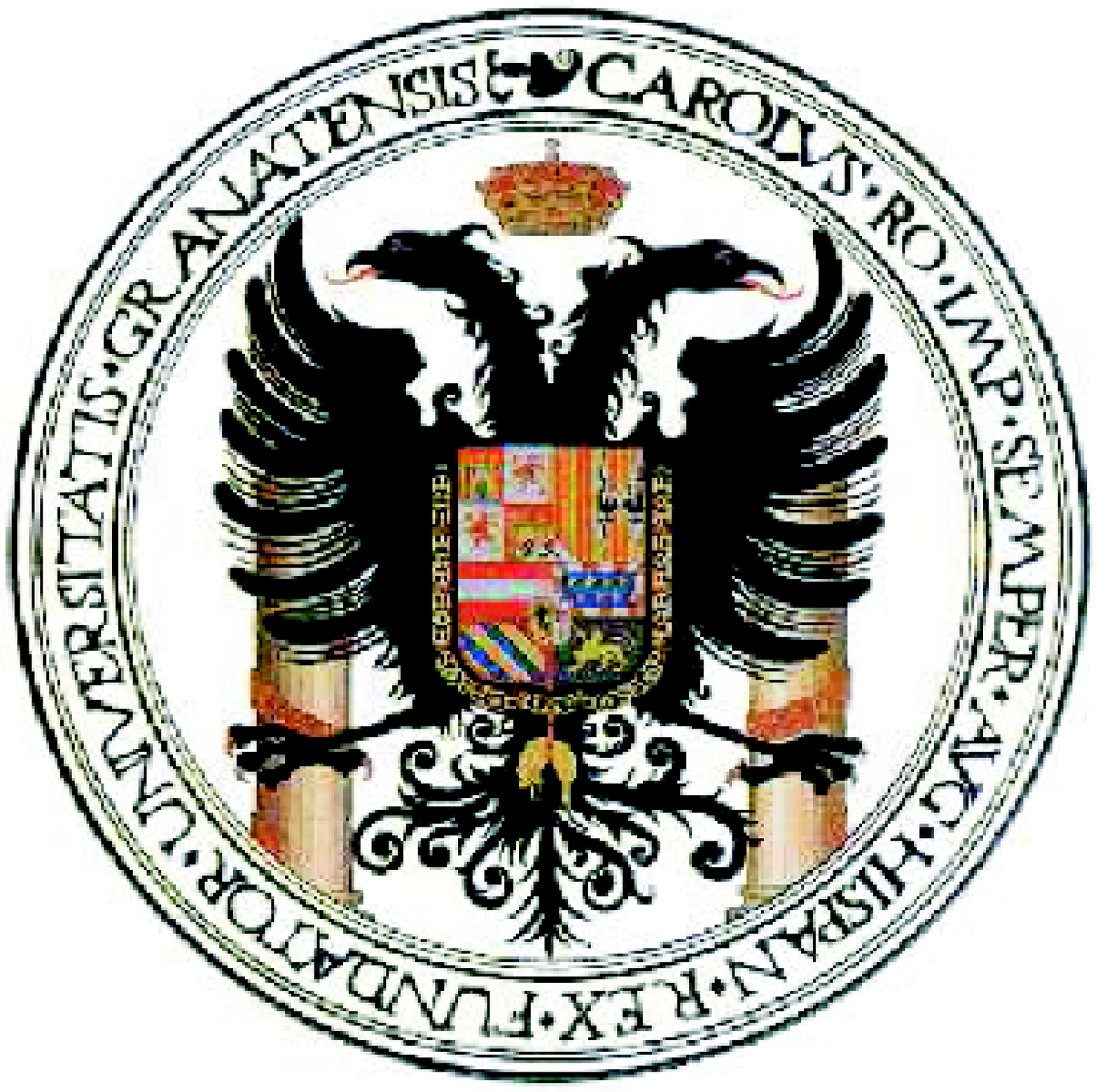}
\end{center}
\end{figure}
\vspace{1.5cm}

{\Huge {\bf{The Higgs Boson and New Physics \\ \vspace{0.5cm} at the TeV Scale}}}\\
\vspace{1.5cm}
{\bf{\Large{Roberto Barcel\'o}}}\\
\end{center}
\vspace{7.7cm}
\begin{flushleft}
{\bf {Ph.D. Advisor: }}\\
{\bf{ Manuel Masip}}
\end{flushleft}
\begin{flushright}
{\bf{--~March, 2012~--}}
\end{flushright}

\thispagestyle{empty}
\newpage


%% file: Contents/Abbreviations/Abb.tex

\vspace{3cm}
\chapter*{Abbreviations}
\addcontentsline{toc}{chapter}{Abbreviations}
\vspace{2cm}

\begin{itemize}
\item c.o.m.: Center of mass.
\item CL: Confidence level.
\item EW: Electroweak.
\item FB: Forward-backward.
\item GB: Goldstone boson.
\item LH: Little Higgs.
\item LHC: Large Hadron Collider.
\item LO: Leading order.
\item MSSM: Minimal Supersymmetric Standard Model.
\item NLO: Next to leading order.
\item PDF: Parton distribution function.
\item SM: Standard Model.
\item SUSY: Supersymmetry.
\item VEV: Vacuum expectation value.
\end{itemize}

\newpage

\thispagestyle{empty}

%% file: Contents/Introduction/Intro.tex

\Introduction

The Standard Model (SM) of particle physics, one of the greatest 
achievements of the 20th century, has proved to be an 
extraordinarily successful
theory to describe the physics at colliders below the TeV scale. 
However, one of the most important pieces of the theory, 
the mechanism responsible
for the electroweak (EW) symmetry breaking, has not been 
confirmed yet. The main objective of the 
Large Hadron Collider (LHC) at CERN
is to reveal the nature of this mechanism, i.e., 
the search for the Higgs boson, the measure of its mass and couplings. 
The discovery of the Higgs and of possible new
particles and symmetries related to the Higgs mechanism
will define a new theoretical 
framework valid at energies above the TeV scale.

The LHC was inaugurated in 2008 and, after solving a 
few problems with the set-up, in 2010 it became the 
most powerful collider in the world, {\it beating} the Fermilab 
Tevatron. Data from its experiments ATLAS and CMS suggest 
the existence of a Higgs boson of 125 GeV compatible with
the one predicted by the SM. Although we are still waiting 
for experimental confirmation from the data collected during 
2012, the mass range where it could be found is already 
very constrained. We are certainly living an exciting time in particle physics.

In addition, so far LHC data do not seem to indicate any 
new physics beyond the SM. As a consequence, 
neutrino masses and the existence of dark matter are the
only experimental evidences currently indicating that the 
model must be completed. Of course, there is the formal 
argument known as the {\it hierarchy problem},
which has been the main motivation for model building during 
the past 30 years and that the LHC should definitively solve. 
Although
not at the level of discovery, there are also experimental 
anomalies that motivate phenomenological studies of different
extensions of the SM. In particular, during the 
preparation of this Thesis one of these anomalies has achieved 
special relevance, the forward-backward
asymmetry in $t \bar t$ production measured at the last stage 
of the Tevatron. This unexpected effect attracted the attention 
of the particle physics community (there are over 100 articles 
appeared during 2011), and it became also my main research line 
during the past year. But this is the
end of the story; let us start from the beginning.

When we planned my Ph.D. project in 2008, we decided that in a 
first phase we would analyze different extensions of the 
SM such as Little Higgs, Supersymmetry
and Extra Dimensions. In a second stage and after the initial 
results from the LHC, we would focus on the phenomenology of 
the model that were favored by the data.
However, due both to the delay in the LHC set-up and to the 
fact that the first observations pointed nowhere besides the 
SM, I had to slightly rethink the final destination of this 
Thesis. Thus, in this work 
two different parts can be distinguished.
In the first one I describe some scenarios for new physics and 
discuss our contributions in each of them. In the second part
I focus on the experimental hint of new physics that I consider 
most interesting, the forward-backward asymmetry in 
$t \bar t$ production measured at the Tevatron, analyzing 
in detail its possible implications at the LHC.

As it is mandatory in any Ph.D. Thesis in particle physics, 
I start reviewing the SM. I pay special attention 
to the Higgs sector and to the main
motivation for this Thesis: the hierarchy problem and the need 
for new physics. Theoretical and experimental constraints on 
the Higgs boson mass are also discussed. 

The second chapter is focused on Little Higgs models, in particular, 
on the so-called
{\it simplest} model. After a short review I discuss our main 
contribution, the possibility that the model accommodates in a 
consistent way a vectorlike $T$ quark relatively light, 
of mass around 500 GeV.
We show that this is possible by slightly changing the 
{\it collective} symmetry breaking principle in the original model 
for an {\it approximate} breaking principle. 
We find the anomalous couplings
of the top quark and the Higgs boson in that scenario, and we deduce 
the one-loop effective potential, showing that it implies an 
EW symmetry breaking and a scalar
mass spectrum compatible with the data.

The third chapter is dedicated to the phenomenological implications 
that new Higgs bosons could have in  
$t \bar t$ production.
We study the possible effects caused by the massive scalars 
present in supersymmetric and Little Higgs models. 
After a review of the Higgs sector in supersymmetric models, 
we analyze the effect 
of these scalars on $t \bar t$ when they
are produced in the s-channel of gluon fusion. We first study 
the cross sections at the parton level and then
we analyze $pp$ collisions at the LHC.

The last chapter is devoted to the forward-backward
asymmetry in $t \bar t$ production. I start reviewing the 
observations and their compatibility with the SM. 
Then I discuss the effects of a generic massive gluon on 
that observable. We propose a framework with a gluon of mass 
below 1 TeV, with small and mostly axial-vector couplings to 
the light quarks and large coupling to the right-handed 
top quark. The key ingredient to define our {\it stealth} gluon, 
invisible in other observables, is a very large decay width 
caused by new decay channels $qQ$, where $q$ is a standard quark and 
$Q$ a massive excitation with the same flavor. The model requires 
the implementation of energy-dependent widths, something that 
is not common in previous
literature. We check that the model reproduces both the asymmetry 
and the $t\bar t$ invariant-mass distribution at hadron colliders, 
and we study the phenomenological
implications of the quarks $Q$. We study 
how the new $qQ$ channel affects current analyses of 
$t \bar t$ production and of $T \bar T$ searches at the 
LHC. We also discuss the
best strategy to detect the $qQ$ channel at that collider. 
We have included three appendixes with details about 
the event selection and reconstruction, which are along the 
lines of those used by the different LHC experiments.

The results contained in this Thesis have led to several 
publications. The study of the Little Higgs models produced 
two publications 
\cite{Barcelo:2007if,Barcelo:2008je}, the first one  
done during a research fellowship awarded in the last year 
of my undergraduate studies.
During the year 2009 we also published a work 
\cite{Barcelo:2009uy} on models with extra dimensions, 
another scenario proposed to solve the hierarchy problem. 
In particular, we studied the interaction of cosmic rays with 
galactic dark matter in models with strong gravity at the 
TeV scale. Since this work, unlike the rest, does not involve
collider physics, I decided not to include it in 
this Thesis. The results about $t \bar t$ production 
through new Higgs bosons gave one publication \cite{Barcelo:2010bm}. 
Finally, our results on the forward-backward
asymmetry in $t \bar t$ production and the study of new 
massive quarks at the LHC in Chapter 4 have appeared in 
three publications
\cite{Barcelo:2011fw,Barcelo:2011vk,Barcelo:2011wu}. 

In addition, I had the opportunity to make a presentation 
in the conference {\it $2^{\textrm{nd}}$ Young Researchers Workshop: 
Physics Challenges in the LHC Era} in Frascati (Italy), May of 2010 
\cite{Frascati:2010}. I made presentations in the 
{\it Bienal de F\'isica Espa\~nola} in 2009 and 2011.
I could also discuss these results in a Seminar at the 
University of Oxford (July 2011), where I was visiting during 
three months.

\newpage

%% file: Ch1.tex

\ref{•}{\Chapter{The Standard Model and beyond} \label{Ch1}}

Although the Standard Model (SM) of particle physics can not be considered as a complete theory of fundamental interactions, due to
its success in explaining a wide variety of experiments it is sometimes regarded as a theory of `almost' everything.
The SM summarizes our understanding of the microscopic world and,
in this sense, it is the result of a long story that involves
the efforts of many people. In the 19th century John Dalton, through his work on stoichiometry, concluded that each given element of nature was composed of
a single type of extremely small object. He believed that 
these objects were the elementary constituents of matter 
and named them atoms, after the Greek word $ \alpha \tau o \mu o $, meaning `indivisible'. Near the end of that century J.J. Thompson
showed that, in fact, atoms were not the fundamental objects of nature but that they had structure and are conglomerates of even smaller particles.
Early 20th-century explorations (Fermi, Hahn, Meitner) culminated in proofs of nuclear fission, a discovery that gave  
rise to an active industry of generating one atom from another. Throughout the 1950s and 1960s, the first accelerators revealed a bewildering variety of 
particles that could be produced 
in scattering experiments. With all these data, in 1961 Sheldon Glashow \cite{GlashowSM} proposed a model combining 
the electromagnetic and the weak interactions which was completed in 1967 by Steven Weinberg \cite{WeinbergSM} and, independently in 1968, by Abdus
Salam \cite{SalamSM}. Together with Quantum Chromodinamics (QCD) for the description of the strong interactions, it defines what is called the SM of particle physics. An important  ingredient of this theory was proposed in 1964 by Peter Higgs \cite{Higgs}, and it has still
to be fully confirmed at the LHC Collider.

In this chapter we will briefly review the SM, paying special
attention to the Higgs sector: present bounds on the Higgs mass,
the hierarchy problem, and the need for physics beyond the SM.
Good reviews about the SM (and beyond) are \cite{Novaes,Pokorski,Hollik,Pich,Langacker1,Langacker2,Amsler,DjouadiSM,IllanaSM}.
 
\Section{The Standard Model}

The SM is a renormalizable relativistic quantum field theory describing three of the four fundamental interactions observed in
nature:
electromagnetic, weak and strong interactions. 
This description is based on a local gauge symmetry 
$SU(3)_C\times SU(2)_L\times U(1)_Y$ spontaneously 
broken to $SU(3)_C\times U(1)_{EM}$.
The fourth fundamental interaction, gravity, is outside
this framework. In this sense, 
the SM could be considered an effective theory only valid
at energies below the Planck scale, where gravitational 
interactions between elementary particles would be unsuppressed.

The gauge principle describes the SM interactions in terms  
of one spin-1 massless boson per each unbroken 
symmetry: 8 gluons ($g$) corresponding to the strong interactions 
and 1 photon ($\gamma$) 
for the electromagnetism. In addition, the 3 broken symmetries 
imply the presence of 3 massive bosons ($W^{\pm}$ and $Z$) 
as mediators of the weak interactions. 

The fermionic matter should be accommodated in multiplets (irreducible representations) of the group transformations. 
The number of copies (or families) of different mass 
is not fixed by the gauge symmetry,
being minimality and the absence of quantum anomalies 
the guiding principles. The SM includes three families of quarks (triplets under $SU(3)_C$) and three of leptons (color singlets). 
All these particles are chiral under the SM symmetry:
the left-handed components define doublets under $SU(2)_L$, while the right-handed components 
are singlets. Therefore, they only get masses after the electroweak (EW) symmetry is broken.
Each lepton family 
consists of a neutrino ($\nu_l$) and a charged lepton ($l^-$)
and for the quarks, each family contains an up ($u$) and a down ($d$)
type quark of electric charge $+2/3$ and $-1/3$, respectively, 
plus the corresponding antiparticles.
A (vectorlike) right-handed neutrino is not 
necessary, so it is not included (see Table.~\ref{SMparicles}).
The electric charge ($Q$), the isospin ($T_3$) and the hypercharge ($Y$) 
are related by the expression $Q = T_3 + Y$ \cite{Higgs,Englert,Gural,Kibble}.

\begin{table}[!h]
\begin{center}
\begin{tabular}{|c|c|c|c|c|}
\hline
\textbf{Multiplets} & $\mathbf{SU(3)_c \times SU(2)_L \times U(1)_Y}$ & \textbf{I} & \textbf{II} & \textbf{III} \\
\hline \hline
\textbf{Quarks} & (\textbf{3},\textbf{2},$-\frac{1}{6}$) & $\left(\begin{array}{c} u_L \\ d_L \\\end{array} \right)$ & $\left(\begin{array}{c} c_L \\ s_L \\\end{array} \right)$ & $\left(\begin{array}{c} t_L \\ b_L \\\end{array} \right)$\\
\cline{2-5}
$s=1/2$ & (\textbf{3},\textbf{1},$\frac{2}{3}$) & $u_R$ & $c_R$ & $t_R$\\
\cline{2-5}
& (\textbf{3},\textbf{1},$-\frac{1}{3}$) & $d_R$ & $s_R$ & $b_R$\\
\hline
\textbf{Leptons} & (\textbf{1},\textbf{2},$-\frac{1}{2}$) & $\left(\begin{array}{c} {\nu_e}_L \\ e_L \\\end{array} \right)$ & $\left(\begin{array}{c} {\nu_\mu}_L \\ \mu_L \\\end{array} \right)$ & $\left(\begin{array}{c} {\nu_\tau}_L \\ \tau_L \\ \end{array} \right)$\\
\cline{2-5}
$s=1/2$ & (\textbf{1},\textbf{1},$-1$) & $e_R$ & $\mu_R$ & $\tau_R$\\
\hline
\begin{tabular}{c} \textbf{Higgs} \\ $s=0$ \end{tabular} & (\textbf{1},\textbf{2},$\frac{1}{2}$) & \multicolumn{3}{|c|}{$H= \left(\begin{array}{c} h^+ \\ h^0 \end{array} \right)$} \\
\hline
\end{tabular}
\label{SMparicles}
\caption{Field multiplets in the SM.}
\end{center}
\end{table}

An important aspect, sometimes taken as a measure of its 
elegance, is the number of free parameters in the SM. 
It has 18 free parameters: 9 fermion masses, 3 CKM mixing angles plus 1 phase, 3 gauge couplings, and 2 parameters 
(the vacuum expectation value of the Higgs field and its physical mass) in
the scalar sector. 
The $CP$-violating $\theta$ angle of QCD can be considered the
19th free parameter of the SM, but it is constrained experimentally
to be very close to zero. 
Finally, neutrino masses can be
obtained through dimension-5 operators and/or adding right-handed
neutrinos, and will define another 7 or 9 new free parameters.
However, strictly speaking they are not part of the SM but
physics beyond it. 

\Section{The Higgs mechanism in the SM}

The Higgs mechanism is a process by which gauge bosons can get a mass. It was first proposed in 1962 by Anderson \cite{Anderson}, who discussed its 
consequences for particle physics but did not work out an explicit relativistic model. Such a model was developed in 1964 by Higgs \cite{Higgs} and, 
independently, by Brout \& Englert \cite{Englert} and Guralnik, Hagen \& Kibble \cite{Gural}. The mechanism is closely analogous to 
phenomena previously discovered by Nambu involving the vacuum structure of quantum fields in superconductivity \cite{SuperNambu}. 

In the SM, the addition by hand of vector-boson and fermion
masses leads to a manifest breakdown of the local
$SU(2)_L\times U(1)_Y$ gauge invariance. The Higgs mechanism,
instead, assumes that the symmetry is broken spontaneously:
while the theory (the Lagrangian) is $SU(2)_L\times U(1)_Y$
invariant, the vacuum (the field configuration with minimum
energy) is not, it breaks the symmetry to 
$U(1)_{EM}$. As a consequence, the 
$W^{\pm}$ and $Z$ gauge bosons and all the matter fields will
acquire masses through interactions with the vacuum.
From a formal point of view, such procedure 
will preserve the gauge principle
and, most important, will keep all the 
properties (renormalizability, unitarity) that make gauge
theories consistent. From a phenomenological point of view,
it will imply a relation between the $Z$ and $W$ masses and
will explain that the photon is massless and the electric
charge conserved.

Let us see more explicitly how this happens. We need to 
introduce a scalar field with, at least, 
3 degrees
of freedom. The simplest choice is a complex $SU(2)$ doublet,
\begin{equation}
\Phi =
\left(\begin{array}{c} \phi^+ \\ \phi^0  
\end{array}\right)_{Y = +{1 \over 2}}\;.
\end{equation}
Imposed gauge invariance and renormalizability
its Lagrangian reads 
\begin{equation}
\mathcal{L_S} = (D^\mu \Phi)^\dagger  (D_\mu \Phi) - V(\Phi)\,, 
\end{equation}
with
\begin{equation}
V(\Phi) = \mu^2 \Phi^\dagger \Phi + \lambda (\Phi^\dagger \Phi)^2\,.
\label{potential}
\end{equation}
If the mass parameter $\mu ^2$ is positive 
($\lambda$ must be positive as well to make the potential bounded from below) then $\mathcal{L}_S$ is simply the Lagrangian of
4 spin-zero particles  ($\Phi^\pm$, $\Phi^0$ and 
$\bar \Phi^0$) of equal mass $\mu$.
\begin{figure}[!h]
\begin{center}
\begin{tabular}{ccc}
\includegraphics[width=0.4\textwidth]{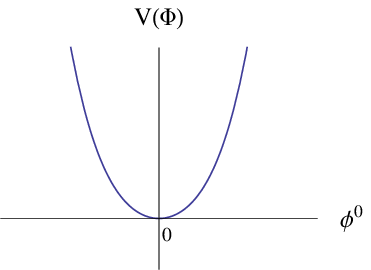} &$\;\;\;$ $\;\;\;$ $\;\;\;$ $\;\;\;$ &
\includegraphics[width=0.4\textwidth]{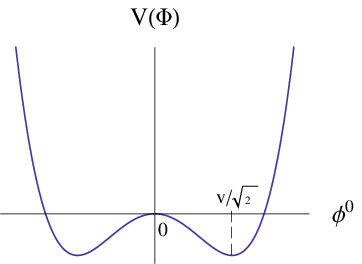}
\end{tabular}
\caption{Projection in the plane $\phi^+$ = 0 of the  potential V($\Phi$) in the cases $\mu^2 > 0$ (left) and (`Mexican hat') 
$\mu^2 < 0$ (right).
\label{potencial}}
\end{center}
\end{figure}
However, if $\mu ^2 < 0$ the origin becomes unstable and the 
minimum of the effective potential will define a vacuum expectation value (VEV) 
\begin{equation}
| \langle \Phi \rangle |={1 \over \sqrt 2}v\,,
\end{equation} 
with $v=\sqrt{-\mu^2/\lambda}$ at the tree level.
Although the minimum is degenerated, all the minima are
equivalent (related by the gauge symmetry), and with all generality we can take
the real component of the neutral component as the only one 
developing a non-zero VEV,
\begin{equation}
\langle \Phi \rangle_0 \equiv \langle0|\Phi|0\rangle = {1 \over \sqrt 2}\left(\begin{array}{c} 0 \\ v  \end{array}\right)\,.
\end{equation} 
With this choice the vacuum is neutral under $Q=T_3+Y$, which
generates the unbroken $U(1)_{EM}$, whereas the rest of the
 $SU(2)_L\times U(1)_Y$ symmetry is spontaneously broken. 
The scalar excitations along the flat directions of the potential
will then define three Goldstone bosons $\theta_i$ 
\cite{Nambu,Goldstone}. We may parametrize
the scalar doublet as
\begin{equation}
\Phi = {1 \over \sqrt 2} \exp{\left( i \theta_i (x) {\sigma_i \over 2}\right)} \left(\begin{array}{c} 0 \\ v + h(x) \end{array}\right),
\end{equation}
being $\theta_i(x)$ and $h(x)$ real fields and $ \sigma_{i=1,2,3.}$ the Pauli matrices. 
Now, 
we can move to the 
so-called unitary gauge by rotating away
the three Goldstone bosons, that are `eaten' by the $W^\pm$ and
$Z$ bosons:
\begin{equation}
\Phi (x) \rightarrow  \Phi (x) = {1 \over \sqrt 2}  \left(\begin{array}{c} 0 \\ v + h(x) \end{array}\right).
\end{equation}
Operating algebraically in the Lagrangian we see that the $W$ and $Z$ bosons get masses while the photon remains massless,
\begin{equation}
M_W = {1 \over 2} v g, \; M_Z = {1 \over 2} v \sqrt{g^2 + g'^2}, \; M_\gamma = 0\,,
\end{equation}
where $g$ and $g'$ are the $SU(2)_L$ and $U(1)_Y$ coupling constants, respectively.

Let us see now how the fermions become massive. We will 
use their Yukawa interactions with the Higgs doublet $\Phi$ 
and its conjugate 
$\tilde{\Phi} = i \sigma_2 \Phi^*$, with hypercharge $Y = -{1 \over 2}$ (it transforms the same way as $\Phi$ under $SU(2)_L$). 
For one fermion generation we obtain gauge invariant 
interactions combining the left-handed doublets with
the right-handed singlets and the Higgs doublet,
\begin{equation}
\mathcal{L_\textrm{Yukawa}} = -y_e \bar{L} \Phi e_R -y_u \bar{Q} \tilde{\Phi} u_R - y_d  \bar{Q} \Phi d_R + h.c.\;.
\end{equation}
After the Higgs VEV we obtain
\begin{equation}
m_e= { y_e v \over \sqrt{2}}, \; m_u= { y_u v \over \sqrt{2}}, \; m_d= { y_d v \over \sqrt{2}}.
\end{equation}
For three generations 
$y_{e,u,d}$ must be replaced by $3\times 3$ matrices 
of Yukawa couplings.

\begin{figure}[htb]
\centerline{ {\includegraphics[width=3.4in]{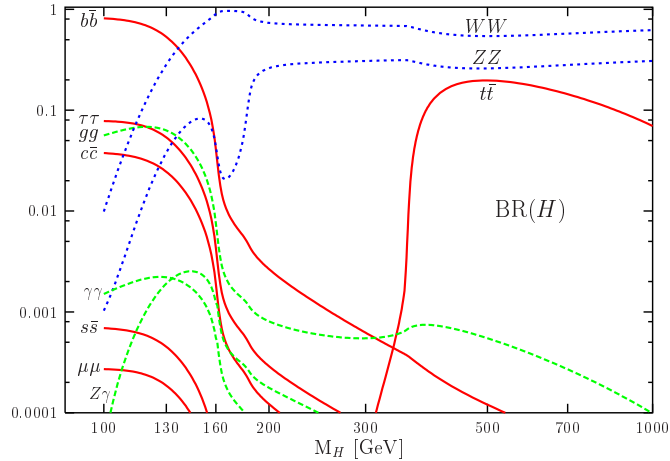}} }
\caption{Decay branching ratios of the Standard-Model Higgs as a function of its mass \cite{DjouadiSM}.} 
\label{branching}
\end{figure}

Since the Higgs couples to gauge bosons and fermions 
proportional to their masses, it will decay into the heaviest ones accessible at the  
phase space. In Fig.~\ref{branching} we show its decay branching ratios
as a function of its mass.

\section{The Higgs particle in the SM}

In the unitary gauge there is only one degree of freedom, $h$,  
which corresponds to the {\it physical} Higgs boson. 
Its kinetic term  and gauge
interactions 
come from the covariant derivative $|D_\mu \Phi|^2$, 
while its mass and self-interactions derive from the scalar potential.
In particular, in terms of the couplings in Eq.~(\ref{potential}) one has
\begin{equation}
m_h^2 = 2 \lambda v^2 = -2 \mu^2,
\end{equation}
with the VEV 
$v=246$ GeV in order to reproduce the $W$ boson mass.
In the SM a measurement of $m_h$ would also fix all its self-interactions.

Although the mass of the SM Higgs is still unknown, 
there are interesting theoretical constraints that can be derived from assumptions on the energy range 
in which the SM is valid before perturbation theory breaks down and new phenomena emerge. These include constraints from unitarity in scattering amplitudes,
perturbativity of the Higgs self-coupling and stability of the EW vacuum. In particular, unitarity constraints in 
longitudinally polarized $WW$ scattering imply that the Higgs mass should not be much larger than 1 TeV\footnote{A similar argument was the basis to abandon
the old Fermi theory for the weak interaction.}. Triviality limits are derived 
assuming the SM to be valid
up to some energy scale and requiring that the self-coupling of the Higgs field does not blow up in the running. For a value of the cut-off scale not much larger
than $m_h$ this implies that $m_h \lesssim 800$ GeV, while $m_h \lesssim 150$ GeV if we assume perturbativity up to the reduced Planck
scale ($\sim  10^{18}$ GeV). Theoretical considerations about the stability of the scalar potential under top-quark corrections to $\lambda$
provide lower bounds on $m_h$, also depending on
the cut-off scale of the SM. If, again, we assume the SM is valid up to the reduced Planck scale, vacuum stability requires $m_h \gtrsim 130$
GeV, and $m_h \gtrsim 115$ GeV if we simply require a sufficiently long-lived metastable vacuum.

Later we will discuss the different experimental constraints on $m_h$ coming from direct and indirect 
searches, but before we would like to review the main reason to expect physics beyond the 
SM associated to the Higgs boson.

\Section{The hierarchy problem}

Despite the extraordinary strength of the SM from an experimental point of 
view, it has been emphasized during the past 30 years the 
need for additional physics that provides {\it formal} consistency to
its Higgs sector and solves the so-called hierarchy problem. 
If we assume the SM valid up to an energy scale $\Lambda$ (its cut-off) 
then the Higgs mass
parameter ($\mu^2$) receives one-loop corrections that grow proportional 
to $\Lambda^2$. The three most significant quadratic 
contributions come from one-loop diagrams with the top quark, the gauge bosons and Higgs self-interactions:

\beq
\Delta \mu^2 = -\frac{3}{8 \pi^2} y_t^2 \Lambda^2+ 
{9 \over 64 \pi^2} g^2 \Lambda^2+ {1 \over 16 \pi^2} \lambda \Lambda^2 \,.
\eeq
\begin{figure}[htb]
\centerline{ {\includegraphics[width=3.9in]{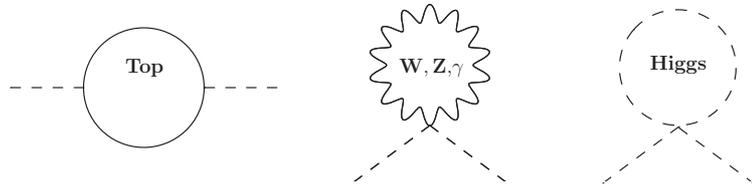}} }
\caption{Most significant quadratic contributions to $\mu ^2$. } \label{fig:smloops}
\end{figure}

\hspace{-0.6cm}The Higgs mass parameter in the effective potential includes then 
one-loop plus (arbitrary) tree-level contributions.
If the SM is all the physics that there is between the EW and the Planck 
or the Grand Unification scale
(i.e., $\Lambda \thicksim 10^{19} \textrm{ GeV}$ or 
$\Lambda \thicksim 10^{16} \textrm{ GeV}$), these different contributions
should cancel around 30 digits to get the observed value of order $v^2$ 
($ \mu^2 =  -v^2 \lambda \thicksim - (100 \textrm{ GeV})^2$). 
Moreover, this cancellation must take place at every order in 
perturbation theory, since another quadratic divergence appears at 
two-loops and so on. This {\it fine tuning} is considered unnatural 
and suggests that the scale $\Lambda$ for new physics (new symmetries
or dynamics) that explains the cancellation is not very high. 

To get one-loop corrections to $\mu^2$ smaller than 10 times its total
value (i.e., less than 10\% of fine tuning)  
the cut-off in each sector should satisfy
\begin{equation}
\Lambda_{top} < 2 \textrm{ TeV,} \qquad \Lambda_{gauge}
< 5 \textrm{ TeV,} \qquad \Lambda_{Higgs} < 12 \textrm{
TeV.}
\label{cutoff}
\end{equation}
Therefore, one could expect that below those energies the new physics becomes
effective and defines a {\it corrected} SM free of quadratic corrections to
$\mu^2$. In this way, the new physics is most needed to cancel the top-quark
contribution, which favors the existence of new {\it top-like}
particles (related to the top quark by symmetries) 
or a special (composite) nature for this quark observable
below this threshold. In addition, one also expects that the 
new physics should couple to the Higgs as strongly as the top quark,
so the Higgs search at hadron colliders could be very correlated with 
the search for new top-quark physics. 
In a similar way, we would expect the extra physics canceling 
gauge-boson corrections below 5 TeV, which suggests a less relevant
role at the LHC but that could conflict the data
on precision EW observables. 
Finally, Higgs self-interactions require new physics below 12 TeV
for $m_h=125$ GeV.

As the center of mass (c.o.m.) energy at the LHC may reach 14 TeV, these 
are compelling arguments to explore there the different
possibilities for the so-called physics beyond the Standard Model. 
If the hierarchy problem is to be solved by new physics, we should
see it at the LHC. Some frameworks for this physics to be analyzed 
in later chapters are:

\begin{itemize}
\item Supersymmetry (SUSY), probably the favorite candidate of the community
over the past 30 years. There is the increasing feeling, however, that it
should have already given {\it any} experimental signal, specially  
in precision experiments (electric dipole moments, flavor physics). The 
discovery of a light Higgs could provide renewed interest. We will discuss
the SUSY Higgs sector in \mbox{Chapter 3}.
\item Models where the Higgs is a pseudo-Goldstone boson of a global symmetry 
broken spontaneously above the EW scale. This includes models
of Little Higgs (in Chapter 2) and models of composite Higgs (the pions of
the global symmetry). Their
objective is to define  consistent models just up to $\approx 10$ TeV 
instead of the Planck scale.
As a consequence, they tend to be simpler than, for example, SUSY.
While SUSY modifies {\it all} the sectors in the SM, these other models
may affect only the Higgs and the top-quark sectors. Given the agreement
of the SM with all the data, this should be considered a big 
{\it advantage}.
\item Models of TeV gravity. The presence of compact extra-dimensions 
opens the possibility that the fundamental scale is at the TeV 
instead of the Planck scale. 
That would make quantum gravity and string theory accessible 
to the LHC. In \cite{Barcelo:2009uy} (not included in this Thesis) we discuss how these gravitational interactions
may affect the propagation of ultrahigh energy cosmic rays.
\item Models in a 5-dimensional anti de Sitter space (AdS). 
They offer multiple possibilities for model building with peculiar 
phenomenologies. Using the AdS/CFT correspondence 
one obtains effective models 
of technicolor or composite Higgs, with the possibility to calculate 
observables pertubatively in the 5-dimensional model. In Chapter 4 we 
use this framework to motivate a model for the top-quark sector.
\end{itemize} 

The discovery of any of these possibilities at the LHC would define a 
model more complete (valid up to higher energies) than the SM. Of course, there is also the 
disturbing possibility that a 125 GeV Higgs is observed but no
genuine new physics is discovered at the LHC. Such final confirmation of the
SM would imply that there is no dynamical explanation to the 
hierarchy problem, and one should probably consider other more 
radical approaches  (an anthropic or accidental explanation \cite{Agrawal:1997gf,Arkani-Hamed:2004fb}).

\Section{Limits on the Higgs mass from precision observables}

With the exception of the Higgs mass, all the SM parameters 
have been determined experimentally: the three gauge coupling constants,
the Higgs VEV and the fermion masses and 
mixing angles. With these parameters it is possible to calculate
any physical observable and compare it with the data. 
Since the strong and the EW coupling constants at high 
energies are relatively small, the tree-level term is usually an
adequate approximation. Precision experiments, however, require 
more accurate predictions, probing the SM at the one-loop level. The
Higgs dependence on these predictions is weak, at its mass appears
only logarithmically.

\begin{figure}
\centerline{ {\includegraphics[width=5.5 in]{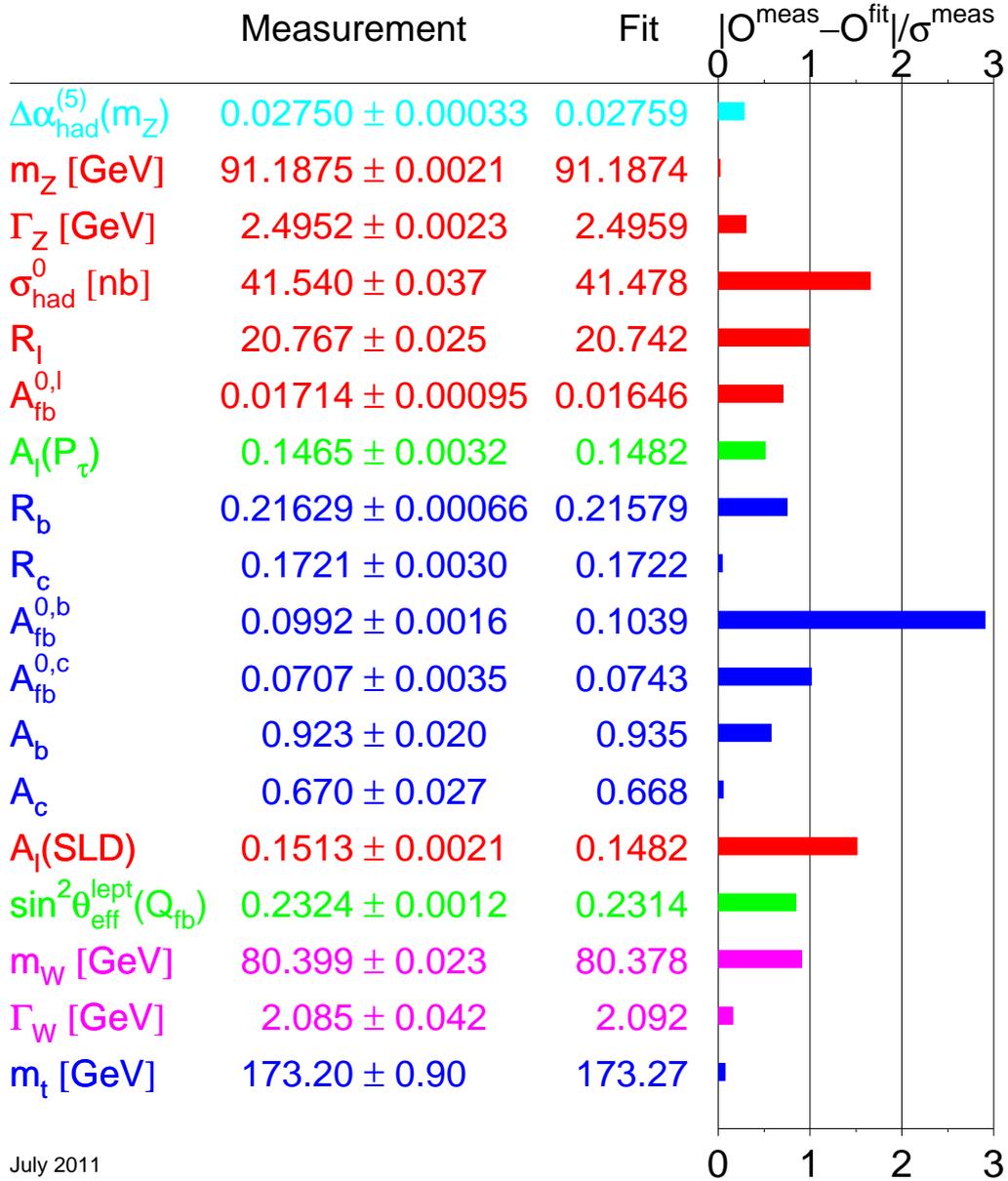}} }
\caption{Summary of electroweak precision measurements at LEP1, LEP2, SLC and the Tevatron. The SM fit results, which have been derived including all known radiative
correction, and the SM deviations are also shown \cite{LEPEWWG}.} 
\label{EWPO}
\end{figure}

\begin{figure}[t]
\centerline{ {\includegraphics[width=3.2in]{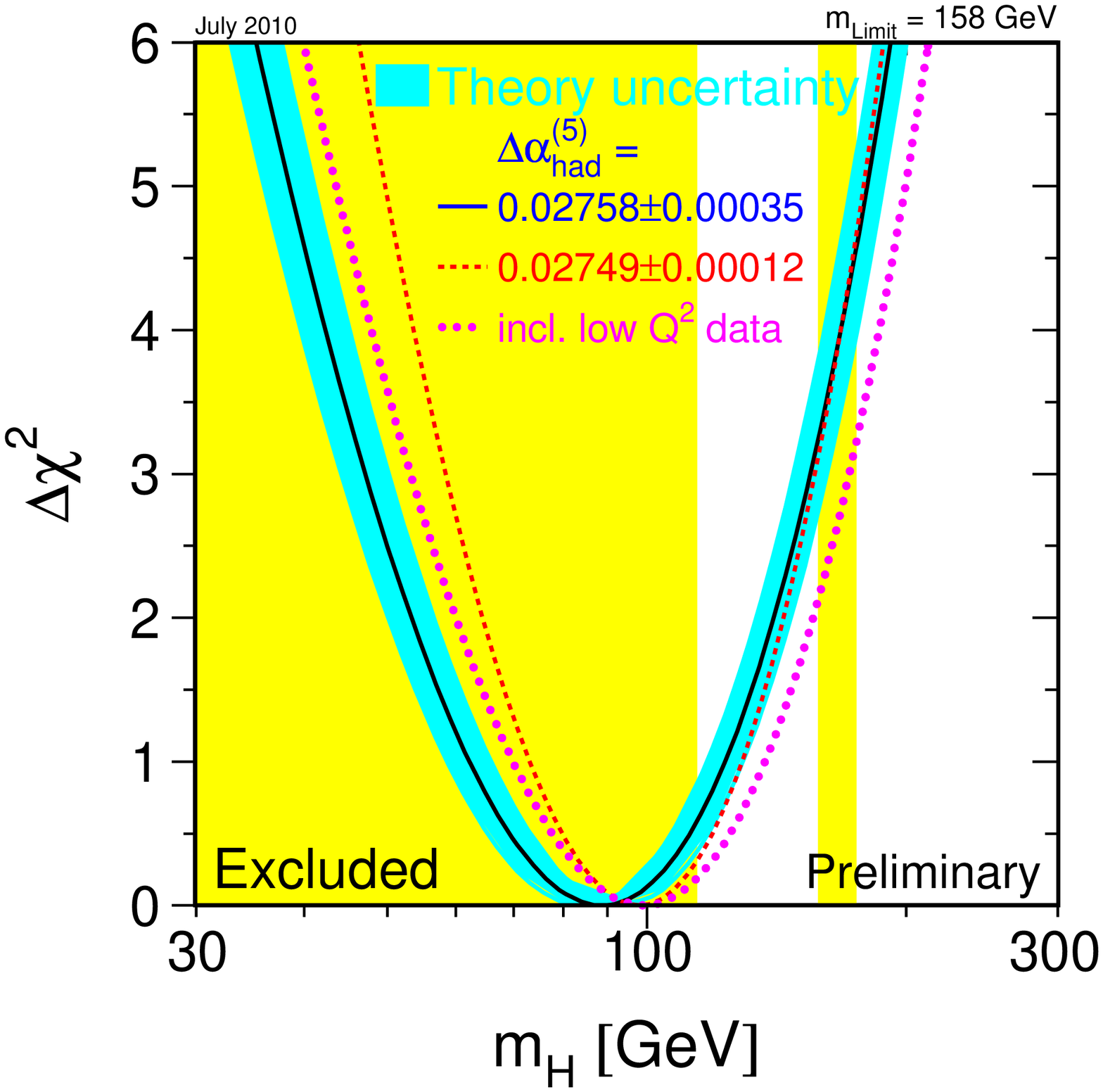}} }
\caption{The $ \Delta \chi^2$ of the fit to the electroweak precision data as a function of the Higgs mass. The solid line results when all
data are included and the blue/shade band is the estimated theoretical errors from unknown higher order corrections. The effect of
including low-$ Q^2 $ data and the use of a different value for $ \Delta \alpha_{had} $ are also shown \cite{LEPEWWG}.} 
\label{bestfit}
\end{figure}

The SM is in good agreement with a very large volume of data. 
Precision measurements 
have found no significant deviations from its predictions, with
preference for a light Higgs. 
For instance, EW precision tests performed during the first run of LEP and
at SLC have provided accurate measurements of the properties 
of the neutral current sector. On the other hand, during the second run
of LEP, LEP2, the $W$ mass, its decay width and branching ratios were measured with an accuracy of a few parts per ten thousand.
Since two of the three tree-level diagrams contributing to 
$W$ pair production contain triple gauge boson couplings, the non-abelian 
nature of the EW interactions was also tested there. Further 
measurements of the $W$ parameters were performed at the Tevatron 
where, in addition, the $W$ boson can be single produced with a large 
cross section. Up to now, with the only exception of neutrino oscillations 
(which require neutrino masses and an extension of the SM), the 
dark matter problem, and despite a few discrepancies of limited 
statistical significance, all experimental data are 
consistent within the SM. Fig.~\ref{EWPO} illustrates the 
agreement/disagreement between the experimental measurements 
and the theoretical predictions for
the most significant EW precision observables \cite{LEPEWWG}. It is given 
in terms of the {\it pulls}: the absolute value of the difference 
between measurement and prediction normalized to the experimental 
error. The theoretical predictions of the SM for the best fit of 
$m_h$ (Fig.~\ref{bestfit})
have been obtained including all known radiative corrections and 
using the measured central values of the top quark mass ($m_t$), 
the strong coupling constant ($\alpha_s$), etc. 
The agreement should be considered excellent even if a few 
measurements show discrepancies between 1 and 3 standard deviations 
($ \sigma $), which are expected when a large number of observables
are included in a fit. 
On the other hand, these few discrepancies have motivated many 
analyses that have tried to favor or disregard them as 
hints for new physics. The most significant of them is probably 
the anomalous magnetic moment of the muon (not included in the table 
as it involves hadronic physics and is not usually considered an EW precision observable).
It has been computed within the SM at four loops. Combining the
data available it is found 1 to 3.2$\sigma$ away from the
SM prediction, depending on whether we use $\tau$ decay data or 
electron-positron data, 
respectively. This deviation could be accounted by contributions 
from new physics such as SUSY with
large $\tan \beta$ \cite{Moroi:1995yh,Carena:1996qa}. Another significant deviation is the bottom
forward-backward asymmetry $A^b_{FB}$ measured at LEP. The best-fit value
gives a prediction which is 2.8$ \sigma $ above its experimental value. 
This deviation has sometimes been interpreted as a hint of
new physics strongly coupled to the third quark family \cite{Malkawi:1996fs,Djouadi:2006rk}. 
Supporting this idea, a $3\sigma$ deviation has also been observed in 
the forward-backward asymmetry in the $t\bar t$ production at the 
Tevatron. We will dedicate the whole Chapter 4 to this point.

As we mentioned before, the Higgs boson mass is an unknown 
parameter that enters logarithmically in the calculation of 
these observables at the loop level. Therefore, precision data will put 
indirect bounds on the value of this parameter. 
It should be noticed, however, that constraints on $m_h$ 
from EW precision measurements are controversial, as they arise from 
a combination of measurements that {\it push} the SM in different directions.
The most constraining observables, besides the $W$ boson mass,
are LEP and SLC measurements of leptonic asymmetries ($A_{LR}$) and $A^b_{FB}$. 
While the former favors a light Higgs boson (as it is also the case for the 
value of the $W$ boson mass), the hadronic asymmetries prefer 
a heavier Higgs. Taking into account all the precision EW
data given in Fig.~\ref{EWPO} in
a combined fit (Fig.~\ref{bestfit}) it results into $m_h\approx 92$ GeV \cite{LEPEWWG}, 
with an experimental uncertainty of $+34$ and $-26$ GeV at 68\% confidence 
level (CL) derived from $ \Delta \chi^2$ = 1. At 95\% of CL 
precision EW measurements tell us that the mass of the SM Higgs boson 
is lower than about 161 GeV (including both the experimental and the 
theoretical uncertainty). This limit increases to 185 GeV when including 
the LEP2 
direct search limit of 114 GeV (shown in light-grey/yellow shaded in
Fig.~\ref{bestfit}). 

\Section{Constraints on the Higgs mass from direct searches}

Before LEP (started in 1989), Higgs masses below 5 GeV were
thought to be unlikely. 
The main probes for $m_h\le 20$ MeV were 
nuclear-physics experiments with large theoretical uncertainties.
Some of these searches were \cite{Kado}:

\begin{enumerate}
\item For very low Higgs boson masses, the SINDRUM 590 MeV proton 
cyclotron spectrometer experiment at the PSI (Villigen) investigated the decay
of the pion to electron, electron neutrino and Higgs boson 
that in turn decays into a pair of electrons. 
It excluded masses in the interval $10\; {\rm MeV} < m_h < 110\; {\rm MeV}$ \cite{Egli:1989vu}. 
\item The CERN-Edinburgh-Mainz-Orsay-Pisa-Siegen collaboration at the SPS (CERN)
also searched for the decay of a Higgs boson into a pair of electrons in 
$K^0_L \rightarrow \pi^0 h$. These searches severely constrained 
$m_h$  below 50 MeV  \cite{Barr:1989pv}.
\item The CLEO experiment at CESR (Ithaca) searched for decays of the Higgs boson into
a pair of muons, pions, and kaons produced through the 
decay $B \rightarrow K^0 h$. It excluded the mass range 
0.2--3.6 GeV \cite{CLEO1989}. 
\item Finally, the CUSB collaboration at CESR investigated the radiative
decay of various states of the $ \Upsilon$ 
into a Higgs boson \cite{CUSB1982}. The search for
a monochromatic photon sample from the decay $\Upsilon \rightarrow \gamma + X$ 
 excluded the range from 2$m_\mu$ up to 5 GeV \cite{CUSB1988}. 
\end{enumerate}

\subsubsection*{Searches at LEP1}

\begin{figure}[h]
\centerline{ {\includegraphics[width=3.1 in]{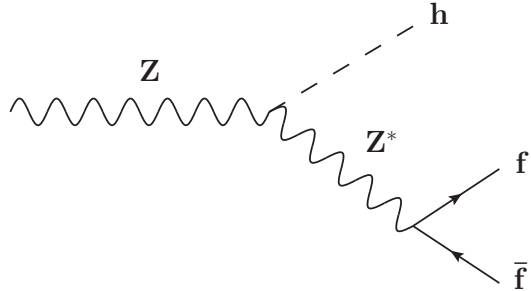}} }
\caption{The most relevant production channel for the Higgs boson in $Z$ decays  
at LEP1.} 
\label{Bjorken}
\end{figure}

\begin{figure}[h]
\centerline{ {\includegraphics[width=5.5 in]{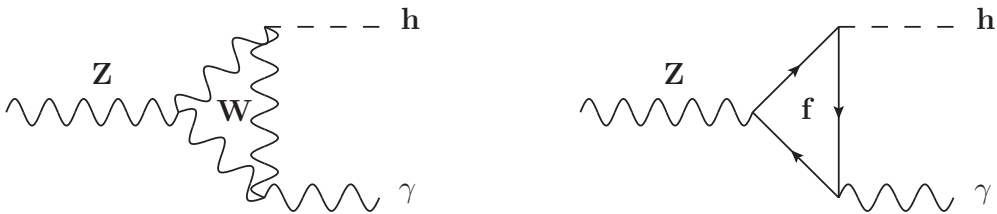}} }
\caption{Diagrams for the one-loop induced decay mode $Z \rightarrow h \gamma$ in the SM.}
\label{fotonloop}
\end{figure}

The first run of LEP, also called LEP1, covers from 1989 to 1995 
at energies close to the Z resonance ($ \sqrt s \thicksim M_Z$). 
Because of the 
large production cross section for a low-mass Higgs boson in Z 
decays, LEP1 provided a very good environment to further exclude 
small values of $m_ h$. 
The dominant production mode is the Bjorken process (in 
Fig.~\ref{Bjorken}), 
where the $Z$ boson decays into a real Higgs boson and an 
off-shell $Z$ boson that then goes into two light fermions, 
$Z \rightarrow h Z^* \rightarrow h f \bar{f}$.
The Higgs boson can also be produced in the decay 
$Z \rightarrow h \gamma$, which occurs through triangular 
loops built-up by heavy fermions and the $W$
boson (in Fig.~\ref{fotonloop}). Relative to $Z \rightarrow h f\bar f$, 
this loop decay process becomes 
important for masses $m_h \ge 60$ GeV, 
although in this case only a handful of events are expected.

As shown in Fig.~\ref{branching}, the Higgs boson in the  
LEP1 (and also LEP2) mass range decays predominantly into hadrons 
(mostly $b\bar{b}$ 
for $m_h \gtrsim 10$ GeV) and with a 8\% branching ratio  
into $ \tau$-lepton pairs. Thus, to avoid the large 
$e^+e^- \rightarrow$
hadron background, the Higgs boson has been searched for in the 
two topologies 
$Z \rightarrow $($h \rightarrow$ hadrons)($Z^* \rightarrow \nu \bar{\nu}$), 
leading to a final state consisting of two acoplanar jets and missing energy, 
and $Z \rightarrow $($h \rightarrow$ hadrons)($Z^* \rightarrow e^+e^-,
\mu^+\mu^-$), with two energetic leptons isolated from the hadronic system. 
The absence of any Higgs boson signal by the four collaborations at 
LEP1 \cite{LEP11,LEP12,LEP13,LEP14} set the 95\% CL limit $m_h \gtrsim 65.2$ GeV.
In these channels the Higgs mass
will simply be the invariant mass of the system recoiling against the 
lepton pair. The bounds are independent of Higgs decay modes provided 
that its coupling to the $Z$ boson is the one predicted in the SM.

\subsubsection*{Searches at LEP2}

\begin{figure}[h]
\centerline{ {\includegraphics[width=3.2 in]{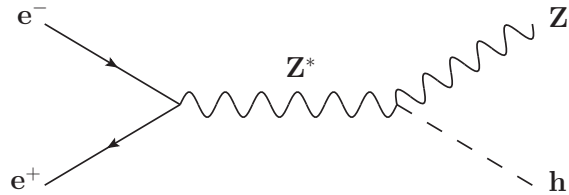}} }
\caption{Production channel for the SM Higgs boson in $e^-e^+$ collisions at LEP2.} 
\label{Higgstrahlung}
\end{figure}

The second run of LEP, with a c.o.m.~energy of
$ \sqrt s = 209$ GeV, covered from 1995 to 2000, when the experiment 
was closed.
In this energy regime the main production channel is 
Higss-strahlung, with the $e^-e^+$ pair going into an off-shell $Z$ boson 
that then decays into a 
Higgs particle and a real $Z$ boson, $e^+e^- \rightarrow Z^* \rightarrow hZ$ 
(in Fig.~\ref{Higgstrahlung}). 

\begin{figure}[htb]
\centerline{ {\includegraphics[width=3.2in]{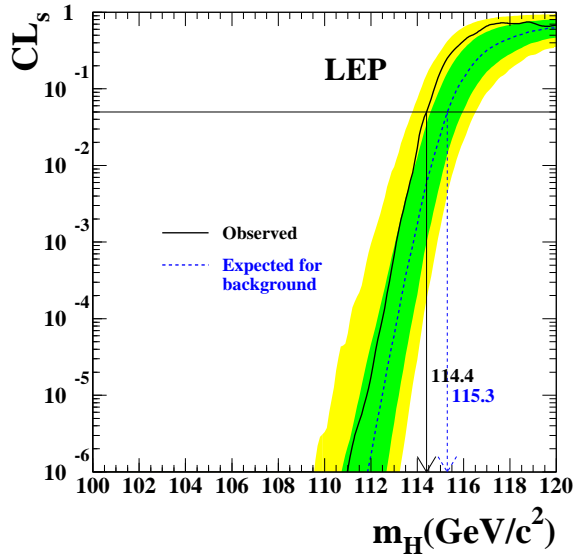}} }
\caption{CL for the signal plus background hypothesis in Higgs production at LEP2. The solid line is for 
the observation, the dashed line is the median background expectation and the dark-grey/green (light-grey/yellow) shaded band
around the median expected line correspond to the 68\% (95\%) simulated probability band. The intersection of the horizontal
line at CL$_s$ = 0.05 with the observed curve defines the 95\% CL lower bound for $m_h$ \cite{LEP2}.} 
\label{CL}
\end{figure}

The searches by the different LEP collaborations have been made in 
many topologies: $e^+e^- \rightarrow $($h \rightarrow b \bar{b}$)($Z^* 
\rightarrow \nu \bar{\nu}$) and $e^+e^- \rightarrow $($h \rightarrow 
b \bar{b}$)($Z^* \rightarrow l^+ l^-$), like at LEP1, and also 
$e^+e^- \rightarrow $($h \rightarrow \tau^+ \tau^-$)($Z^* \rightarrow  
b \bar{b}$) and $e^+e^- \rightarrow $($h \rightarrow b \bar{b}$)($Z^* 
\rightarrow \tau^+ \tau^- $).
Combining 
the results they obtained an 
exclusion limit \cite{LEP2}
\begin{equation}
m_h > 114.4 \textrm{ GeV}
\end{equation}
at 95\% CL (Fig.~\ref{CL}). The upper limit was 
expected to reach
$m_h > 115.3 \textrm{ GeV}$, with the discrepancy coming from  
a 1.7$ \sigma $ excess (reported initially as a 2.9$ \sigma $ excess) 
of events that could be indicating a Higgs boson in the vecinity of 
$m_h = 116 \textrm{ GeV}$ \cite{LEP2}. This anomaly was considered 
not significant enough to keep taking data 
and the experiment was terminated as scheduled
(a discovery requires a 
5$ \sigma $ deviation).

\begin{figure}[htb]
\centerline{ {\includegraphics[width=3.2in]{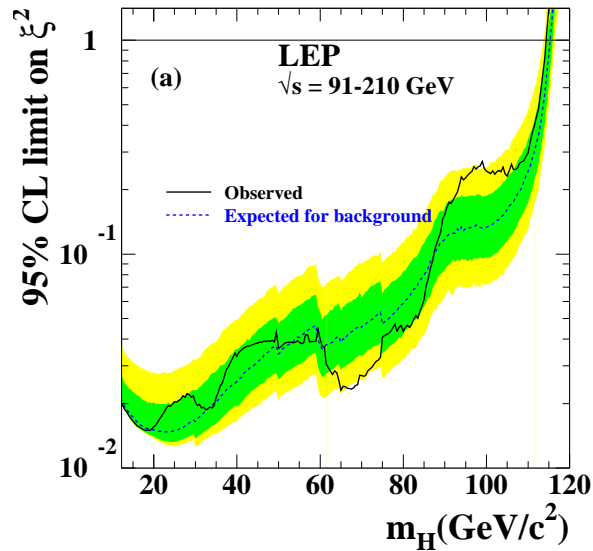}} }
\caption{Upper bound on the coupling $ \xi^2= (g_{hZZ} / g^{SM}_{hZZ})^2 $ as a function of the Higgs mass. The solid line represents the observed limit while the
dark (light) shaded band is for the 68\% (95\%) probability band \cite{LEP2}.} 
\label{Lightcoupling}
\end{figure}

The bound $m_h > 114.4 \textrm{ GeV}$ can be evaded if, for example, 
the Higgs boson has a weaker couplings $ g_{hZZ} $ to the $Z$ boson 
than in the SM. 
This would suppress the $ e^+e^- \rightarrow hZ$
cross section and then the number of Higgs events (proportional to 
$ g_{hZZ}^2 $). In Fig.~\ref{Lightcoupling} we show the 95\% CL 
bound on $m_h$ as a function of  $ \xi= (g_{hZZ} / g^{SM}_{hZZ})$. 
Thus, Higgs bosons
with $m_h \lesssim 80$ GeV and couplings to the $Z$ boson an order 
of magnitude smaller than in the SM have also been discarded \cite{LEP2}.
On the other hand, a non-standard Higgs with half the SM coupling 
would relax the bound to 95 GeV. We will see in Chapter 2 that
in Little Higgs models the Higgs doublet mixes with a singlet, 
reducing the value of its couplings to the EW gauge bosons.  

\subsubsection*{Searches at Tevatron}

\begin{figure}[htb]
\centerline{ {\includegraphics[width=3.8in]{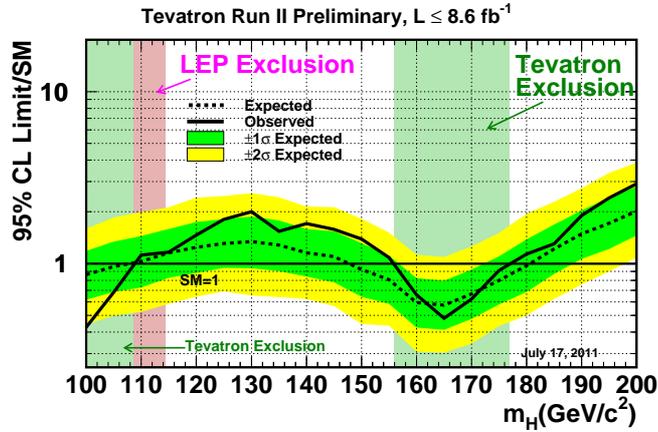}} }
\caption{Observed and expected (median, for the background-only hypothesis) 95\% CL upper limits on the ratios to the SM
cross section, as functions of the Higgs boson mass for the combined CDF and D$\emptyset$ analyses. The limits are expressed as a
multiple of the SM prediction for test masses (every 5 GeV) for which both experiments have performed dedicated searches
in different channels. The points are joined by straight lines for better readability. The region for which the solid curve dips below the horizontal line at the value of 1 is excluded with a 95\% CL. The dashed curve shows the expected limit in the absence of the Higgs boson, based on simulations. The green/dark-shaded and yellow/light-shaded bands correspond (respectively) to 68\%, and 95\% CL regions from the expected limits. The limits displayed in this figure are obtained with
the Bayesian calculation \cite{Tevatron}.} 
\label{Tevatron}
\end{figure}

The Tevatron data taking covers from 1987 to 2011. It was a $p\bar p$ 
collider with a c.o.m.~energy $\sqrt{s}$ = 1.96 TeV. 
Here we will briefly discuss the most recent results on Higgs searches. 
In particular, analyses \cite{Tevatron} sought Higgs bosons 
produced with a vector boson ($q \bar{q} \rightarrow hW/Z$), through 
gluon-gluon fusion ($ gg \rightarrow h$), and through vector boson 
fusion ($q \bar{q} \rightarrow q' \bar{q}' h$) with an integrated 
luminosity up to 8.2 fb$ ^{-1}$ at CDF and 8.6 fb$ ^{-1}$ at D$\emptyset$.
The decay modes under study were $h \rightarrow b \bar{b}
$, $h \rightarrow W^+ W^-$, $h \rightarrow ZZ$, $h \rightarrow \tau^+ \tau^-$ and $h \rightarrow \gamma \gamma$. The results of these
analyses are summarized in Fig.~\ref{Tevatron}.

At 95\% CL the 
upper limits on Higgs boson production are a factor of 1.17,
1.71, and 0.48 times the values of the SM cross section for $m_h$ =115 GeV,
140 GeV, and 165 GeV, respectively. The corresponding median upper limits expected in the
absence of Higgs boson production are 1.16, 1.16, and 0.57. There is a small ($\thickapprox$ 1$\sigma$) excess of data
events with respect to the background estimation in searches for the Higgs boson in the mass range
125 GeV $<$ $m_h $ $<$ 155 GeV. At the 95\% CL the region 156 GeV $<$ $m_h$ $<$ 177 GeV is excluded.

\subsubsection*{Searches at LHC}

The LHC data taking started in 2010. It is a $pp$ collider with a c.o.m.~energy $\sqrt{s}$ = 7 TeV.
Here we will discuss the most recent results with the data collected by ATLAS and CMS during 2010 and 2011.

\begin{figure}[tb]
\centerline{ {\includegraphics[width=3.5in]{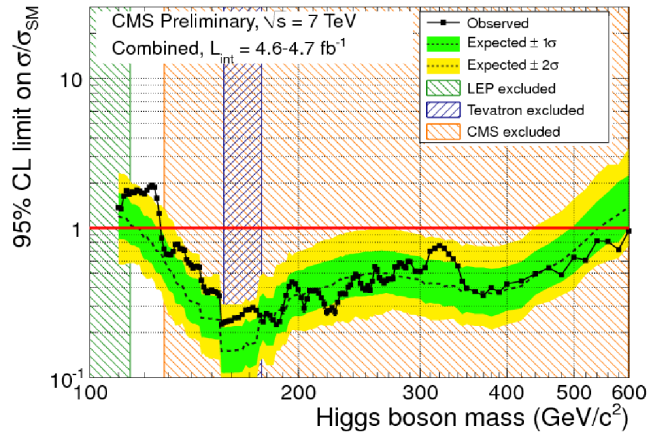}} }
\caption{Exclusion limit on the mass of the SM Higgs boson at 95\% CL (below red line). The analysis is based on 4.7 fb$^{-1}$ of proton-proton data collected by CMS in 2010 and 2011. The hatched bands show the mass regions previously excluded by LEP, the Fermilab Tevatron, and now by CMS \cite{LHCCMS}.} 
\label{CMS2}
\end{figure}

In December of 2011 the CMS collaboration presented their latest results on
SM Higgs boson searches \cite{LHCCMS}. The data amounted to 4.7 fb$^{-1}$, 
meaning that CMS can explore almost the entire mass range above the 114 GeV
limit from LEP up to 600 GeV. 
They combined searches in a number of Higgs decay channels: 
$h \rightarrow b \bar{b}
$, $h \rightarrow W^+ W^-$, $h \rightarrow ZZ$, $h \rightarrow \tau^+ \tau^-$ and $h \rightarrow \gamma \gamma$ (the same as in Tevatron).
The preliminary results exclude the existence of a SM Higgs boson 
in a wide range of masses: 127--600 GeV at 95\% CL 
and 128--525 GeV at 99\% CL (Fig.~\ref{CMS2}).
A 95\% CL exclusion means that the SM Higgs boson with that mass 
would yield more evidence than that observed in the data at least 
95\% of the time in a set of repeated experiments. 

\begin{figure}[htb]
\centerline{ {\includegraphics[width=3.6in]{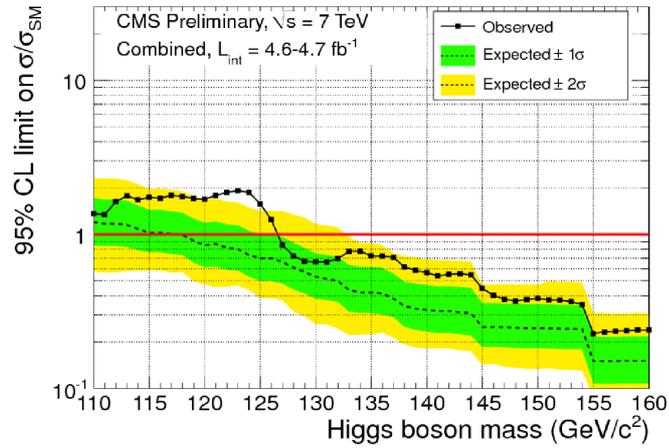}}}
\caption{SM Higgs exclusion limit at 95\% CL for 4.7 fb$^{-1}$ proton-proton data collected by CMS in 2010 and 2011, showing the lower mass region \cite{LHCCMS}.} 
\label{CMS}
\end{figure}

A SM Higgs boson mass between 115 GeV and 127 GeV is not excluded 
(Fig.~\ref{CMS}). Actually, compared to the SM prediction there is an excess 
of 2$\sigma$ in this mass region that appears, quite consistently, 
in five independent channels. With the amount of data collected so far, 
it is difficult to distinguish between the two hypotheses of 
existence versus non-existence of a Higgs signal in this low mass region. 
The observed excess of events could be a statistical fluctuation of the 
known background processes, either with or without the existence of the 
SM Higgs boson in this mass range. The larger data samples to be collected in 2012 will reduce the statistical 
uncertainties and reveal the
existence (or not) of the SM Higgs boson in this mass region.

\begin{figure}[h!]
\centerline{ {\includegraphics[width=3.6in]{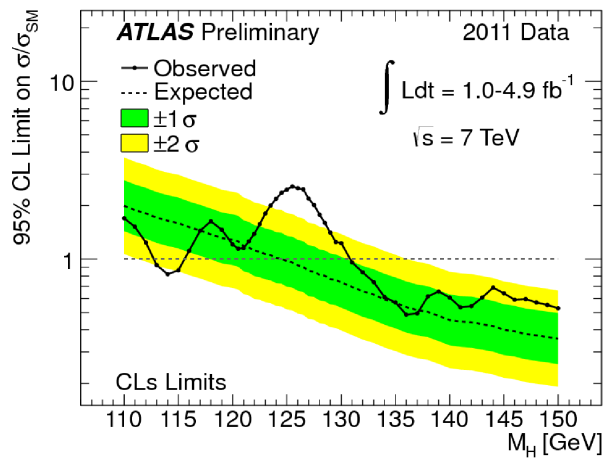}} }
\caption{Experimental limits from ATLAS on SM Higgs production in the mass range 110--150 GeV. The solid curve reflects the observed experimental limits for the production of Higgs of each possible mass value (horizontal axis) \cite{LHCATLAS}.} 
\label{Atlas}
\end{figure}

The results presented the same day by the ATLAS collaboration \cite{LHCATLAS} seem to reach
an analogous conclusion. They restrict, with up to 
4.9 fb$^{-1}$, the search for the Higgs boson to the mass range 
115--130 GeV (Fig.~\ref{Atlas}). Like CMS, 
they find an excess in several independent channels compared to the 
SM prediction, but this is still not conclusive. 
As the ATLAS spokesperson Fabiola Gianotti said: 
`This excess may be due to a fluctuation, but it could also be 
something more interesting. 
We cannot conclude anything at this stage. We need more study and 
more data. Given the outstanding performance of the LHC this year, 
we will not need to wait long for enough data and can look forward to resolving this puzzle in 2012'.

%% file: Ch2.tex

{\Chapter{Little Higgs with a light $T$ quark}\label{Ch2}}


The hierarchy problem has been
the main reason to expect new physics below 
$\Lambda\approx 1$ TeV. This new physics would extend
the range of validity of the model up to larger scales. 
In particular, 
extensions like supersymmetry,
technicolor or, more recently, extra dimensions could do
the job and rise the cut-off of the SM up to the 
the Planck or the Grand Unification scale.

This point of view, however, has become increasingly uneasy 
when facing the experimental evidence. Flavor physics, electric 
dipole moments and other precision EW observables 
suggest that, if present, the sfermion masses, the technicolor
gauge bosons, or the Kaluza-Klein excitations of the standard
gauge fields are above 5 TeV \cite{Erler:2004nh}. 
To be effective below the TeV and consistent with the data
these models require a {\it per cent} 
fine tuning, whereas the presence of these new particles 
above 5 TeV implies that nature deals with 
the Higgs mass parameter first using a mechanism to 
cancel 30 digits and then 
playing {\it hide and seek} with the last two digits.
It may be more consistent either to presume that 
there should be {\it another} reason explaining this 
{\it little hierarchy} between the EW and the scale of 
new physics or that there is no 
dynamical mechanism that explains {\it any} fine tuning 
in the Higgs mass parameter \cite{Agrawal:1997gf,Arkani-Hamed:2004fb}. 
This second possibility has been
seriously considered after recent astrophysical and
cosmological data have confirmed a non-zero vacuum energy density 
(the preferred value does not seem to be explained by 
any dynamics at that scale), and it will be clearly 
favored if no physics beyond the SM is
observed at the LHC. 

Little Higgs (LH) models would release this tension by providing an explanation
for the gap between the EW scale and the
scale of new physics. It is not that LH
does not imply physics beyond the SM (it does), but being its
objective and its structure more simple
it tends to be more consistent with the data than these other
fundamental mechanisms. The LH idea of the Higgs as a pseudo-Goldstone boson
of an approximate global symmetry 
broken spontaneously at the TeV scale could be
{\it incorporated} into
a SUSY \cite{Roy:2005hg,Berezhiani:2005pb,Csaki:2005fc,Bai:2007tv} or a strongly
interacting theory \cite{Contino:2003ve1,Contino:2003ve2,Contino:2003ve3,Giudice:2007fh}
to explain the {\it little} hierarchy between
the Higgs VEV and the SUSY breaking
scale or the mass of the composite states.
Thus, LH ideas provide an interesting framework with natural 
cancellations in the scalar sector and new symmetries that 
protect the EW scale and define consistent
models with a cut-off as high as $\Lambda$ $\approx$ 10 TeV, 
scale where a more fundamental mechanism would manifest.
The most important consequence  would be that  
all the new physics to be explored at the LHC would
be described by the LH framework.

More precisely, in LH models the scalar sector has a (tree-level) 
global symmetry that is broken spontaneously at
a scale $f\le 1$ TeV. The SM Higgs doublet is then a Goldstone boson (GB) of the broken
symmetry, and remains massless and with a flat potential at that scale. 
Yukawa and gauge
interactions break explicitly the global symmetry. However, the models 
are built in such a
way that the loop diagrams giving non-symmetric contributions must 
contain at least two
different couplings. This collective breaking keeps the Higgs sector 
free of one-loop quadratic
top-quark and gauge contributions (see \cite{Perelstein:2005ka,Schmaltz:2005ky}
 for a recent review). Two types of models
have been extensively considered in the literature: the {\it littlest} based on $SU(5)$ \cite{Arkani-Hamed:2001nc,delAguila:2008zu,delAguila:2010nv} 
and the {\it simplest} based on $SU(3) \times SU(3)$ \cite{Schmaltz:2004de,delAguila:2011wk}. We will work on 
the latter in this chapter.

These LH models, however, {\it suffer} tensions that push the
value of $f$ in different directions.
A scale low enough to solve the little hierarchy problem
suggests $f\le 1$ TeV. On the other hand, LH models introduce new gauge
vector bosons of mass proportional to $f$ that mix with the EW bosons. 
This mixing implies $f\ge 3$ TeV in order to be consistent with precision data.
The generic solution to this problem 
proposed in the literature 
is the presence of a discrete symmetry known as $T$ parity. 
Such symmetry forbids the mixing of particles of different
parity and implies acceptable corrections to the EW observables even
for $f\approx 700$ GeV. 

In this chapter we will propose an alternative solution. 
We will see that in the simplest model it 
is possible to {\it separate} the mass of the extra $T$ quark,
which is responsible for the cancellation of the dominant top-quark 
corrections to $\mu^2$, from the mass of  
the extra vector bosons. A scenario with a 600 GeV $T$ quark and
2 TeV extra gauge bosons would be efficient both to cancel dangerous
quadratic corrections and to avoid excessive mixing with the standard
vector bosons.
 
After a short review of the simplest LH model, we show how to 
accommodate such scenario, and we 
analyze its phenomenological implications in Higgs and top-quark physics. 
In addition, we show that the collective breaking of 
the global symmetry in the original model
is not adequate to give an acceptable mass for the Higgs boson. We study
the one-loop effective potential and show that an {\it approximate}
symmetry (as opposed to the one broken {\it collectively}) implies
the separation between the $T$ and the $Z'$ and $W'$ masses and 
could also solve the insufficient value of the Higgs mass in the usual model. 
This chapter is based on the two 
publications \cite{Barcelo:2007if,Barcelo:2008je}.

\Section{Global symmetries and Goldstone bosons}

The generic idea is that the Higgs appears as a 
GB of a global
symmetry broken spontaneously above the EW scale. If the symmetry is exact, 
the GBs are massless and 
their couplings are only derivatives. 
Let us see a few examples as an introduction.

\subsection*{The U(1) case}

Consider a theory with a single complex scalar field $\phi$
with potential $V(\phi^* \phi)$ invariant 
under the global $U(1)$ symmetry transformations
$\phi \rightarrow e^{i \alpha} \phi\,.$
If the minimum of the potential is at some distance $f$ from the origin we may express $\phi$ as
\begin{equation}
\phi(x)=\left( f + {r(x)\over \sqrt{2}}\right) e^{i\theta(x)/f}\,.
\end{equation}
Rewriting the Lagrangian in terms of these new fields we find
that the radial mode $r$ (with mass of order $f$) is invariant under $U(1)$ whereas $\theta$ shifts 
\beq 
\theta \rightarrow \theta +\alpha\,. 
\eeq 
The potential does not change with $\theta$ (it has a flat direction), implying that $\theta$ is massless and without 
self-interactions: it is the GB of the broken global symmetry.

\begin{figure}[htb]
\vskip 0.2truein \centerline{
{\includegraphics[width=2in]{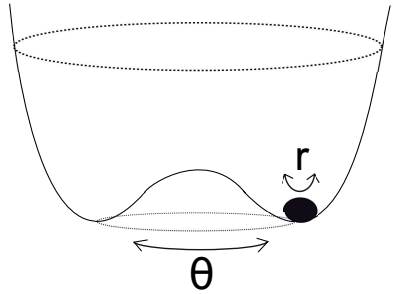}}} \vskip- 0.1truein
\caption{The `Mexican hat' potential for $\phi$. The black dot represents the
vacuum expectation value $f$, $r$ is the radial mode and $\theta$ the Goldstone
boson.} \label{fig:mexhat}
\end{figure}

\subsection*{The SU(N) case}

Generalizing to the non-Abelian case, we will
find one GB for each independent broken symmetry. The VEV of a single fundamental field
$\phi$ of $SU(N)$ will always break  
$SU(N)$ to $SU(N-1)$. Since all the possibilities are equivalent (related by the $SU(N)$ symmetry),
we may take the lowest component as the only one with a nontrivial VEV. The number of broken generators
is the dimension of $SU(N)$ minus the dimension of $SU(N-1)$, 
\beq
[N^2-1]-[(N-1)^2-1]=2N-1\,. 
\eeq
The GBs can be parametrized 
\begin{equation}
 \phi= e^{i {\pi^a(x)T^a}/f}\phi_0= \exp \left\{{i\over f} \left(\begin{array}{c|c} & \pi_1\\
& \vdots \\ & \pi_{N-1} \\ \hline \pi^*_1 \cdots \pi_{N-1}^* &
\pi_0/\sqrt{2}
\end{array}\right)
\right\} \left(\begin{array}{c} 0 \\ \vdots \\ 0 \\ f +{r(x)\over
\sqrt{2}} \end{array}\right) \equiv e^{i \Pi/f} \phi_0 \;,
\label{BNG}
\end{equation}
where $\pi^a(x)$ are real fields and $T^a$ are the generators of the broken symmetries.
The field $\pi_0$ is real while the fields $\pi_1 \cdots
\pi_{N-1}$ are complex. The last
equality defines a convenient short-hand notation.
Under an unbroken $SU(N-1)$ transformation the GBs go to
\begin{equation}
\phi \rightarrow U_{N-1}\ \phi = (U_{N-1}\ e^{i \Pi}\
U_{N-1}^\dagger)\, U_{N-1}\ \phi_0 = e^{i (U_{N-1}  \,\Pi\,
U_{N-1}^\dagger)}\, \phi_0\;, 
\end{equation}
where we have used that $\phi_0$
is invariant under the unbroken $U_{N-1}$. Therefore, the GBs
transform in the usual {\it linear} way,
\begin{equation} 
\Pi \rightarrow U_{N-1} \, \Pi\, U_{N-1}^\dagger\;.
\end{equation}
More explicitly, the unbroken $SU(N-1)$ transformations are
\begin{equation} 
U_{N-1} = \left(\begin{array}{cc} \hat U_{N-1} & 0 \\  0 & 1 
\end{array}\right)\;.
\end{equation}
The real GB $\pi_0$ is then a singlet, whereas the $N\!-\!1$ complex
GBs transform as
\beqa 
\left(\begin{array}{cc|c} 0& & \vec \pi \\ &&\\ \hline \vec
\pi^\dagger& & 0
\end{array}\right) \rightarrow U_{N\!-\!1} \left(\begin{array}{cc|c}
0& & \vec \pi \\ &&\\ \hline \vec \pi^\dagger& & 0
\end{array}\right) U_{N\!-\!1}^\dagger =
\left(\begin{array}{cc|c} 0& & \hat U_{N\!-\!1} \vec \pi \\
&&\\ \hline \vec \pi^\dagger \hat U_{N\!-\!1}^\dagger& & 0
\end{array} \right) \;,
\eeqa
where we have used a vector notation $\vec \pi$ to represent the $N\!-\!1$ complex GBs as
a column vector. Therefore, $\vec \pi$ transforms in the fundamental
representation of $SU(N-1)$,
\begin{equation}
\vec \pi \rightarrow \hat U_{N-1} \vec \pi\;.
\end{equation}
Under the broken symmetry we have
\beqa 
\phi
\rightarrow U \, e^{i \Pi} \, \phi_0 &=& \exp \left\{i
\left(\begin{array}{cc} 0 & \vec \alpha \\ \vec \alpha^\dagger & 0
\end{array}\right) \right\} \exp\left\{i \left(\begin{array}{cc} 0 &
\vec \pi \\ \vec \pi^\dagger & 0 \end{array}\right) \right\}
\phi_0  \nonumber \\
&\equiv& \exp\left\{i \left(\begin{array}{cc} 0 & {\vec \pi'} \\
{{\vec \pi'}}\,\!^\dagger & 0 \end{array}\right) \right\}
U_{N-1}(\vec\alpha, \vec\pi) \
\phi_0  \nonumber \\
&=& \exp\left\{i \left(\begin{array}{cc} 0 & {\vec \pi'} \\ {{\vec
\pi}'}\,\!^\dagger & 0 \end{array}\right) \right\} \phi_0\,,
\label{eq:brokentrafo} 
\eeqa
where we have used that any  $SU(N)$ transformation
can be written as the product of a transformation in the \textit{coset} $SU(N)/SU(N-1)$ times
an $SU(N-1)$ transformation. The $U_{N-1}(\vec \alpha, \vec \pi)$ transformation,
which depends on $\vec \alpha$ and $\vec \pi$, leaves $\phi_0$ invariant and can therefore be removed.
For infinitesimal transformations one obtains
\beqa 
\vec \pi \rightarrow \vec \pi' =
\vec \pi + \delta \vec \alpha\,. 
\eeqa
Like in the $U(1)$ case, the shift symmetry ensures that GBs define flat directions in the potential and can
only have derivative interactions.

\Section{The simplest Little Higgs model}

The scalar sector of the simplest LH model \cite{Schmaltz:2004de} contains two triplets, $\phi_1$ and $\phi_2$, under a global
$SU(3)_1\times SU(3)_2\times U(1)_\chi$\footnote{See the next section to understand the reason to add the $U(1)_\chi$ symmetry.} symmetry:
\beq
\phi_1\rightarrow e^{i\theta^a_1 T^a} \phi_1\;,\;\;\;\;
\phi_2\rightarrow e^{i\theta^a_2 T^a} \phi_2\;,
\eeq
where $T^a$ are the $SU(3)$ generators. 
It is then assumed that these triplets get 
VEVs $f_{1,2}$ and break the global symmetry
to $SU(2)_1\times SU(2)_2\times U(1)_Y$. The spectrum of scalar fields
at this scale will consist of 10 massless modes (the GBs of the 
broken symmetry) plus two massive fields (with masses
of order $f_1$ and $f_2$). 
In particular, if the two VEVs are 
\beq
\langle \phi_{1}\rangle = 
\left(\begin{array}{c} 0 \\ 0 \\ f_1  \end{array}\right)\; , \;\;\;\;
\langle \phi_{2}\rangle = 
\left(\begin{array}{c} 0 \\ 0 \\ f_2  \end{array}\right)\;,
\eeq
it is possible to parametrize the fields like
\beq 
\phi_{1}=e^{+i \Theta'}
e^{+ i\; {f_2\over f_1} \Theta } \left(\begin{array}{c} 0 \\  0 \\
f_1+ \displaystyle{r_1\over \sqrt{2}}
\end{array}\right)\,,\;\;\;
\phi_{2}= e^{+i \Theta' }e^{- i\;{f_1\over f_2} \Theta} \left(\begin{array}{c} 0 \\
0 \\ f_2+ \displaystyle{r_2\over \sqrt{2}}
\end{array}\right)\,,
\label{paramet} \eeq where \beq
\Theta'= {1\over f}\,
\left(\begin{array}{ccc} \eta'/\sqrt{3} & 0 &h' \\
0& \eta'/\sqrt{3} & \hat{h} \\
h^{'\dagger} & \hat{h}^\dagger & -2\eta' /\sqrt{3} \end{array}\right)\,,\;\; \Theta= {1\over f}\;
\left(\begin{array}{ccc} \eta/\sqrt{2} & 0 &h^0 \\
0& \eta/\sqrt{2} & h^- \\
h^{0\dagger} & h^+ & \eta /\sqrt{2} \end{array}\right)\,,
\label{theta1} 
\eeq  
$f=\sqrt{f_1^2+f_2^2}$ and $r_{1,2}$ are massive scalars
of order $f_{1,2}$, respectively. \\

If now, in order to {\it remove} some of the GBs, the diagonal combination of  
$SU(3)_1\times SU(3)_2$ and the $U(1)_\chi$ are made local, i.e.,
\beq
\phi_{1(2)}\rightarrow e^{i\theta^a(x) T^a} \phi_{1(2)}\;, 
\eeq
then the VEVs also break the local symmetry $SU(3)_L \times
U(1)_\chi$ to the SM group $SU(2)_L\times U(1)_Y$.
In the unitary gauge, 5 GBs, the ones in $\Theta'$, disappear. 
The other 5 GBs (the complex doublet $(h^0\; h^-)$ and the CP-odd $\eta$) are
parametrized as 
\beq 
\phi_{1}=
e^{+ i\; {f_2\over f_1} \Theta (x) } \left(\begin{array}{c} 0 \\  0 \\
f_1+ \displaystyle{r_1\over \sqrt{2}}
\end{array}\right)\;,\;\;\;\;
\phi_{2}= e^{- i\;{f_1\over f_2} \Theta(x)} \left(\begin{array}{c} 0 \\
0 \\ f_2+ \displaystyle{r_2\over \sqrt{2}}
\end{array}\right)\,,
\label{paramet} 
\eeq
with $\Theta(x)$ defined in Eq.(\ref{theta1}).

\subsection*{Color and hypercharge}
 
The addition of the gauge $SU(3)_C$ symmetry is straightforward. The hypercharge results from the combination
\beq 
Y={-1\over \sqrt{3}}T^8+X \,,
\eeq
where $X$ is the generator of $U(1)_\chi$ and $T^8$ the charge
\beqa
T^8={1\over 2\sqrt{3}} \left( \begin{array}{ccc} 1 &  & \\ & 1 & \\
& & -2
\end{array} \right) \,
\eeqa
in $SU(3)_L$.
With this embedding of the SM symmetry the components in a 
$SU(3)$ triplet are
\beq
\phi_i={\bf 3}_{-1/3}= \left(\begin{array}{c} \phi^0 \\  \phi^- \\
s^0
\end{array}\right)\,.
\eeq

\subsection*{Exotic matter content}

The fermion content must be accommodated in complete multiples of the local $SU(3)_C \times SU(3)_L \times U(1)_\chi $ gauge symmetry. In particular,
the $SU(2)_L$ doublets become triplets:
\beq
\left(\begin{array}{c} t \\ b \end{array}\right) , \;
\left(\begin{array}{c} \nu
\\ e  \end{array}\right) \;\; \longrightarrow \;\; \Psi_Q=\left(\begin{array}{c} t \\ b \\
T \end{array}\right) , \; \Psi_L=\left(\begin{array}{c} \nu 
\\ e \\ N   \end{array}\right)\;,
\eeq
where $T$ and $N$ are singlets under $SU(2)_L$ with electric charge
$Q_T=\frac{2}{3}$ and $Q_N=0$, respectively. 
The Yukawa sector must give 
these extra fields a mass at the scale $f$. To see how it works, we will just discuss 
the third family, being the arguments for the lighter generations analogous. Instead of a quark singlet $t^c$ we will add two of them,
$t^c_1$ and $t^c_2$, together with a neutral lepton singlet $n^c$.

$(i)$ In the top-quark sector one assumes 
\beq 
-{\cal L}_t =
\lambda_1\; \phi_1^\dagger \Psi_Q t_1^c + \lambda_2 \;\phi_2^\dagger
\Psi_Q t_2^c + {\rm h.c.}\;, \label{topcolectivo} 
\eeq 
where, from now on, all fermions are two-component 
spinors. As we will see in the next section, 
these two Yukawas are necessary to give mass both to the top and the $T$ quarks.
Moreover, if $\lambda_1$ or $\lambda_2$ were zero the global symmetry protecting the Higgs mass
would be exact in this sector. Therefore, diagrams containing only one of the two couplings 
do not break the symmetry and do not contribute to $\mu^2$. 
\begin{figure}[!h]
\begin{center}
\includegraphics[width=65mm]{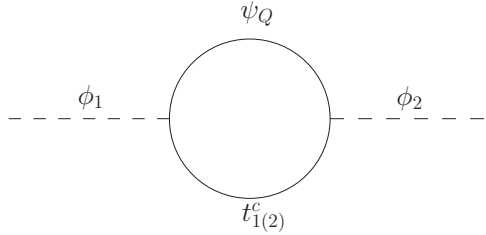}
\caption{Forbidden diagrams in the top quark sector.}
\label{yukawaeps}
\end{center}
\end{figure}
This mechanism is known as
\textit{collective symmetry breaking}. The one-loop
diagram in Fig.~\ref{yukawaeps} breaks the global $SU(3)_1\times
SU(3)_2$ symmetry and would introduce quadratic contributions
to $\mu^2$, but it 
is absent if $\lambda_{1,2}$
are the only non-zero couplings. At one loop the dominant contribution to $\mu^2$ comes from the diagram in Fig.~\ref{top2}, which introduces a logarithmic correction 
$ \sim m_T^2\ln\Lambda$ that would be acceptable.
\begin{figure}[!h]
\begin{center}
\includegraphics[width=55mm]{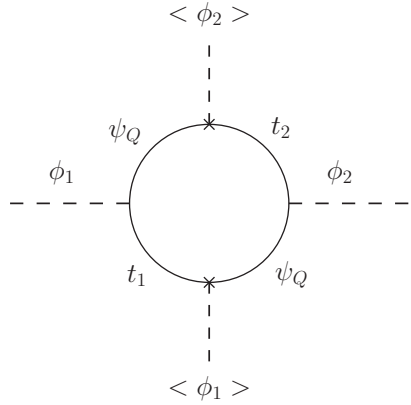}
\caption{Diagram connecting $\phi_1$ and $\phi_2$.}
\label{top2}
\end{center}
\end{figure}

\begin{figure}[!h]
\begin{center}
\includegraphics[width=65mm]{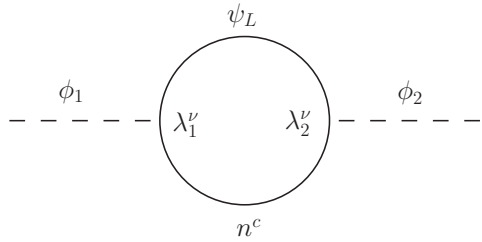}
\caption{Forbidden diagram in the lepton sector.}
\label{leptonica.eps}
\end{center}
\end{figure}

$(ii)$ The lepton sector with a massive $N$ reads,
\beq 
-{\cal L}_\nu = \lambda^\nu_1\; \phi_1^\dagger \Psi_L n^c
+{\rm h.c.}\;. 
\eeq
Diagrams with either $\phi_1$ or $\phi_2$ are allowed,
but not both simultaneously. If both terms appear we would find diagrams
of the kind of Fig.~\ref{leptonica.eps} giving quadratic contributions 
to $\mu^2$.

Summarizing, the matter content is three families with quantum numbers under $SU(3)_C \times SU(3)_L \times U(1)_\chi$:
\beqa
\begin{array}{cc}
\Psi_Q=({\bf 3},{\bf 3})_{1\over 3} & \Psi_L=({\bf 1},{\bf 3})_{-{1\over 3}} \\
b^c=({\bf \bar{3}},{\bf 1})_{1\over 3}& e^c=({\bf 1},{\bf 1})_1 \\ 2 \times
t^c=({\bf \bar{3}},{\bf 1})_{-{2\over 3}} & n^c=({\bf 1},{\bf 1})_0
\end{array}    \eeqa

In the gauge-boson sector we have 8 massive particles plus the photon as the
$SU(3)_L \times U(1)_X$ symmetry breaks into $U(1)_{EM}$. Five of these massive vector fields are neutral and the other four are charged. Of course, the
EW gauge bosons are included among them.

\Section{A light $T$ quark}

The purpose of Little Higgs models is to protect the mass squared 
of the Higgs from one-loop quadratic 
contributions using a global symmetry, 
rising the {\it natural} cut-off of the theory up 
to $\approx 10$ TeV.
This requires an extra $T$ quark lighter than 1 TeV. On the other hand,  
the extra vector bosons that have been introduced to absorb 
the extra GBs of the 
global symmetry must be heavy enough to avoid  unacceptable four 
fermion operators and, specially, a too large mixing with the 
EW gauge bosons.
The value of this mixing is very constrained by LEP data and 
other precision experiments
and requires $f= \sqrt{f_1^2+f_2^2} \geq 3$ TeV \cite{Schmaltz:2004de,Casas:2005ev,Han:2005dz,Marandella:2005wd}. 
However, this condition is not necessarily fulfilled with
two VEVs of the same order; one could have, 
for example, $f_2\geq 3$ TeV and the other VEV significantly smaller. 
It turns out that in order
to keep  a light 
$T$ quark the latter choice is more convenient.
At the scale $f$ and neglecting the EW VEV
Eq.(\ref{topcolectivo}) implies a massless top quark and a $T$ quark 
with mass $m_T$:
\beq 
-{\cal L}_t \supset \lambda_1\; f_1 T t_1^c +
\lambda_2\; f_2 T t_2^c + {\rm h.c.} = m_T\; T T^c
 + {\rm h.c.}\;,
\label{eq6}
\eeq 
where $m_T=\sqrt{\lambda_1^2 f_1^2 + \lambda_2^2
f_2^2}$ and $\;T^c= s_\alpha\; t_1^c+c_\alpha\; t_2^c$, with
$s_\alpha=\lambda_1 f_1/\sqrt{\lambda_1^2 f_1^2+\lambda_2^2 f_2^2}$.
To obtain a mass for the $T$ quark smaller than $f$ is possible if
$f_1\ll f_2$ and $\lambda_2\ll \lambda_1$.
Once the Higgs field gets a VEV, the top quark
mixes with the extra $T$ quark defining 
the mass eigenstates 
\beq
t'=c_\theta\; t-s_\theta\; T \;,\;\;\;\;
T'=s_\theta\; t+c_\theta\; T \;,
\eeq
where we denote $V_{Tb}\equiv s_\theta$.
 
The mixing of $T$ with the top quark introduces gauge couplings 
with the $W$ and $Z$ bosons for the extra $T$ quark:
\beqa
{\cal L}_W &=& -{g\over \sqrt{2}}\;
\overline t \sigma^\mu b \;
W^+_\mu + {\rm h.c.}\nonumber \\
&=& -{g\over \sqrt{2}}
\left( \sqrt{1-V^2_{Tb}} \;\overline t'\sigma^\mu b
+ V_{Tb}\; \overline T'\sigma^\mu b \right)
W^+_\mu + {\rm h.c.}\;,
\eeqa

\beqa
{\cal L}_Z &\supset & -{g\over 2 c_W}
\left( {\begin{array}{cc}
\overline t & \overline T \\
\end{array}} \right)
\sigma^\mu
\left( {\begin{array}{cc}
1 & 0 \\
0 & 0 \\
\end{array}} \right)
\left( {\begin{array}{c}
t \\
T \\
\end{array}} \right)
Z_\mu \nonumber \\
&\approx &
-{g\over 2 c_W}
\left( {\begin{array}{cc}
\overline t' & \overline T' \\
\end{array}} \right)
\sigma^\mu
\left( {\begin{array}{cc}
1-V^2_{Tb} & V_{Tb} \\
V_{Tb} & V^2_{Tb} \\
\end{array}} \right)
\left( {\begin{array}{c}
t' \\
T' \\
\end{array}} \right)
Z_\mu \;.
\eeqa
As we see, we obtain top-quark flavor-changing neutral currents coupled 
to the $Z$ boson. This same kind of currents also appear in Yukawa couplings 
with the Higgs boson. In this regard, it is important to mention
that in models with a vectorlike $T$ quark it is usually 
assumed that 
$T$ couples to the Higgs boson only through mixing 
with the top quark. This implies a Yukawa coupling 
$V_{Tb} m_T/v$ \cite{Aguilar-Saavedra:2002kr}. In our model, however, 
the Higgs couples
to $T$ and $T^c$ even if the mixing term $V_{Tb}$ is zero, 
since $h$ includes
an order $m_t/m_T$ singlet component (see next section).\\

\begin{figure}
\begin{center}
\vspace{0.8cm}
\includegraphics[width=80mm]{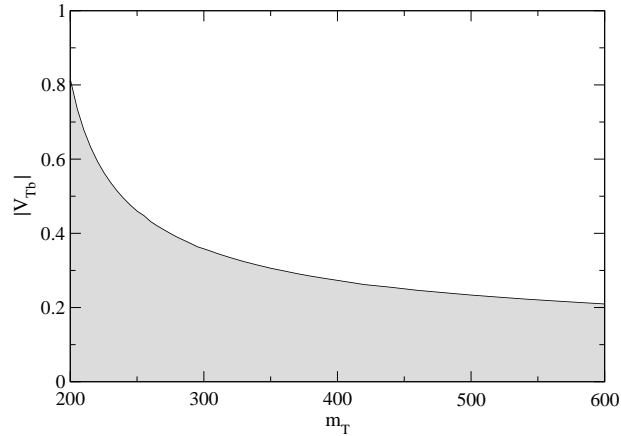}
\caption{Allowed values of the coupling $|V_{Tb}|$ of the extra $T$ quark
(shaded) as a function of its mass \cite{Aguilar-Saavedra:2002kr}. \label{VTb}}
\end{center}
\end{figure}

The presence of a $T$ quark has phenomenological consequences.
In \cite{Aguilar-Saavedra:2002kr}
Aguilar-Saavedra discusses the constraints on $V_{Tb}$ from precision 
observables as a function of the $T$-quark mass (see Fig.~\ref{VTb}). 
It is remarkable that the limits vanish  
when $m_T \rightarrow m_t$; a heavier $T$ quark would take (through
mixing) part of the $t$ quark interaction to the EW bosons changing
the standard corrections to EW observables.

\Section{Little Higgs or extra singlet model?}

To break the EW symmetry the Higgs field $h^0$ has to get a 
potential (see next sections) defining a non-zero VEV
\beq
\langle h^0 \rangle = u/\sqrt{2}\;. \eeq
This would translate into the following triplet VEVs:
\beq
\langle \phi_{1}\rangle = e^{+ i\; {f_2\over f_1} \langle \Theta \rangle }
\left(\begin{array}{c} 0 \\ 0 \\
f_1
\end{array}\right) \; , \;\;\;\; \langle \phi_{2}\rangle =
e^{- i\; {f_1\over f_2} \langle \Theta \rangle }\left(\begin{array}{c} 0 \\ 0 \\ f_2
\end{array}\right)\;,
\eeq
where
\beq \langle \Theta \rangle= {1\over f}\;
\left(\begin{array}{ccc} 0 & 0 &u/\sqrt{2} \\
0& 0 & 0 \\
u/\sqrt{2} & 0 & 0 \end{array} \right)\ \;.
\label{theta}
\eeq
Working out the expression we obtain
\beq
\langle \phi_{1}\rangle = \left(\begin{array}{c} i f_1 s_1 \\ 0 \\
f_1 c_1
\end{array}\right) \; , \;\;\;\; \langle \phi_{2}\rangle =
\left(\begin{array}{c} -i f_2 s_2 \\ 0 \\ f_2 c_2
\end{array}\right)\;, \eeq
with 
\beq s_1\equiv \sin {u f_2\over
\sqrt{2} f f_1} \; , \;\;\;\; s_2\equiv \sin {u f_1\over \sqrt{2} f
f_2} \;.
\eeq
In previous literature the way to carry out this point has usually been
through a first order approximation
\beq
\langle \phi_{1}\rangle \sim \left(\begin{array}{c} i u f_2\over
{\sqrt{2} f} \\ 0 \\
f_1
\end{array}\right) \; , \;\;\;\; \langle \phi_{2}\rangle \sim
\left(\begin{array}{c} -i u f_1\over{ \sqrt{2} f
} \\ 0 \\ f_2
\end{array}\right)\;,
\eeq
that may not be good if $f\approx f_2\gg f_1 > u / \sqrt{2}$
and that would make our argument below less apparent.  

Since the upper components of the triplets transform as 
a $SU(2)_L$ doublet, it is clear that in order 
to get the mass of the $W$ and $Z$
bosons we need
\beq \sqrt{f_1^2 s_1^2 + f_2^2s_2^2} = {v\over
\sqrt{2}} \approx 174\;{\rm GeV}\;. \eeq 
If $f\approx f_2\gg f_1$ we obtain
\beq s_1\approx \sin {u \over
\sqrt{2} f_1} \; , \;\;\;\; s_2\approx  {u f_1\over \sqrt{2}
f^2}\;, \eeq 
and the EW scale would just be
\beq {v\over
\sqrt{2}}\approx f_1 s_1\;. \eeq

The Higgs $h$ is obtained expanding $h^0$ (in the unitary gauge) 
as $h^0=(u+h)/\sqrt{2}$. We find that $h$ has a singlet and a doublet component:
\beq
\phi_{1} =
\left(\begin{array}{c} if_1 (s_1 \cos \displaystyle{h \over
\sqrt{2}f_1}
+ c_1\sin \displaystyle{h\over\sqrt{2} f_1}) \\ 0 \\
f_1 (c_1 \cos \displaystyle{h \over \sqrt{2} f_1} - s_1 \sin
\displaystyle{h\over\sqrt{2} f_1}) \end{array}\right) \approx
\displaystyle{1 \over \sqrt{2}}
\left(\begin{array}{c} i c_1 h \\
0 \\ - s_1  h \end{array}\right) + ... \eeq
If $f_1$ is much larger than the EW scale, $s_1$ will be small and $h$
mainly a doublet. However, if $f_1$ is  $ \sim v/\sqrt{2}$, the singlet
component $s_1$ grows large. 
In this case, the doublet component 
lost by $h$ becomes part
of the radial scalar $r_1$ whose mass is $\sim f_1$. \\

This is a generic issue of the LH models. The scale $f$ of a broken global
symmetry is always defined by the VEV of a $SU(2)_L$ singlet, 
implying
a massive mode (a radial singlet of mass $m_r\sim f$) 
in addition to the GBs (the Higgs). 
Then, the EW symmetry breaking mixes the Higgs 
with this massive singlet. This mechanism is at work in any framework
were the Higgs is a pseudo-GB of a global symmetry, including 
composite Higgs models: the resulting Higgs will always include 
a singlet
component of order $v/(\sqrt{2}f)$. In our model the effect is
even more relevant because the scale of global symmetry breaking is
split in two, $f_1$ and $f_2$, and it is the lighter VEV the one
defining the singlet component.

\begin{figure}
\begin{center}
\includegraphics[width=90mm]{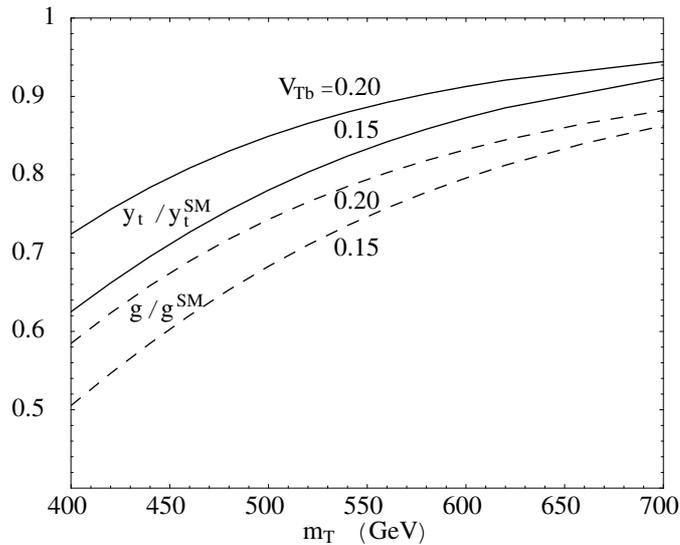}
\caption{Suppression of the top quark (solid) and the gauge 
(dashes) couplings versus the SM values for 
$V_{Tb}=0.20,0.15$ and different values of $m_T$.
\label{fLH2}}
\end{center}
\end{figure}

Let us now see the consequences of having a singlet component 
in the Higgs field.
We consider a Higgs field $h$ and a massive scalar $r_1$ that
include both $SU(2)_L$ doublet $d$ and  singlet $s$ components:
\beqa d & = & c_1h+s_1r_1 \nonumber \\ s & = & -s_1h+c_1r_1
\eeqa
The scalar $h$ gets a VEV $u$, but the 
$Z$ boson only couples to the doublet component in $h$ (suppressed 
by a factor of $c_1^2$), getting a mass
\beq
{\cal L}\supset\frac{g^2 \langle h \rangle^2c_1^2}
{8\cos^2\theta_W}Z_\mu Z^\mu
\eeq
To fit the measured value of $M_Z$ the Higgs VEV $u$
must be larger than the one ($v$) in the SM, to compensate 
the appearance of the cosine multiplying:
\beq
\langle h \rangle = u = \frac{v}{c_1}
\eeq
The gauge coupling of the $Z$ boson with the Higgs boson appears suppressed (see Fig. \ref{fLH2}):
\beq
{\cal L}\supset\frac{g^2h^2c_1^2}{8cos^2\theta_W}Z_\mu Z^\mu \;,
\eeq
i.e., 
\beq
\frac{g}{g^{SM}}=c_1 \;.
\eeq
The same consequence is deduced from the mass of the $W$ 
boson (the $\rho=M_W^2/M_Z^2\cos^2\theta_W$ parameter does not change) 
and the coupling to the Higgs boson.

In the Yukawa top-quark sector we have
\beqa
{-\cal L}&\supset& \frac{y_t}{\sqrt{2}}d t t^c+h.c.\nonumber\\
&=&\frac{y_t}{\sqrt{2}}\left( c_1\left( u+h\right) + s_1 r_1\right) +h.c. \;,
\eeqa
which implies 
\beq
m_t=\frac{y_t c_1 u}{\sqrt{2}}
\eeq
and 
\beq
{-\cal L}\supset 
m_ttt^c+\frac{m_t}{v}c_1htt^c\,.
\eeq
Therefore, the top-quark Yukawa coupling 
is also suppressed by a factor $c_1$ 
respect to the value in the SM.

Summarizing, since the singlet component of $h$ does not couple,  
both its gauge ($g$ and $g'$) and Yukawa ($\sqrt{2} m_f/v$)
couplings will appear suppressed by a 
factor of $c_1$.
These anomalous LH couplings have nothing to do with the 
non-linear realization
of the scalar fields, they just reflect the mixing with the 
scalar singlet
massive at the scale of the global symmetry breaking.
As we mentioned above, in our case $f_1$ can be close to $v/\sqrt{2}$
while consistent with all precision data, and 
the effect may be large enough to be observable 
at the LHC.

Notice also that the Little Higgs $h$, not being a pure
doublet, only unitarizes {\it partially} the SM cross sections 
involving massive vector bosons. In particular, the 
cut-off at $\approx 1$ TeV set by $WW$ elastic scattering would be
{\it moved} up to (1/$s_1$) TeV. Below that scale the massive
scalar $r_1$ (or a techni-$\rho$ in composite models) should complete the unitarization.

A final comment concerns the limit $f_1\rightarrow v/\sqrt{2}$.
The GB $h$ becomes there a pure $SU(2)_L$ singlet, 
and the (unprotected) field $r_1$, massive at the scale $f_1$,  
becomes a doublet and is the {\it real} Higgs that breaks the 
EW symmetry. 
In this limit, the natural cut-off would be the same as in  
the SM, whereas in the general case
with $f_1>v/\sqrt{2}$ it is at $\approx 4\pi f_1$.

\Section{Phenomenology}

The three basic ingredients of the LH models under study here are 
the presence of heavy vector bosons, 
a relatively light $T$ quark, and a sizable singlet component 
in the Higgs field. A fourth ingredient, the presence of the
$CP$-odd light scalar $\eta$ (in Eq.~(\ref{theta1})), also a GB of the global symmetry, 
seems more model dependent (see next section).
Let us briefly analyze the 
phenomenological consequences of these ingredients.

\begin{enumerate}

\item Effects on EW precision observables.

$(i)$ The massive gauge bosons would introduce mixing 
with the standard bosons and four fermion operators. This
could manifest as a shift in the $Z$ mass and other precision data. 
However, none of these effects is observable 
if $f_2 \ge 3$ TeV \cite{Schmaltz:2004de}.

$(ii)$ The effects on EW precision observables due to 
the singlet component of the Higgs field are also 
negligible even if the extra scalar $r_1$ is heavier than the Higgs boson. 
Although the Yukawa coupling of the top 
with the neutral Higgs is here smaller than in the SM, it is
the coupling with the {\it would be} GBs (the scalars 
eaten by the $W$ and $Z$ bosons) what determines
the large top quark radiative corrections, and these 
are not affected by the presence of singlets.

$(iii)$ The bounds on a vectorlike $T$ quark from
precision EW data  
have been extensively studied in the literature, 
we will comment here the results in 
\cite{Aguilar-Saavedra:2002kr} as they apply
to LH models in a straightforward way.

The mixing of the top quark with the $T$ singlet reduces
its coupling with the $Z$ boson. This, in turn, affects the top 
quark radiative corrections (triangle diagrams) to the $Zbb$ 
vertex, which is measured in the partial width $Z$ to  
$b\overline b$ [$R_b=\Gamma(Z\rightarrow b\overline b)/
\Gamma(Z\rightarrow {\rm hadrons})$]
and forward-backward asymmetries.
The heavier $T$ quark also gives this 
type of corrections to the $Zbb$ 
vertex, and for low values of $m_T$ 
both effects tend to 
cancel ({\it i.e.}, if $m_T=m_t$ the vertex $Zbb$ 
is the same as in the SM). For large values of $m_T$ (above
500 GeV) the upper bound on $V_{Tb}$ from precision $b$ 
physics is around 0.2 \cite{Aguilar-Saavedra:2002kr}.

The $T$ quark would also appear in vacuum polarization
diagrams, affecting the oblique parameters $S$, $T$, and 
$U$. For degenerate masses ($m_T=m_t$) the corrections to
$T$ and $U$ vanish for any value of the mixing $V_{Tb}$ 
and the correction to $S$ is small 
($\Delta S\approx -0.16 V_{Tb}$). 
For large values of 
$m_T$ the only oblique parameter with a 
sizable correction is $T$ ($\Delta T\approx 2.7 V_{Tb}$ for 
$m_T=500$ GeV), but the limits on $V_{Tb}$ 
are in this case smaller than
the ones from $R_b$ \cite{Aguilar-Saavedra:2002kr}.

\item The phenomenological impact of these models on Higgs physics
at hadron colliders may be important. The main effects can be 
summarized as follows.

$(i)$ Suppression of the $gg\rightarrow h$ cross section (Fig.~\ref{gg}).
This effect is due to the suppression of the top-quark 
Yukawa coupling relative to the SM value (see Fig.~\ref{fLH2})
and also to the contribution of the extra $T$ quark. 
Although this second factor is numerically less important, it is remarkable that
always interferes destructively in the amplitude: the relative
minus sign versus the top-quark contribution follows from the
cancellation of quadratic corrections to the Higgs 
mass parameter.

\begin{figure}[!h]
\begin{center}
\includegraphics[width=0.3\textwidth]{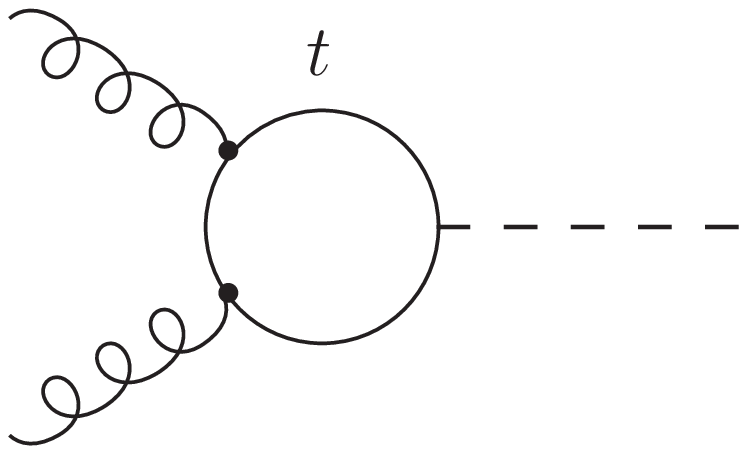}
\includegraphics[width=0.3\textwidth]{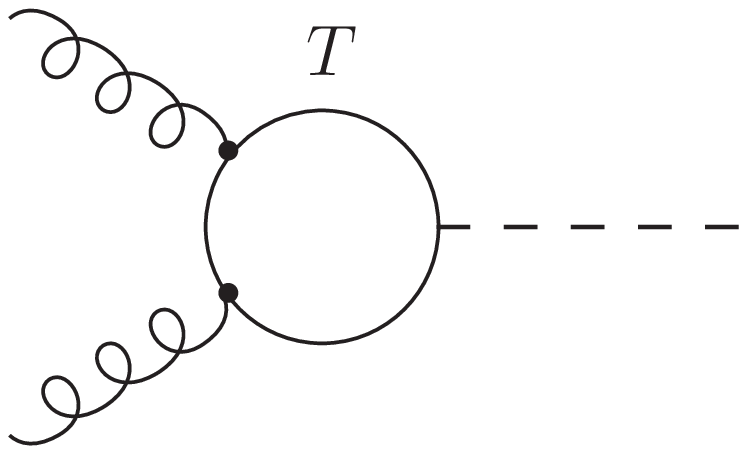}
\caption{Diagrams contributing to $gg\rightarrow H$.}
\label{gg}
\end{center}
\end{figure}

It is easy to obtain approximate expressions for this suppression
factor in the limit of $m_h\ll m_t,m_T$ \cite{Georgi,Rizzo:1979mf}:
\beqa
R_{gg}\equiv{\sigma(gg\rightarrow h)\over \sigma^{SM}(gg\rightarrow h)}
& \approx & \left( {y_t\over y_t^{SM}}+{y_T v\over m_T}\right)^2
\nonumber \\
& \approx & \left( c_1 c_\theta + s_1 s_\theta
-t_1 s^2_\alpha (s_1 c_\theta-c_1 s_\theta)\right)^2
\nonumber \\
& \approx & c_1^2
\eeqa

\begin{figure}
\begin{center}
\includegraphics[width=90mm]{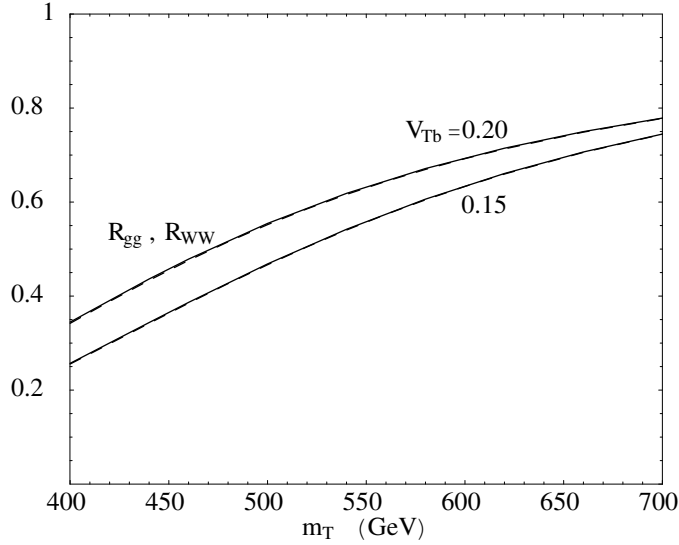}
\caption{Ratios $R_{gg}\equiv\sigma(gg\rightarrow h)/
\sigma^{SM}(gg\rightarrow h)$ (solid)
and $R_{WW}\equiv
\sigma(WW\rightarrow h)/\sigma^{SM}(WW\rightarrow h)$
(dashes) for 
$V_{Tb}=0.20,0.15$ and different values of $m_T$.
$R_{gg}$ and $R_{WW}$ coincide at the $1\%$ level.
\label{f3}}
\end{center}
\end{figure}

In Fig.~\ref{f3} we plot the ratio 
$R_{gg}\equiv\sigma(gg\rightarrow h)/\sigma^{SM}(gg\rightarrow h)$
for different values of $V_{Tb}$ and $m_T$. For $m_h=150$ GeV
the approximation above is good at the $1\%$ level.
This effect, which could {\it hide} the Higgs at the LHC, 
has been recently discussed in general models with scalar
singlets \cite{O'Connell:2006wi,Bahat-Treidel:2006kx} and 
also in the framework of LH models with $T$ parity \cite{Chen:2006cs}.

$(ii)$  Suppression in the production cross sections 
that involve gauge interactions: 
$WW\rightarrow h$, $q \overline q\rightarrow W h$, etc. 
(Fig~\ref{ww}). In Fig.~\ref{f3} we also plot the ratio $R_{WW}\equiv
\sigma(WW\rightarrow h)/\sigma^{SM}(WW\rightarrow h)
\approx c_1^2$. It is remarkable that for $m_h=150$ GeV the suppression
in these cross sections coincides with the one in
$\sigma(gg\rightarrow h)$ at the $1\%$ level.

\begin{figure}[!h]
\begin{center}
\includegraphics[width=0.3\textwidth]{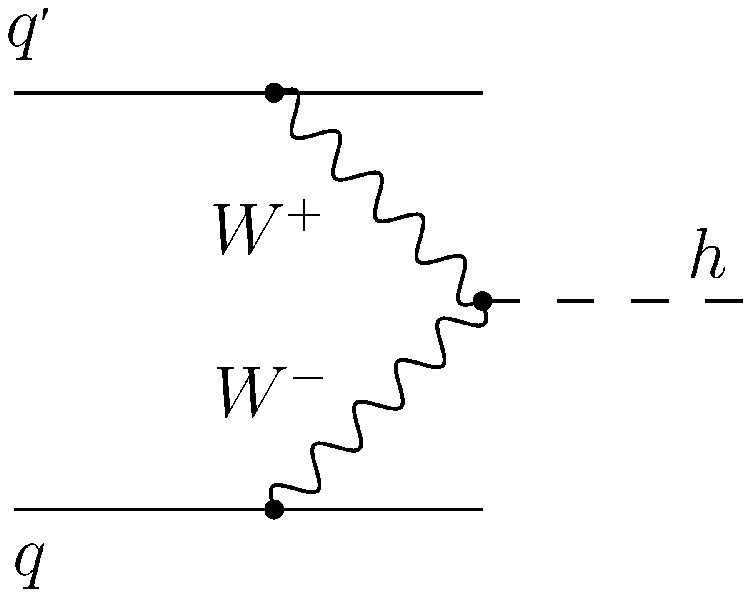}
\hspace{0.6cm}
\includegraphics[width=0.35\textwidth]{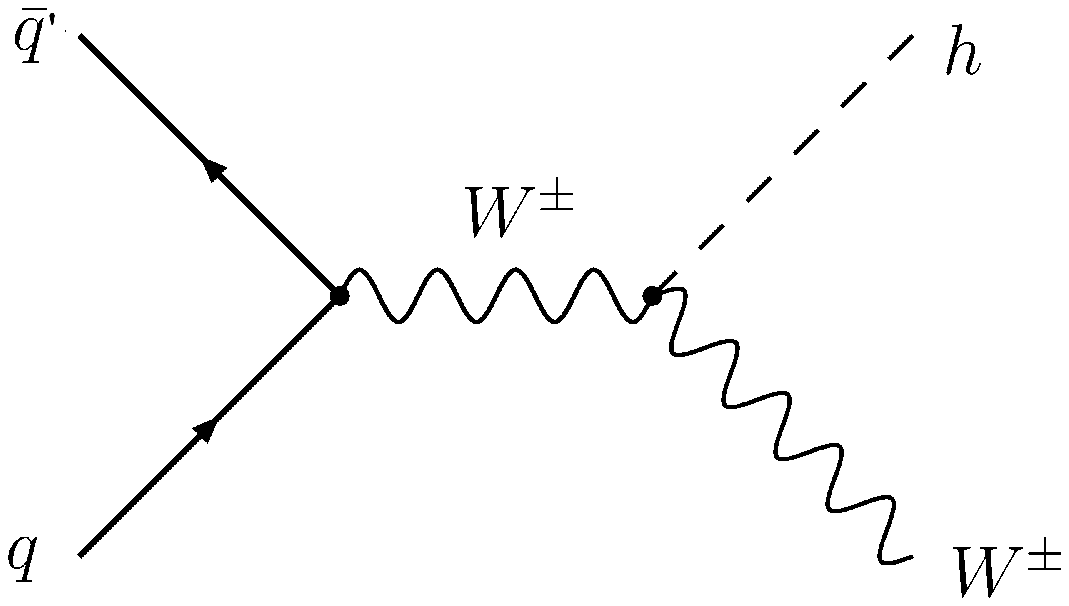}
\caption{Diagrams $WW\rightarrow
h$ and $q \overline q\rightarrow W h$.}
\label{ww}
\end{center}
\end{figure}

$(iii)$ New production channels through $T$-quark 
decay \cite{Han:2005ru}.
A $T$ quark of mass below 600 GeV will be copiously produced
at the LHC. In particular, the cross section to produce 
$T\overline T$ pairs in $pp$ collisions goes from 
$10^4$ fb for $m_T=400$ GeV to $10^3$ fb for $m_T=600$ GeV
\cite{Aguilar-Saavedra:2006gw1,Aguilar-Saavedra:2006gw2}.
Once produced, a $T$ quark may decay into $Wb$, $Zt$, $ht$ and
$\eta t$ \cite{Kilian:2004pp1,Kilian:2004pp2}. 
We find an approximate relation among the
partial widths in the limit of $m_T$ much larger than the 
mass of the final particles:
\beqa
&&\Gamma(T\rightarrow Wb)\approx {\alpha\over 16 s_W^2}\;
V_{Tb}^2 \;{m_T^3\over M_W^2}\nonumber \\ 
&&\Gamma(T\rightarrow Zt)\approx 
{1\over 2}\;\Gamma(T\rightarrow Wb)\nonumber \\ 
&&\Gamma(T\rightarrow ht)\approx  {1\over 2}\;
(c_1^2+{s_1^2\over t_\alpha^2})\;
\Gamma(T\rightarrow Wb)\nonumber \\ 
&&\Gamma(T\rightarrow \eta t)\approx  {1\over 2}\;
(s_1^2+{c_1^2\over t_\alpha^2})\; \Gamma(T\rightarrow Wb)
\eeqa

Notice that the $T$ quark will decay through the 4 channels
with branching ratios that are independent of $V_{Tb}$.
$T\rightarrow W^+ b$ gives the best discovery potential 
for the $T$ quark, whereas the Higgs $h$ will be produced
with a branching ratio close to the $20\%$. The detailed 
signal and background study at the LHC 
in \cite{Aguilar-Saavedra:2006gw1,Aguilar-Saavedra:2006gw2} 
shows that $T\overline T
\rightarrow W^+ b \overline t h\rightarrow W^+b W^- \overline b h$ 
and $T\overline T \rightarrow h t h \overline t
\rightarrow W^+ b W^- \overline b hh$
give a very high statistical significance for the Higgs
(around $10\sigma$ for 30 fb$^{-1}$). We expect similar
results in this model, although the presence of the scalar 
$\eta$ can open new decay channels for the Higgs.
In particular, if $m_h>2m_\eta$ the 
coupling $h\eta\eta$ opens the interesting channel 
$h\rightarrow \eta\eta \rightarrow 4b$ \cite{Cheung:2006nk}
that, together with the suppression in the $hZZ$ coupling, could
loosen considerably the 114 GeV LEP bound on the Higgs mass.

\end{enumerate}

\Section{Global symmetry breaking effects and 
effective potential}

In LH models the global symmetry is not exact, it is broken
at the one-loop level. These loop corrections generate a 
potential for the GBs that should imply the {\it right} 
EW VEV and an acceptable mass for the Higgs boson and for
the extra $\eta$ scalar. In this section we study the
one-loop effective potential for these fields in  
the usual simplest LH model with collective 
symmetry breaking. We will show that such model is not able to 
accommodate the splitting of $f_1$ and $f_2$ together with
an acceptable Higgs potential, and in the next section we
will propose a variation of the model.

The one-loop effective potential
(or Coleman-Weinberg potential) \cite{Coleman:1973jx} for $h$ and $\eta$ 
can be expressed in terms of 
the masses that the different fields would get if these scalars
grow a VEV. Therefore, we will start finding the mass-eigenstates
of both the fermion and the gauge boson sectors.

\subsection*{Fermion masses}

We will just consider the top-quark sector, as the lighter 
fermions give negligible contributions.  
In the case with collective breaking the Yukawas read 
\beq -{\cal L}_t = \lambda_1\; \phi_1^\dagger
\Psi_Q t_1^c + \lambda_2 \;\phi_2^\dagger \Psi_Q t_2^c + {\rm
h.c.}\;, \label{eq5}
\eeq
Once the Higgs gets a VEV we find the mass matrix
\beq -{\cal L}_t \supset \left(
\begin{array}{cc} t & T \end{array} \right) \left(\begin{array}{cc}
\lambda_1 f_1 s_1 &
-\lambda_2 f_2 s_2 \\
\lambda_1 f_1 c_1& \lambda_2 f_2 c_2
\end{array} \right)
\left(\begin{array}{c} t_1^c \\
t_2^c
\end{array} \right)\,, \eeq
with 
\beq s_1(h)= \sin {h f_2\over
\sqrt{2} f f_1} \; , \;\;\;\; s_2(h)= \sin {h f_1\over \sqrt{2} f
f_2} \;.
\eeq
This expression is independent of the VEV of the $\eta$ 
field, which appears as a phase that can be absorbed by the fields
and thus remains massless.

Working out the diagonalization we find the masses of the top and 
the $T$ quarks: 
\beq
m_{t(T)}^2(h)={M^2\over 2} \left( 1 - (+) \sqrt{1-s_{2\alpha}^2\;
s_{12}^2(h)}\right)\;, \eeq where \beqa s_{12}(h)& = & \sin {h
f\over \sqrt{2} f_1 f_2}
\;,\nonumber \\
M^2 & = & \lambda_1^2 f_1^2 + \lambda_2^2 f_2^2
\;,\nonumber \\
s_{2\alpha}& = & {2\lambda_1 \lambda_2 f_1 f_2\over \lambda_1^2
f_1^2 +\lambda_2^2 f_2^2} \;. 
\eeqa
The Coleman-Weinberg expression for the fermion contribution to
the one-loop potential is 
\beq 
V_{top}=-{3\over 16\pi^2} \Lambda^2\;
{\rm Tr}\; [m^\dagger m]+ {3\over 16\pi^2}\; {\rm Tr}\; [(m^\dagger
m)^2 \log \left( \Lambda^2\over m^2 \right) ] \;. \label{vtop} 
\eeq
The global symmetry has some important implications
on $V_{top}$. 
First of all, ${\rm Tr}\; [m^\dagger m]=m_t^2+m_T^2$ is a constant 
(it does not depend on $h$), 
so the quadratic contribution is zero. 
Moreover, up to a constant we can write
\beq 
V_{top}= {3\over
16\pi^2} m_t^4 \log \left( m_T^2\over m_t^2 \right) + {3\over
16\pi^2} \left( m_t^4+m_T^4\right) \log \left( \Lambda^2\over m_T^2
\right)\,. 
\eeq 
This potential can be understood as the usual top quark quartic 
correction below $m_T$ plus a contribution proportional to
\beq 
m_t^4+m_T^4={M^4\over 2}
\left( 2 - s^2_{2\alpha} \right) \; s_{12}^2(h)
\eeq
above that scale. The latter contribution is logarithmically divergent and
it will redefine (renormalize) the quartic coupling to 

\begin{minipage}{0.4\textwidth}
\includegraphics[width=50mm]{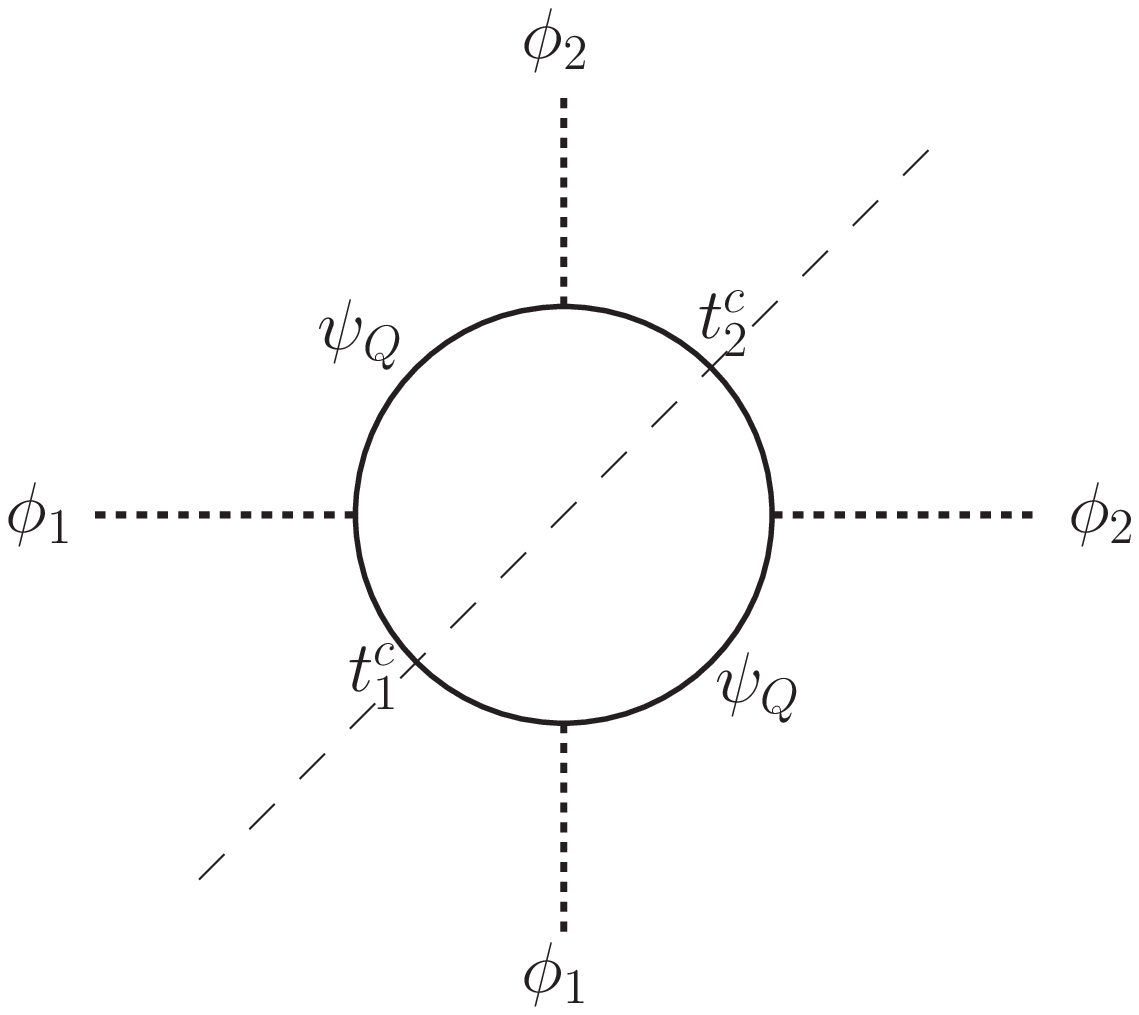}
\label{bah}
\end{minipage}
\begin{minipage}{0.4\textwidth}
\beq
 a\; \left(
\phi_1^\dagger \phi_2 \right)_{\bf 1} \left( \phi_2^\dagger \phi_1
\right)_{\bf 1}\supset a\;  f_1^2 f_2^2 \;\left( 1-
s_{12}^2(h)\right)\;. \label{q11}\nonumber
\eeq
\end{minipage}
\\
where the subindex $1$ indicates that the triplets are 
combined into a $SU(3)$ singlet.
The sensibility of the potential to the ultraviolet 
physics can be taken into account considering $a$
as a free parameter or, equivalently, taking $a=0$ and varying 
the cut-off freely.

\subsection*{Gauge boson masses}

All the masses come from
the covariant kinetic terms 
$(D^\mu\Phi_i)^\dagger (D_\mu\Phi_i)$ of the triplets 
$\Phi_{1,2}$. We will separate the study of the charged and the 
neutral gauge bosons.

$(i)$ In the charged sector we have
\beq
D_\mu\phi_1 \supset -ig \sum_{i=1,2,6,7} A^i_\mu T^i \phi_1
={g f_1\over \sqrt{2}}
\left( \begin{array}{c} 0 \\ 
s_1W_\mu-c_1W'_\mu \\ 0
\end{array} \right)\;,
\eeq
where we have defined
\beq
W={1\over \sqrt{2}}\left( A^1_\mu+i A^2_\mu \right)\;,\;\;
W'={1\over \sqrt{2}}\left( A^7_\mu+i A^6_\mu \right)\;,
\eeq
with an analogous expression for $D_\mu\phi_2$. 
In this basis the mass matrix reads
\beq
{g^2\over 2}
\left( \begin{array}{cc} f_1^2 s_1^2+f_2^2 s_2^2&
f_2^2 s_2 c_2 - f_1^2 s_1 c_1\\ 
f_2^2 s_2 c_2 - f_1^2 s_1 c_1 & f_1^2 c_1^2+f_2^2 c_2^2
\end{array} \right)\;, 
\eeq
and has the eigenvalues
\beq
M^2_{W_1(W_2)}={g^2f^2\over 4} \left(
1 -(+) \sqrt{1- s_{2\beta}^2\; s_{12}^2}\right)\;,
\eeq
where
\beq
s_{2\beta}\equiv 2 {f_1 f_2 \over f^2}\;.
\eeq

Notice that, again, the two masses add to a constant
independent of the Higgs VEV (in $s_{1,2}$), 
which will imply no quadratic
divergences in the potential at the one-loop level.

$(ii)$ In the neutral sector we find
\beqa
D_\mu\phi_1 &\supset & \left(-ig \sum_{i=3,4,5,8} A^i_\mu T^i \phi_1
+ {i g_X \over 3} A^X_\mu \right)\nonumber \\
&=&{g f_1\over 2}
\displaystyle 
\left( \begin{array}{c} s_1 \sqrt{1+t^2} Z_\mu + s_1 
\displaystyle{1-t^2\over
\sqrt{3-t^2}} Z'_\mu - c_1 A^5_\mu -i c_1 A^4_\mu\\ 
0 \\ 
s_1 A^4_\mu + i s_1 A^5_\mu + i 2 c_1 
\displaystyle {1\over \sqrt{3-t^2}} Z'_\mu 
\end{array} \right)\;,
\eeqa
where
\beq
t={g'\over g}= \sqrt{3}\; {g_X\over \sqrt{3g^2+g_X^2}}
\eeq
and
\beqa
Z'_\mu&=&\sqrt{1-{t^2\over 3}}\; A^8_\mu +
{t\over \sqrt{3}}\; A^X_\mu\;,\nonumber  \\
Z_\mu&=&{1\over \sqrt{1+t^2}}\left( A^3_\mu 
+{t^2\over \sqrt{3}}\; A^8_\mu
-t \sqrt{1-{t^2\over 3}}\; A^X_\mu \right)\;.
\eeqa
The mass matrix of $(Z_\mu\;,Z'_\mu\;,A^4_\mu\;,A^5_\mu)$
is then 

\beq
{g^2\over 2}
\left( \begin{array}{cccc} (1+t^2) (f_1^2 s_1^2+f_2^2 s_2^2)&
{(1-t^2)\sqrt{1+t^2}\over \sqrt{3-t^2}}(f_1^2 s_1^2+f_2^2 s_2^2)&
0&-\sqrt{1+t^2}(f_1^2 s_1c_1-f_2^2 s_2 c_2)\\
{(1+t^2)\sqrt{1+t^2}\over \sqrt{3-t^2}}(f_1^2 s_1^2+f_2^2 s_2^2)&
\begin{array}{c} \\
{(1-t^2)^2\over 3-t^2}(f_1^2 s_1^2+f_2^2 s_2^2)+\\
{4\over 3-t^2}(f_1^2 c_1^2+f_2^2 c_2^2)\\ 
\end{array}&
0&{1+t^2\over \sqrt{3-t^2}}(f_1^2 s_1c_1-f_2^2 s_2 c_2)\\
\begin{array}{c} 0 \\ \; \end{array}&0&f^2&0\\
-\sqrt{1+t^2} (f_1^2 s_1c_1-f_2^2 s_2 c_2)&
{1-t^2\over \sqrt{3-t^2}}(f_1^2 s_1c_1-f_2^2 s_2 c_2)&
0& f^2
\end{array} \right)\;.
\eeq

It is easy to see that, again, the trace of this matrix does not
depend on the Higgs VEV:
\beq
{\rm Tr}\;[ M^2 ] = {g^2\over 2} \left( {4\over 3-t^3}+2 \right) f^2\;,
\eeq
implying the absence of one-loop quadratic divergences.
The mixing terms of $A^5_\mu$ with $Z_\mu$ and $Z'_\mu$ were
overlooked in \cite{Schmaltz:2004de}. Although they cancel
at the lowest order 
in $v/(\sqrt{2} f)$, we will see that they are essential to obtain the right 
ultraviolet dependence of the effective potential. 

The total one-loop contribution from the gauge fields
to the effective potential is then
\beq
V_{gauge}={3\over 64\pi^2} \Lambda^2_g {\rm Tr}\; [M^2]+
{3\over 64\pi^2} {\rm Tr}\; [ M^4 \log \left( 
\Lambda^2_g\over M^2 \right) ] \;.
\eeq

As explained before, the quadratic divergence vanishes.
The second term gives the usual
$W^\pm$ and $Z$ corrections up to the scale 
$\approx g f$ where the extra vector bosons get mass
and, above it, there is an $SU(3)$-symmetric 
logarithmic divergence proportional to the sum of all 
vector bosons masses to the fourth power. 
In particular, in the charged sector we have
($W_i$ carries particle plus antiparticle)
\beq
M^4_{W_1}+M^4_{W_2}={g^4 f^4\over 4} \left(
1 - {1\over 2} s_{2\beta}^2\; s_{12}^2(h)\right)\;, 
\eeq
whereas the four neutral vectors give
\beq
\sum_{i=1}^4 M_{Z_i}^4  = {g^4 f^4\over 2} 
\left( 1+{8\over (3-t^2)^2} - {1+t^2\over 3-t^2} s_{2\beta}^2\; 
s_{12}^2(h) \right)\;.
\eeq
We could redefine (renormalize) these divergent 
terms into a term identical to the one obtained 
in the fermion sector 
plus the operator
\beq
b\; \left( \phi_1^\dagger
\phi_2 \right)_{\bf 8}
\left( \phi_2^\dagger \phi_1 \right)_{\bf 8}\supset
{2b\over 3}\;  f_1^2 f_2^2 \;\left( 2+ s_{12}^2(h)\right)\;.
\label{q88}
\eeq
We find that the ultraviolet physics (quartic terms proportional to
$a$ and $b$ {\it or} the cut-offs $\Lambda_{t,g}$ in the 
top-quark and gauge sectors) can only define 
the coefficient of a term proportional to $s_{12}^2(h)$. 
This single arbitrary parameter from the ultraviolet completion will 
not be enough (see below) to obtain a Higgs mass
above 114 GeV.

\subsection*{Higgs mass from collective symmetry breaking}

\begin{figure}[!h]
\begin{center}
\includegraphics[width=80mm]{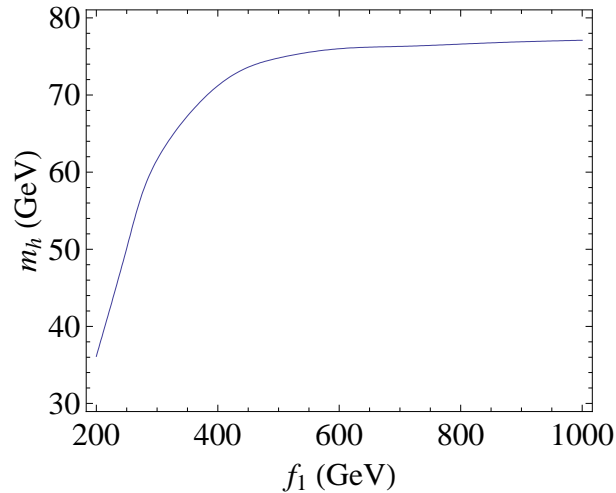}
\caption{Maximum value of the Higgs mass in the model with 
collective symmetry breaking for
$f_2=3$ TeV and different values of $f_1$.}
\label{fighiggs}
\end{center}
\end{figure}

The effective potential obtained in 
the previous subsection only depends on
$\lambda_{1,2}$ and the cut-off $\Lambda_t$ (it actually 
depends on a combination of $\Lambda_t$ and
$\Lambda_g$, so we fix $\Lambda_g=5$ TeV). These three 
parameter must reproduce the $Z$ boson mass ($M_Z=91$ GeV) (i.e.,
the right Higgs VEV) and the top quark mass ($m_t=171$ GeV).
Fig.~\ref{fighiggs} shows the maximum value of the Higgs 
mass in the model with collective symmetry breaking for
$f_2=3$ TeV and values of $f_1$ between $200$ GeV and $1$ TeV. \\

For the plot we have required a mass of the extra $T$ 
quark below $2$ TeV (since we look for natural cancellations
of the quadratic corrections) and a mixing with the top quark 
$V_{Tb}$ smaller than $0.25$. Thus, Fig.~\ref{fighiggs}
shows the maximum value of the Higgs mass in the model with 
collective symmetry breaking for
$f_2=3$ TeV and different values of $f_1$, with 
values for the other parameters
satisfying the conditions exposed before. 
Increasing $f_2$ does not change significantly the results.

\Section{Effective potential in a modified model}

To get an acceptable effective potential we will change
the selecting principle for Yukawa couplings
and thus the way the global symmetry is broken in the
top-quark 
sector. Instead of two similar couplings $\lambda_{1,2}$
implying collective breaking, we will assume an 
{\it approximate} symmetry suppressing some of the
couplings by a factor of $\epsilon\approx 0.1$. Such a 
framework seems more natural in order to separate $f_1$ from
$f_2$ and obtain heavy gauge bosons together with a lighter
$T$ quark.

The Lagrangian may contain now four couplings,
\beqa -{\cal L}_t &=& \lambda_1\; \phi_1^\dagger \Psi_Q t_1^c +
\lambda_2
\;\phi_2^\dagger \Psi_Q t_2^c + \nonumber \\
&& \lambda'_1\; \phi_1^\dagger \Psi_Q t_2^c + \lambda'_2
\;\phi_2^\dagger \Psi_Q t_1^c + {\rm h.c.}\;.\label{yt} \eeqa
In the original simplest LH model only $\lambda_{1,2}$ are not zero
and diagrams must contain  
both couplings simultaneously to break the $SU(3)_1 \times SU(3)_2$ symmetry. 
In our model, the global $SU(3)_1\times SU(3)_2$ symmetry 
is approximate in this sector: $\Psi_Q$ 
is a triplet under
$SU(3)_1$ and a singlet under $SU(3)_2$, which implies that 
the couplings $\lambda_1$ 
and $\lambda'_1$ are order 1. On the other hand, after this
assignments 
the couplings $\lambda_2$ and $\lambda'_2$ will break the
symmetry, and  
we will take them one order of magnitude
smaller. In addition, making a redefinition of the fields
we can take, with all generality, $\lambda'_1=0$:
\beqa  \lambda_1\; \phi_1^\dagger \Psi_Q t_1^c +
\lambda'_1\; \phi_1^\dagger \Psi_Q t_2^c= \sqrt{ \lambda_1^2 + 
\lambda'^2_1}\phi_1^\dagger \Psi_Q \left( \frac{\lambda_1}
{\sqrt{ \lambda_1^2 + \lambda'^2_1}}t_1^c+\frac{\lambda'^2_1}
{\sqrt{ \lambda_1^2 + \lambda'^2_1}}t_2^c \right)\,,
\eeqa
defining 
\beqa \tilde{t}_1^c &=&\frac{\lambda_1}{\sqrt{ \lambda_1^2 + 
\lambda'^2_1}}t_1^c+\frac{\lambda'^2_1}{\sqrt{ \lambda_1^2 + 
\lambda'^2_1}}t_2^c \;, \\ \tilde{\lambda_1}&=& \sqrt{ \lambda_1^2 
+ \lambda'^2_1} \;,
\eeqa
we obtain $\tilde{\lambda'_1}=0$.

While the usual model includes  one-loop 
logarithmic corrections of type $\lambda^2\log\Lambda$ to
the Higgs mass, in our model there will be  quadratic
corrections, 
although suppressed by factors of $\epsilon$. 
These {\it harder} corrections 
will provide quartic couplings that will rise the mass of the 
physical Higgs without introducing fine tuning for 
a cut-off at 5--10 TeV.

Let us deduce the new contribution from the Yukawa sector
to the effective potential of the two physical GBs $h$ and $\eta$.
If these two fields get a VEV 
($\langle h^0 \rangle=\frac{u}{\sqrt 2}$ and 
$\langle \eta \rangle=y$), after a proper
redefinition of the fermion fields
they imply the following fermion mass matrix in the top quark
sector: 
\beq
-{\cal L}_t \supset \left(
\begin{array}{cc} t & T \end{array} \right) \left(\begin{array}{cc}
\lambda_1 f_1 s_1-e^{i\theta} \lambda'_2 f_2 s_2&
-\lambda_2 f_2 s_2+e^{-i\theta} \lambda'_1 f_1 s_1\\
\lambda_1 f_1 c_1+e^{i\theta} \lambda'_2 f_2 c_2& \lambda_2 f_2
c_2+e^{-i\theta} \lambda'_1 f_1 c_1
\end{array} \right)
\left(\begin{array}{c} t_1^c \\
t_2^c
\end{array} \right)\;
\eeq
where 
\beq 
\theta= {y f\over \sqrt{2}f_1 f_2} \;,
\eeq 
and
$\lambda_{1,2}$ are real and positive. 

Some comments are here in order. 

\begin{itemize}

\item The mass 
of the extra $T$ quark is
\beq m_T^2=(\lambda_1^2+{\lambda'_1}^2)\; f_1^2 +
(\lambda_2^2+{\lambda'_2}^2)\; f_2^2
+2(\lambda_1\lambda'_2+\lambda_2{\lambda'}_1)\;f_1f_2 \;c_{12}
c_\theta -m_t^2\;,
\eeq
with
\beq c_{12} \equiv \cos {u f\over
\sqrt{2} f_1 f_2}\;.
\eeq
As all the couplings except for $\lambda_1$ are small,
and this coupling only contributes to $m_T$ multiplied by the smaller 
VEV $f_1$, the approximate symmetry justifies a light $T$ quark.

\item Unlike in the case of collective
symmetry breaking (i.e., $\lambda'_{1,2}=0$) the fermion 
masses depend on the VEV $y$ of the singlet $\eta$ and 
will imply an acceptable mass for this field.

\item The smaller up and charm quark masses could appear if the
assignments for the quark triplets under the 
approximate symmetry are different: 
triplets under the second $SU(3)_2$ and 
singlets under the first one. In particular,
the only large Yukawas (one per family) should couple 
these triplets with $\phi_2$. This means that the approximate symmetry
would increase by a factor of $\epsilon^{-1}$ the extra quarks corresponding to the first 
two families ($m_{U,C}\approx f_2$) while suppressing the mass of the standard
fermions ($\lambda^{u,c} \sim \epsilon \lambda^t$).

That would make the 
extra up-type
quarks very heavy ($m_{U,C}\approx f_2$), whereas the 
up and the charm fields would couple to the Higgs with
suppressed Yukawa couplings.

\item Down-type quarks (and also charged leptons) 
may get their mass through 
dimension 5 operators \cite{Schmaltz:2004de} like:
\beq
-{\cal L}_b \approx
{y_b\over f}\; \phi_1 \phi_2 \Psi_Q b^c + {\rm h.c.}\;,
\eeq
but they neither require extra fields nor large couplings.

\item Finally, here
the lepton doublets become triplets that include a $SU(2)_L$ 
singlet, $\psi_L^T=(\nu\;e\;N)$. This forces 
the addition of a fermion singlet $n_c$ per family and the Yukawa couplings
\beq -{\cal
L}_\nu = \lambda_1^\nu\; \phi_1^\dagger \Psi_L n^c +
\lambda_2^\nu\; \phi_2^\dagger \Psi_L n^c + {\rm h.c.}\;.\label{ynu}
\eeq
Once the Higgs gets a VEV we obtain
\beq -{\cal
L}_\nu \supset i( \lambda_2^\nu f_2s_2 - \lambda_1^\nu f_1 s_1) \nu n^c +
(\lambda_1^\nu f_1 c_1 + \lambda_2^\nu f_2 c_2)Nn^c + {\rm h.c.}\;.\label{ynu}
\eeq
Redefining $i\nu\rightarrow\nu$:
\beq -{\cal
L}_\nu \supset m_{N'}(s_\theta\nu+c_\theta N)n^c + 
{\rm h.c.}=m_{N'}N'n^c + {\rm h.c.}\;,\label{ynu}
\eeq
where $N'=(s_\theta\nu+c_\theta N)$,
$s_\theta=\displaystyle{ \frac{\lambda_2^\nu f_2 s_2 - \lambda_1^\nu f_1 s_1}{m_{N'}}}$ and \\ $m_{N'}=\sqrt{{\lambda_1^\nu}^2 f_1^2 +{\lambda_2^\nu}^2 f_2^2 + \lambda_1^\nu \lambda_2^\nu f_1 f_2 c_{12}}$. If $\Psi_L$ is a triplet under $SU(3)_2$ and a singlet 
under $SU(3)_1$, then the approximate symmetry implies 
$\lambda^\nu_2\approx 1 \gg \lambda_1^\nu$. 
The two fermion singlets $N'$
(approximately $N$ because $s_\theta\sim s_2$) and $n^c$ will 
get masses $\sim \lambda_2^\nu f_2$. 
We are introducing quadratic divergences proportional to 
$\lambda_1^\nu\lambda_2^\nu$, which are acceptable since
 $\lambda_2^\nu$ is small.
Moreover, in these models there is an alternative to the 
usual \textit{see-saw} mechanism.
The massive Dirac neutrino together with a small lepton 
number violating mass term:
\beq
 -{\cal L}_\nu \supset {1\over 2} m\; n^c n^c + {\rm h.c.}\label{ynu}
\eeq
of order 0.1 keV will generate standard Majorana neutrinos 
at one-loop of mass 0.1 eV \cite{delAguila:2005yi}.

\end{itemize}

Summarizing, in these models all the extra \textit{right-handed} 
neutrinos and the up-type quarks excluding the one cancelling the 
quadratic corrections of the
top quark can be very heavy, with masses $\sim f_2\approx 3$ TeV.

\subsection*{The Higgs mass in the modified model}

Let us find the Higgs mass in the model that we propose,
with $f_1\approx$ 0.1$f_2$,
$\lambda_1\approx 1$, and  the rest of the couplings in 
the top-quark sector
at least one order of magnitude smaller. We 
found new operators breaking the global 
symmetry in the one-loop effective potential that 
were not allowed in the model with collective 
symmetry breaking. These couplings may imply a 
Higgs boson in agreement with LEP2 constraints and 
also an acceptable mass for the singlet $\eta$. We will also
quantify the amount of fine tuning implied in the model.

The basic new feature introduced by 
$\lambda'_{1,2}$ in the top-quark sector 
is the following quadratic divergence:
\begin{equation}
\begin{split}
\Delta V_{top}=-{3\over 16\pi^2} \Lambda_{top}^2 \Bigg( & f_1^2 \left(
\lambda_1^2 + {\lambda'}_1^2 \right) + f_2^2 \left( \lambda_2^2 +
{\lambda'}_2^2 \right) + \\
& 2 f_1 f_2 \left( \lambda_1 \lambda'_2
+\lambda_2 \lambda'_1\right) \ \cos {h f\over \sqrt{2}
f_1 f_2} \ \cos {\eta f\over \sqrt{2}f_1 f_2} \Bigg).
\end{split}
\end{equation}
This term is crucial to achieve an acceptable potential as
it is proportional to $c_{12}(h)$, while in the model with 
collective breaking 
all the ultraviolet contributions are 
proportional to $s_{12}^2(h)$ (i.e., $c_{12}^2(h)$).

Let us consider a particular case.
We will fix $f_1=400$ GeV
and $f_2=3$ TeV. We will take
$\lambda_1=1.19$, $\lambda_2=-0.25$, and $\lambda'_2=0.03$
($\lambda'_1=0$ since it disappears after a redefinition of 
the fields). For the cut-off we take
$\Lambda_{t}=\Lambda_{g}=5$ TeV and we add an ultraviolet term $a \sim 1/16\pi^2$
(this is equivalent to take different cut-offs in each sector).

\begin{figure}[t]
\begin{center}
\includegraphics[width=89mm]{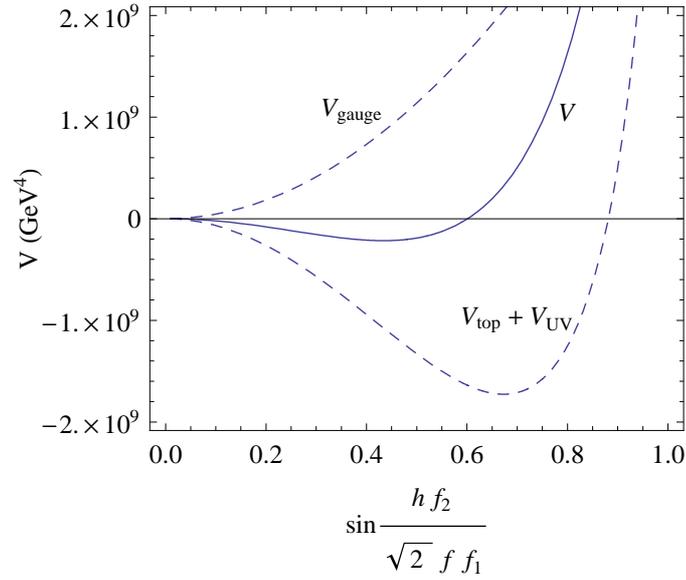}
\caption{One-loop Higgs potential as a function of $s_1(h)$ 
for the choice of parameters given in the text.
\label{fig2LH}}
\end{center}
\end{figure}

Fig.~\ref{fig2LH} shows the different contributions to the 
effective potential
and its total value as functions of $s_{1}(h)$. The minimum, 
$s_{1}=0.43$ (i.e., $u=259$ GeV), reproduces $m_Z=91$ GeV and 
$m_t=171$ GeV with
a mass of the extra $T$ quark of 920 GeV. The potential 
implies $m_h=156$ GeV and
$m_\eta=107$ GeV. Playing with $f_1$, $\lambda'_2$ and $a$ it 
is possible to get masses of the Higgs boson both smaller and larger. 

\begin{figure}
\begin{center}
\includegraphics[width=89mm]{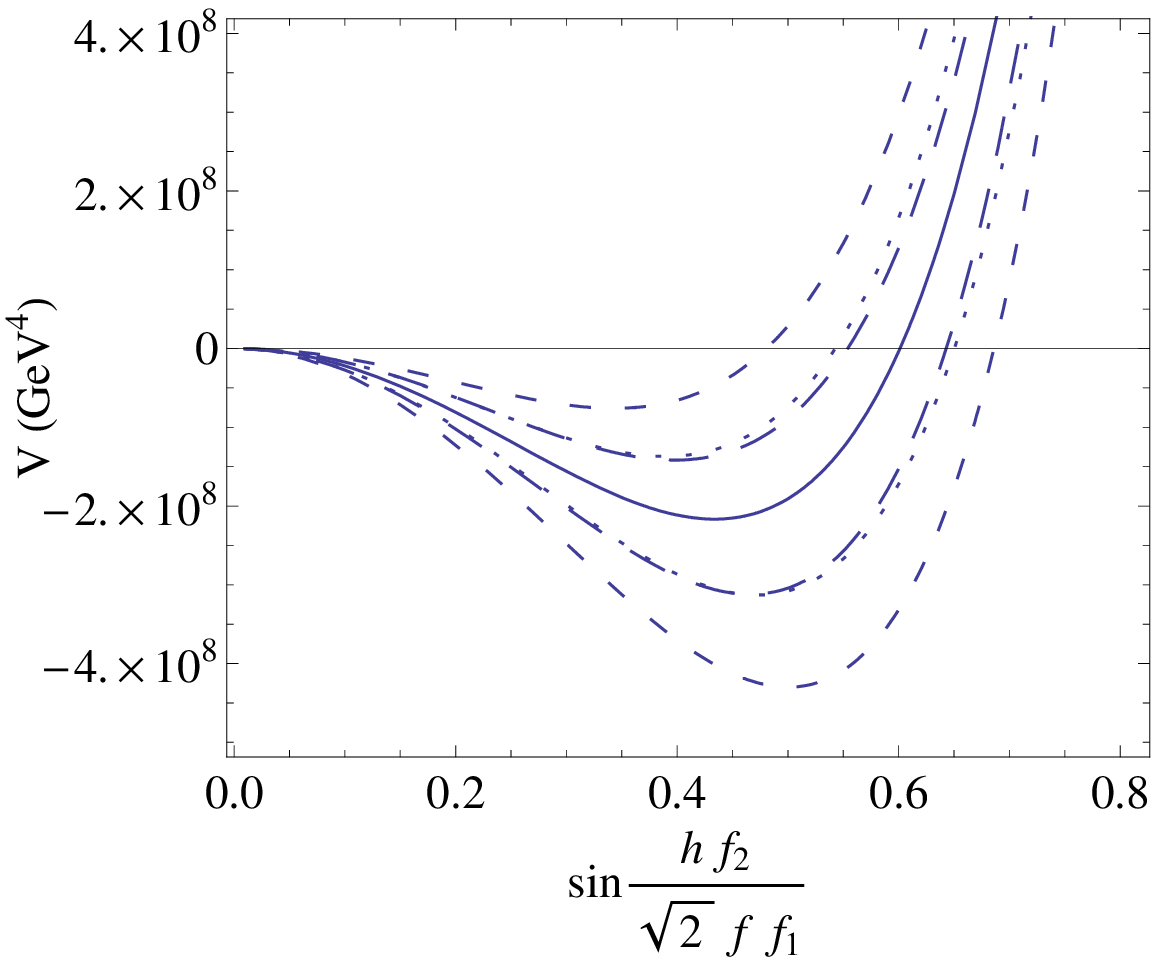}
\caption{Variation of the Higgs potential for a 
$ \pm $5\% variation of $f_1$ (dashes), $\lambda'_1$
(long dashes) and $a$ (dots) and the potential for the 
central values given in the text (solid line). 
The EW scale changes in up to a +20\% or a -25\%, 
whereas $m_h$ varies 
between 126 GeV and 178 GeV.}
\label{fig3}
\end{center}
\end{figure}

In order to check if our solution needs fine tuning 
we have varied in a 5\% the VEV $f_1$, the coupling
$\lambda'_1$ and the ultraviolet coupling constant $a$ respect to the 
values taken before. Fig.~\ref{fig3} shows the change on the 
Higgs potential
caused by the variation of each parameter. We find that the 
EW scale $v/\sqrt{2}$ changes up and down in a 20\% and 25\% 
respectively, while $m_h$ varies 
between 126 GeV and 178 GeV. This is a check of the absence 
of fine tuning in the scalar sector.

Finally, note that we have obtained an acceptable Higgs mass 
with no need for a $\mu$ term $\phi_1^\dagger \phi_2$ put by hand
in the scalar potential, as it is sometimes assumed in the
simplest LH model.

%% file: Ch3.tex

{\Chapter{Top-pair production through extra Higgs bosons}\label{Ch3}}

The main objective of the LHC is to reveal the nature of the
mechanism breaking the EW symmetry. Of course, the first step should
be to determine the Higgs mass and to check whether the Higgs couplings
to fermions and vector bosons coincide with the ones predicted within
the SM. A second step, however, 
would be to search for additional particles 
that may be related to new dynamics or symmetries present at the
TeV scale and that are {\it the solution} to the hierarchy 
problem. 

The top-quark sector appears then as a promising place 
to start the search, as it is there where the EW symmetry 
is broken the most (it contains the heaviest fermion of the SM).
Indeed, the main source of quadratic corrections destabilizing
the EW scale is the top quark, so the new physics {\it should}
manifest at least in that sector at low energies. This is the case,
for example, studied in the previous chapter, where a vectorlike
$T$ quark is introduced by a global symmetry \cite{Barcelo:2007if,Barcelo:2008je}.

In addition, the new symmetry solving the hierarchy problem
may also define a non-minimal
Higgs sector. In that case, the large 
Yukawa coupling of the top quark with the Higgs boson ($h$) will also 
imply large couplings with the extra Higgses. For example, in
SUSY extensions $h$ will come together with heavier Higgs 
fields: a 
neutral scalar ($H$) 
and a pseudoscalar ($A$) \cite{Martin:1997ns,Djouadi:2005gj}. 
Also in the LH models discussed before, 
the breaking of the global symmetry requires a massive 
scalar singlet (the radial field $r$) that tends to be strongly
coupled to the extra $T$ quark. In general, 
the large couplings of these 
scalar fields to colored particles could 
imply both a sizeable production rate in hadron collisions and a
dominant decay channel into $t\bar t$. 

It is well known that the top quark is crucial to understand the
physics of the Higgs boson at hadron colliders. In particular,
the leading Higgs 
production channel goes through a top-quark loop in gluon fusion. 
In this chapter we will explore the {\it opposite} effect: how 
the Higgs may affect the production of top-quark pairs observed 
at the Tevatron and, specially, at the LHC. 
We will show that in the SM the Higgs effects are 
not relevant because of a simple reason: a standard Higgs heavy enough to
be resonant in $t\bar t$ production would have very strong couplings
to itself ({\it i.e.}, 
to the GBs eaten by the $W$ and $Z$ bosons), much stronger
than to the top quark. As a consequence, its large width would 
dilute all the effects.
In contrast, we will show that in models with extra Higgses below
the TeV scale the new fields can be heavy enough (above the 
$2m_t$ threshold) with no need for a large self-coupling
and thus a dominant decay mode into $t\bar t$. These fields
may provide observable anomalies in the $t\bar t$ invariant mass
distribution ($m_{t\bar t}$) measured at colliders.

\begin{figure}[htb]
\begin{center}
\includegraphics[width=1\linewidth]{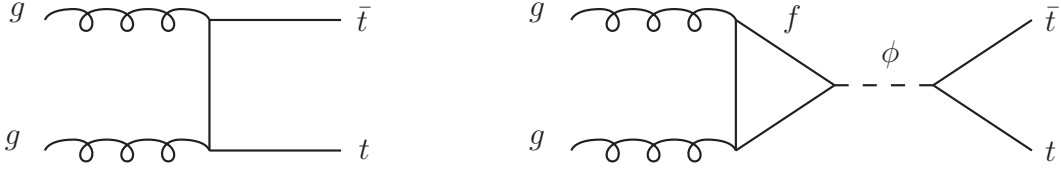} 
\end{center}
\caption{Diagrams that interfere in $t\bar t$ production.
\label{figS1}}
\end{figure}

We will first review the analytical 
expressions \cite{Gaemers:1984sj,Dicus:1994bm} for the 
parton-level process $gg\rightarrow t\bar t$ 
in QCD and mediated by a 
generic scalar or pseudoscalar field produced at one loop
(see Fig.~\ref{figS1}--right).
In the loop we will put the top or a heavier $T$ quark.
Then we will use these expressions to study the 
possible effect  of a heavy standard Higgs
with $m_h>2m_t$. We will introduce SUSY and 
apply the same type of analysis both to SUSY and LH Higgs bosons, studying
in each case the
parton-level cross sections and 
the possible signal at the LHC. Finally, we will also comment on
the possible relevance at the Tevatron. The results in this 
have been published in~\cite{Barcelo:2010bm}.

\vspace{0.5cm}

\Section{Top quarks from scalar Higgs bosons}

The energy and the luminosity to be achieved at the LHC make
this collider a top-quark factory, with around $1.5\times 10^5$ 
pairs at $\sqrt{s}=7$ TeV and 1 fb$^{-1}$. This quark 
will certainly 
play a special role in the LHC era. The potential to 
observe new physics in $m_{t\bar t}$ at hadron colliders has 
been extensively discussed in previous literature 
\cite{Dicus:1994bm,Frederix:2007gi,Barger:2006hm,
:2007dia,Abazov:2008ny,Cabrera:2009zza,Baur:2007ck,
Hioki:2009hm,Kumar:2009vs}. In general, any heavy $s$--channel 
resonance with a significant
branching ratio to $t\bar t$ will introduce distortions:
a {\it bump} that can be evaluated in the narrow-width
approximation or more complex structures (a {\it peak} followed
by a {\it dip}) when interference effects are 
important \cite{Berdine:2007uv}. 
These structures are produced by the interference between the 
diagrams depicted in Fig.~\ref{figS1} (plus crossings). 
Notice that the final state of Fig.~\ref{figS1}--right is
a color singlet (it is mediated by a colorless $\phi$) and, hence, there is no interference between this 
and the octet $gg\rightarrow g \rightarrow t\bar t$ contributions. Although the scalar contribution 
involves a fermion loop,  
the gauge and the Yukawa couplings are both strong, and at 
$\sqrt{\tilde s}= m_{t\bar t}\approx m_\phi$ it may
give an observable contribution.

Let us write the general expressions for a scalar $\phi$ 
coupled to the top quark and (possibly)
to a heavy fermion $T$.
The leading-order (LO) 
differential cross section for $gg\rightarrow t\bar t$ will 
include the squares of the scalar and the QCD amplitudes as well
as their interference,
\beqa
{{\rm d} \sigma \over {\rm d} z} &=& 
{{\rm d} \sigma_{QCD} \over {\rm d} z}+ {\alpha_s^2 \;
y_{\phi  t \bar t}^2\; s^2\; \beta^3\over 1536\; \pi^3}\;
\left| {N(s)\over s-m_\phi^2+i\; m_\phi\Gamma_\phi(s)} \right|^2  \nonumber  \\
  && 
-{\alpha_s^2\; y_{\phi t \bar t}\; m_t\; 
\beta^3\over 48\sqrt{2}\; \pi}\; {1\over 1-\beta^2 z^2}\;
{\rm Re} \left[ {N(s)\over s-m_\phi^2+i\; m_\phi\Gamma_\phi(s)}
\right] \;,
\label{csphi}
\eeqa
where $z=\cos\theta$ is the cosine of the angle 
between an incoming $g$ and
$t$, $m_t$ and $y_{\phi t \bar t}$ are the top-quark mass and 
Yukawa coupling, 
and $\beta=\sqrt{1-4 m_t^2/s}$
is the velocity of $t$ in the center of mass frame.
The function $N(s)$ associated to the fermion loop is
\beq
N(s)=\sum_f {3\; m_f\; y_{\phi  f \bar f}\over \sqrt{2}\;s}
\left[ 1+\left( 1-{4m_f^2\over s} \right) I_f(s) \right]\;,
\eeq
where $f$ may be the top or another quark strongly coupled
to $\phi$, and 
$I_f(s)$ takes a different
form depending on the mass $m_f$:
\beq
I_f(s) = \left\{
\begin{array}{l l} 
\displaystyle
\left( {\rm Arcsin}\sqrt{s\over 4m_f^2} \right)^2 
& s < 4m_f^2\;; \\
\\
\displaystyle
-{1\over 4} \left( \ln {1+\sqrt{1-4 m_f^2/s}\over
1-\sqrt{1-4 m_f^2/s}} - i \;\pi \right)^2
& s > 4m_f^2\;.
\end{array} \right. 
\label{if}
\eeq
If $2m_f>\sqrt{s}$ then $I_f$ is real and the interference 
vanishes at $s=m_\phi^2$. If $f$ is the top or
any fermion with 
$2m_f<\sqrt{s}$, then this contribution can be seen as a 
final-state $f\bar f$ interaction \cite{Dicus:1994bm}. 

The LO differential QCD contribution comes from $gg\rightarrow g \rightarrow t\bar t$, the t-channel
diagram Fig.~\ref{figS1}--left and its crossing u-channel, and their interference. It can be written as
\beq
{{\rm d} \sigma_{QCD} \over {\rm d} z} = {\pi \alpha_s^2 \over 12 s} \beta \left( 
{16 \over (1- \beta^2 z^2)} -9 \right) \left( 
{(1+ \beta^2 z^2) \over 8} + {m_t^2 \over s} - {4 m_t^4 \over s^2 (1- \beta^2 z^2)}  \right)\,.
\eeq
For a pseudoscalar $A$ we have 
\beqa
{{\rm d} \sigma \over {\rm d} z} &=& 
{{\rm d} \sigma_{QCD} \over {\rm d} z}+ {3\;\alpha_s^2 \;
y_{A t \bar t}^2\; s^2\; \beta \over 512\; \pi^3}\;
\left| {P(s)\over s-m_A^2+i\; m_A\Gamma_A(s)} \right|^2  \nonumber  \\
 && 
-{\alpha_s^2\; y_{A t \bar t}\; m_t\; 
\beta \over 16\sqrt{2}\; \pi}\; {1\over 1-\beta^2 z^2}\;
{\rm Re} \left[ {P(s)\over s-m_A^2+i\; m_A\Gamma_A(s)}
\right] \;, 
\label{csA}
\eeqa
with
\beq
P(s)=\sum_f {m_f\; y_{A f \bar f}\over \sqrt{2}\; s}
\;I_f(s)\;.
\eeq
In the expressions above we have used the {\it energy-dependent width} obtained from the
imaginary part of the one-loop scalar 2-point function. In
most studies the width of the resonances is taken constant, but we will see in the next
chapter that the energy dependence may imply important effects. 

To have 
an observable anomaly it is necessary that the width
$\Gamma_\phi$ is small. This is the
reason why a very heavy standard Higgs $h$ would
be irrelevant. As we mentioned before, a 500 GeV Higgs boson would couple strongly
to the top quark but even stronger to itself,  
$\lambda=m_h^2/(2v^2)\approx 2$.
Its decay into would-be GBs (eaten by the 
massive $W$ and $Z$) would then dominate, implying a 
total decay width 
\beq
\Gamma_h \approx 
{3\over 8\pi\; v^2} \left[
{m_t^2 m_h\;\beta_t^3} 
+{m_h^3 \over 4} \left(
\beta_V^3 +{3\over 4} \beta_V(1-\beta_V^2)^2\right)
\right]\;\approx 60\; {\rm GeV}\;,
\eeq
where 
\beq
\beta_{t(V)}=\sqrt{1-{4m_{t(V)}^2\over m_h^2}}\;
\eeq
and we have taken a common $W,Z$ mass $m_V\approx 90$ GeV.

\begin{figure}
\begin{center}
\includegraphics[width=0.45\linewidth]{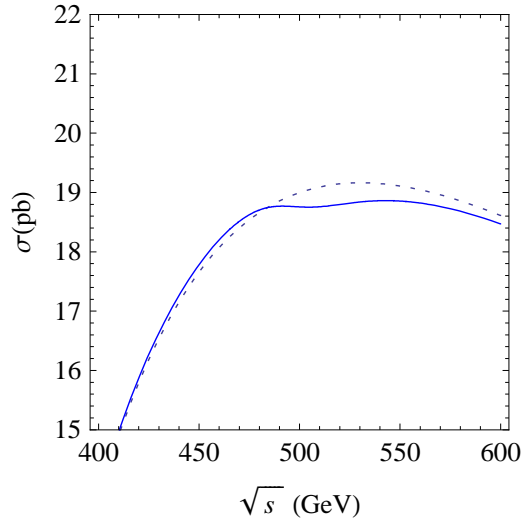} 
\end{center}
\caption{$\sigma(gg\rightarrow t\bar t)$
with a standard Higgs of mass $m_h=500$ GeV.
\label{figS2}}
\end{figure}
The plot in Fig.~\ref{figS2} shows a too small deviation due to the standard Higgs
in the parton-level process $\sigma(gg\rightarrow t\bar t)$. 
In order to have a smaller width and a larger effect
the mass of the resonance must {\it not} be EW.
In particular, SUSY or LH models
provide a new scale and massive Higgses with no need for large
scalar couplings. As we are already familiar with LH models we will
start reviewing SUSY before analyzing this effect.

\Section{Basics of SUSY}

SUSY is basically different from the usual symmetries defining a Lie group
in the fact that it incorporates anticonmuting (spinorial) generators $Q$. 
Its algebra is a {\it graded} generalization of the Poincare algebra,
with $Q$ relating particles of different spin:
\begin{equation}
Q|\textrm{fermion}> = |\textrm{boson}>     \;\;\;\;      
Q|\textrm{boson}> = |\textrm{fermion}>
\end{equation}
It was proposed in the early seventies \cite{Golfand:1971iw,Wess:1973kz,Wess:1974jb,Wess:1974tw}, and since then SUSY has become
a very interesting framework for model building. It is true that SUSY still 
lacks any (genuine) experimental support, and that the non observation
of flavor-changing neutral currents or electron and neutron electric dipole moments requires
an explanation. 
One has also to admit that the 
initial efforts to build a finite (or at least a renormalizable) 
version of supergravity failed. However, SUSY has been able to {\it adapt}
throughout the past forty years,
and its search is currently 
one of the main objectives at the LHC. It is remarkable that SUSY is necessary
to define a consistent string theory. When it is incorporated into a 
field theory, all quadratic divergences cancel. Therefore, the minimal 
SUSY version of the SM (MSSM) \cite{Haber:1984rc,Errata,GunionHaber}
is free of the hierarchy problem. In addition, 
the unification of the three gauge couplings is {\it better} (more precise
and at a higher scale) than in the SM, and it provides a viable dark
matter candidate. Finally, all these features can be kept if SUSY is broken
spontaneously and only soft-SUSY breaking terms of order
$m_{\textrm{susy}}$ are generated. 
SUSY has the {\it virtue} that it can be
decoupled smoothly, with variations versus the SM of order 
$v^2/m^2_{\textrm{susy}}$ to all the observables except for the mass of the
Higgs boson, that is necessarily light (see below). 
It is not our purpose here to review SUSY exhaustively
(see \cite{Martin:1997ns} for a pedagogical introduction), although in
the next section we will discuss the Higgs sector 
of the MSSM \cite{Djouadi:2005gj} in some detail.

A minimal amount of SUSY and of matter fields will imply 
a superpartner $\tilde \phi$ for each particle $\phi$  
of the SM. $\tilde \phi$ and $\phi$ 
have the same quantum numbers and mass but their spin
differs in 1/2. For each fermion we will then find a scalar 
partner or sfermion: one 
complex scalar for
the left-handed fermion and another one for the right-handed fermion.

It is customary to define the theory in terms of left-handed chiral
superfields, and use the same symbol for the bi-spinor and the superfield.
In this way, for example, $e$ indicates
the left-handed electron (its conjugate $\bar e$ is
the right-handed positron), whereas $e^c$ would be the left-handed 
positron. $\tilde e$ and $\tilde e^c$ would be complex scalars
of electric charge $-1$ and $+1$, respectively.
The Higgs sector will be
defined by two doublets of opposite hypercharge, including the scalar
Higgses and the spin $1/2$ Higgsinos. The second doublet is necessary
to give masses to the standard quarks and leptons
through Yukawa interactions (notice
also that the higgsino 
in just one doublet would introduce gauge anomalies).
Finally, the gauge bosons will be part of a vector superfield carrying 
also its fermionic partner or gaugino. 

The different spin in 
a SUSY multiplet 
will imply the cancellation of all quadratic corrections
to the Higgs mass parameter. 
However, since no superpartners have been found, SUSY must be broken 
in such a way that sfermions and gauginos  
get a mass of order $m_{\textrm{susy}}$ non EW ({\it i.e.}, not 
proportional to $v$). This will introduce corrections
proportional to $m_{\textrm{susy}}^2$, suggesting that at least the stops should
be below 1 TeV.
Let us now review some details
of the Higgs sector in the MSSM and other aspects 
that will be needed in our study of how SUSY 
Higgses may affect $t \bar{t}$ production. 

\subsubsection{The scalar potential of the MSSM}

The masses and interactions of the Higgs fields  include
both SUSY preserving and soft-SUSY breaking terms. The former 
are derived from the superpotential $W$ (we will not consider the
possibility of operators of dimension larger than four, {\it i.e.}, 
we take a canonic gauge kinetic function and a minimal Kahler potential). 
$W$ must include the Yukawa interactions required to give
masses and mixings to quarks and leptons.
If one assumes renormalizability and gauge invariance,
the superpotential is given by
\begin{equation}
W =  Y_u\, H_2 Q u^c + Y_d\, H_1 Q d^c + 
Y_l\, H_1 L e^c + \mu H_1 H_2\,,
\end{equation}
where $Y_u$, $Y_d$ and $Y_l$ are $3 \times 3$ Yukawa matrices in flavor space, 
and all gauge, flavor and Lorentz indexes are understood. 
This minimal $W$ includes a discrete symmetry, called $R$-parity, under 
which all SM particles are even and the SUSY partners are odd. 
This symmetry will 
imply the estability of the lightest SUSY particle, providing a 
viable candidate for dark matter. 
$Q$ and $L$ above 
contain the $SU(2)$ quark and lepton doublets, respectively, whereas
the two Higgs doublets are 
\beq
H_1 = \left(\begin{array}{c} H^0_1 \\ H^-_1 \end{array}\right)_{Y =-1/2}\,, 
     \;\;\;\;    
H_2 = \left(\begin{array}{c} H^+_2 \\ H^0_2 \end{array}\right)_{Y= +1/2}\,.
\eeq
The $\mu$ term is a SUSY contribution to the Higgs boson mass
that must be of the same order as $m_{\textrm{susy}}$. The presence of such 
term is sometimes referred
as the $\mu$ problem: why is $\mu$ so much smaller than the Planck mass?
Notice that SUSY would guarantee the stability of $\mu$ under loop 
corrections, but in principle it does not justify its EW value. This
term, however, breaks also a global $U(1)$ symmetry known as 
$R$-symmetry. Such symmetry acts non-trivially on the superspace 
coordinates and on the superpotential, implying different charges for
the different components in a superfield. In particular, all matter
superfields have a charge $+1$ under this symmetry while the Higgs 
superfields are neutral (the scalar component will have the same 
charge as the superfield, and the left-handed fermion that charge
minus one). 
If the total charge of the superpotential is $+2$, then this 
symmetry will forbid the $\mu$ term.
In some models the breaking of SUSY and of the $R$-symmetry are related,
and the $\mu$ term is naturally similar to the soft-SUSY 
masses (in these models the $R$-parity would be a discrete remnant 
of the broken $R$-symmetry).

Once the gauge symmetry and $W$ are fixed, the SUSY-preserving part of 
a scalar potential can be expressed in terms of the auxiliary component
of the chiral and the vector superfields:
\beq
V(\phi)=V_F+V_D\,.
\eeq
The F-terms are derived from the superpotential, 
\beq
V_F= F_i^*  F_i = 
\sum_i  \left|{{\partial{W}} \over{\partial{ \phi_i}}}\right|^2,
\eeq 
whereas the D-terms are fixed by the gauge interactions:
\beq
V_D= {1 \over 2} \sum_a D^a D^a = {1 \over 2} \sum_a g_a^2 (\phi^* T^a \phi)^2,
\eeq 
where $\phi_i$ is the scalar component of a
chiral superfield, $g_a$ accounts for its coupling to the different gauge 
sectors and $T^a$ are the generators. 
For the two Higgs doublets in the MSSM one has 
\beq
V_{\textrm{SUSY}}^{\textrm{Higgs}} = \left| \mu \right|^2 ( \left| H_1 \right|^2 + \left| H_2 \right|^2) + {1 \over 8} (g^2 + g'^2) 
( \left| H_1 \right|^2 - \left| H_2 \right|^2)^2 + {1 \over 2} g^2 \left| 
H^\dagger_1 H_2 \right|^2.
\eeq
To obtain the EW breaking we 
need to add all possible soft-SUSY breaking terms to this
potential. 
These terms fall into two classes: scalar and gaugino
mass terms consistent with the symmetries, 
and (gauge and $R$-parity invariant) 
scalar trilinears. 
The later can be written as a term proportional to the superpotential 
plus their hermitian conjugate. In the Higgs sector the new terms are
just masses $m^2_{1,2}$ for each doublet and 
a bilinear term $m^2_{12}$ mixing $H_1$ and $H_2$, resulting 
into the potential 
\beqa
V_{\textrm{Higgs}} &=& m^2_{H_1} \left| H_1 \right|^2 + m^2_{H_2} \left| H_2 \right|^2
- m^2_{12}( H_1 H_2 + \textrm{h.c.}) \nonumber \\ &+&  {1 \over 8} (g^2 + g'^2) 
( \left| H_1 \right|^2 - \left| H_2 \right|^2)^2 + {1 \over 2} g^2 \left| 
H^\dagger_1 H_2 \right|^2,
\label{VHiggsminimize}
\eeqa
where $m^2_{H_i} \equiv \left| \mu \right|^2 + m_i^2 \; (i=1,2)$, 
$m^2_{12}$ is made positive by a field redefinition, 
and $m_i^2$ can be either positive or negative.

Let us find now under what conditions this (tree-level) Higgs potential 
can produce the EW VEV. First, we need to 
make sure that it is bounded from below.
The quartic interactions 
will stabilize the potential for all  values 
of $H_1$ and $H_2$ except for the particular direction 
with $ \left| H_1 \right| = \left| H_2 \right|$, where 
these quartic contributions cancel. Therefore, we need that the 
quadratic contribution
is positive along that direction:
\beq
2  m^2_{12}  < 2 \left| \mu \right|^2 + m_1^2 + m_2^2\,.
\eeq
On the other hand, the potential will have a non-trivial minimum 
only if 
\beq
\left( m^2_{12} \right)^2 >  
(\left| \mu \right|^2 + m_1^2)(\left| \mu \right|^2 + m_2^2)\,.
\eeq
Note that if $m_1^2=m_2^2$ both conditions cannot be satisfied 
simultaneously and EW symmetry breaking
would not be realized. It is remarkable, however, that in models where 
the SUSY breaking mechanism implies  universal conditions for the soft 
parameters at large scales, top-quark quantum corrections favor the 
appearance of an acceptable VEV \cite{Ibanez:1982fr}.

Let us then suppose that the minimum in the potential 
implies a VEV $v_1/\sqrt{2}$ of $H_1$. This VEV will be degenerate 
(notice that $SU(2)_L\times U(1)_Y$ acts non trivially on it), and
with all generality we can take it positive and along 
the neutral component of the doublet. It is then easy to see 
that the second doublet $H_2$ will also grow a VEV 
$v_2/\sqrt{2}$ along its neutral
component, and that for the potential in Eq.~(\ref{VHiggsminimize}) it will always
be real and positive (a complex VEV would require other couplings 
or/and superfields):
\beq
\langle H_1^0 \rangle = {v_1 \over \sqrt{2}} \;\;\,, 
     \;\;\;    
 \langle H_2^0 \rangle = {v_2 \over \sqrt{2}} \;.
\eeq
The VEVs must then satisfy
\beq
v_1^2 + v_2^2 = v^2 = (246\; \textrm{GeV})^2\,,
\eeq
which can be achieved with different values of the ratio
\beq
\textrm{tan} \beta = {v_2 \over v_1}\,.
\eeq

The four complex fields ({\it i.e.}, eight scalar degrees of freedom) 
in the two Higgs doublets must  be
rearranged to define mass eigenstates. The three
would-be GBs will live
in the actual combination of doublets that have a non-zero VEV. 
In the unitary gauge it reads
\beq
\cos\beta H_1+\sin\beta H^*_2 = {1 \over \sqrt{2}} 
\left(\begin{array}{c} v + H_v \\ 0 \end{array}\right) 
\,.
\eeq
The physical charged Higgses $H^\pm$ and a $CP$-odd neutral scalar
$A$ will lie along the orthogonal combination. One obtains
\beq
-\sin\beta H_1+\cos\beta H^*_2 = {1 \over \sqrt{2}} 
\left(\begin{array}{c} H_0 +i A \\ H^- \end{array}\right)  \,.
\eeq
Finally, the two $CP$-even fields $(h,H)$ will result from a
different combination of the (shifted) real components 
${1 \over \sqrt{2}} \phi_i^0$ of $H_i^0$:
\beq
\left(\begin{array}{c} H \\ h \end{array}\right) = 
\left(\begin{array}{cc} \cos \alpha & \sin \alpha \\ 
-\sin \alpha & \cos \alpha \end{array}\right)  
\left(\begin{array}{c} \phi_1^0 \\ \phi_2^0 \end{array}\right)\,.
\label{massHh} 
\eeq
The angle $\alpha$ results from the diagonalization of 
\beq
\left(\begin{array}{cc} m_{\phi_1}^2  & m_{\phi_1 \phi_2}^2 \\m_{\phi_1 \phi_2}^2 & m_{\phi_2}^2 \end{array}\right) = \left(\begin{array}{cc} M^2_A \sin^2  \beta + M^2_Z \cos^2  \beta  & -(M^2_A + M^2_Z) \sin  \beta \cos  \beta \\-(M^2_A + M^2_Z) \sin  \beta \cos  \beta & M^2_A \cos^2  \beta + M^2_Z \sin^2  \beta \end{array}\right), 
\eeq
and it is related to $\beta$ and to $M_A$ through the
expression
\beq
\alpha = {1 \over 2} \arctan \left( \tan 2 \beta 
{M_A^2 + M_Z^2 \over M_A^2 - M_Z^2 }\right).
\eeq
The mass matrix above yields
\beq
m^2_{H,h}= {1 \over 2} \left[ M^2_A + M^2_Z \pm 
\sqrt{(M^2_A + M^2_Z)^2 - 4M^2_AM^2_Z \cos^2 2 \beta} \right]\,.
\label{eigenmassHh}
\eeq
A celebrated consequence of Eq.~(\ref{eigenmassHh}) is that the mass 
of the lightest $CP$-even Higgs boson is bounded from above,
\beq
m_h \leq M_Z \left| \textrm{cos} 2 \beta \right| \leq M_Z \;.
\label{masaSUSYhiggs}
\eeq
The presence of a light Higgs is arguably 
the most clear prediction of the MSSM, and it is just
slightly modified in more general SUSY scenarios.
This contrasts with the SM: the Higgs boson mass 
is there proportional to the arbitrary self-coupling $\lambda$, while here quartic
couplings are related to the EW gauge couplings, and as a consequence
the tree-level mass can not go above $M_Z$.
Regarding the other neutral Higgs boson $H$, its mass is unconstrained
and  in the limit $M_A >> M_Z$ it becomes heavy, $m_H \simeq m_A$. 
In that {\it decoupling} limit the two rotations are simply
related,  $ \alpha \rightarrow \beta - \pi/2$, 
and the effective low-energy theory 
includes a single scalar $h$ with precisely the same
couplings as those of the SM Higgs boson. Therefore, over the
favored region  in the MSSM parameter space (with $m_{susy}>v$)
the search for the lightest Higgs is equivalent to search for a
light SM Higgs boson. The decoupling limit 
applies for values $m_A\ge 2m_t$ and will be implicit throughout this
chapter. 

Of course, LEP excluded all Higgs masses 
in Eq.~(\ref{masaSUSYhiggs}). If the MSSM has not been ruled out yet 
is due to the importance of the quantum corrections to the Higgs 
effective potential and,
in particular, to $m_{H,h}$. 
These mainly come from the top/stop sector: 
the largest contribution at the one-loop level 
is proportional to $m_t^4/M_Z^2$ and grows logarithmically with
the stop squark masses. Qualitatively, a large Higgs mass requires a
large stop mass, which in turn tends to introduce fine tuning in the
effective potential. 

Since we will need them later, we provide the two-loop result
for the mass of the two fields $h$ and  $H$ in terms of the mass
of the $CP$-odd scalar $A$, the stop masses, and the stop trilinears
\cite{deBoer:1994he}:
\beq
m_{H,h}^2={1\over 2} \left( m_A^2+M_Z^2+\Delta_{11}+
\Delta_{22}\pm \sqrt{\Delta_0^2} \right)
\eeq
where:
\beqa
\Delta_0^2 &=&\left( m_A^2+M_Z^2+\Delta_{11}+
\Delta_{22} \right)^2-4\;m_A^2 M_Z^2 \cos^22\beta 
\nonumber \\
&&-4 \left( \Delta_{11} \Delta_{22}- \Delta_{12}^2\right)
-4 \left( M_Z^2\cos^2\beta +m_A^2\sin^2\beta \right)\Delta_{22}
\nonumber  \\
&&-4 \left( M_Z^2\sin^2\beta +m_A^2\cos^2\beta \right) \Delta_{11} 
-4 \;\sin 2\beta \left( M_Z^2+m_A^2 \right) \Delta_{12} \;,\\
%
\Delta_{11}&=&{3g^2\over 16\pi^2}\; {m_t^4\over M_W^2 \sin^2\beta}
\left[{\mu\left( A_t m_0-\mu \cot\beta\right)\over 
\tilde m_{t1}^2-\tilde m_{t2}^2}\right]^2
d(\tilde m_{t1}^2,\tilde m_{t2}^2)\;,\\
%
\Delta_{22}&=&{3g^2\over 16\pi^2}\; {m_t^4\over M_W^2 \sin^2\beta}
\;\Bigg[\; {2 A_t m_0 \left( A_t m_0-\mu \cot\beta\right)\over 
\tilde m_{t1}^2-\tilde m_{t2}^2}\;
\ln {\tilde m_{t1}^2\over \tilde m_{t2}^2}
\nonumber \\
&&+\ln {\tilde m_{t1}^2 \tilde m_{t2}^2\over m_t^4}
+\left({ A_t m_0 \left( A_t m_0-\mu \cot\beta\right)\over 
\tilde m_{t1}^2-\tilde m_{t2}^2}\right)^2
d(\tilde m_{t1}^2,\tilde m_{t2}^2)\;\Bigg]\;,\\
\eeqa
\beqa
\Delta_{12}&=&-{3g^2\over 16\pi^2}\; {m_t^4\over M_W^2 \sin^2\beta}
\;{\mu\left( A_t m_0-\mu \cot\beta\right)\over 
\tilde m_{t1}^2-\tilde m_{t2}^2}\;
\Bigg[\; \ln {\tilde m_{t1}^2\over \tilde m_{t2}^2}
\nonumber \\
&&+{ A_t m_0 \left( A_t m_0-\mu \cot\beta\right)\over 
\tilde m_{t1}^2-\tilde m_{t2}^2} \;
d(\tilde m_{t1}^2,\tilde m_{t2}^2)\;\Bigg]\;, 
\eeqa
and
\beq
d(m_{1}^2, m_{2}^2)=2-{m_1^2+m_2^2 \over
m_1^2-m_2^2}\;\ln {m_1^2\over m_2^2}\;.
\eeq
Including these corrections one obtains values of the Higgs mass in 
agreement with current experimental bounds. In addition, we observe
that the masses of $H$ and $A$ are in turn very close to each other
(the mass difference is of order $M_Z^2/m_{\textrm{susy}}$). 
Varying the $\mu$ parameter and the stop masses and trilinears between
100 GeV and 1 TeV, 
for $m_A= 500$ GeV we obtain typical values of $m_H-m_A$ between 
$-2$ and $+10$ GeV, as expected in the
decoupling regime. 
This small separation will be important for our study in this chapter.

\subsubsection{Yukawa couplings to fermions}

Let us now deduce the couplings of the Higgs fields to the top 
and the bottom quarks. The fact that $H_1$ only couples to the
singlet $d^c$ and $H_2$ only to $u^c$ 
will automatically forbid 
flavor neutral currents mediated by the extra neutral bosons.
The Yukawa interactions can be derived from
$W$ as
\beq
\mathcal{L}_{\textrm{Yukawa}} = - \sum_{ij} 
\left[ {{\partial^2{W}} \over 
{\partial{\phi_i} \partial{\phi_j}}} \psi_i \psi_j 
+ \textrm{h.c.} \right] \,.
\eeq
This implies
\beq
\mathcal{L}_{Yukawa} \subset - y_t (t t^c H_0 - t b^c H_2^+) - 
y_b ( b b^c H_v - b t^c H_1^-) + \textrm{h.c.} \;.
\eeq
To obtain the observed masses the 
couplings must be
\beqa
y_t &=& {\sqrt{2} m_t \over v_2} = {\sqrt{2} m_t \over v \textrm{ sin} 
\beta}\,, \nonumber \\ y_b &=& {\sqrt{2} m_b \over v_1} = {\sqrt{2} m_b \over 
v \textrm{ cos} \beta}\,.
\eeqa
Expressing then the 
fields $H_1$ and $H_2$ in terms of the physical fields, one 
can obtain the Yukawa couplings in terms of the fermion masses.
For the charged Higgses and the pseudoscalar $A$ it 
results\footnote{A factor of $i$ is understood
for the pseudoscalar $A$.}
\beqa
y_{H^+bt^c} &=& - y_{htt^c}^{SM} {1\over \tan\beta} \;, \nonumber \\
y_{H^-tb^c} &=&   y_{hbb^c}^{SM} \tan\beta \nonumber \;, \\
y_{A tt^c} &=&   y_{htt^c}^{SM} {1\over \tan\beta} \;, \nonumber \\
y_{A bb^c} &=&  y_{hbb^c}^{SM}  \tan\beta \,. \\
\label{gHff}
\eeqa
For the neutral $CP$-even scalars, in the decoupling limit one
gets 
\beqa
y_{htt^c} &=&  \frac{\sqrt{2} m_t}{v}  \frac{\cos \alpha}{\sin\beta} 
\approx  y_{htt^c}^{SM} \;, \nonumber \\ 
y_{hbb^c} &=& \frac{\sqrt{2} m_b}{v}  \frac{\cos \alpha}{\cos\beta} 
\approx  y_{hbb^c}^{SM} \;,  \nonumber \\ 
y_{Htt^c} &=&  \frac{\sqrt{2} m_t}{v}  \frac{\sin\alpha}{\sin\beta} 
\approx - y_{htt^c}^{SM} {1\over \tan\beta} \;, \nonumber \\ 
y_{Hbb^c} &=&  \frac{\sqrt{2} m_b}{v}  \frac{\cos\alpha}{\cos \beta} 
\approx  y_{hbb^c}^{SM} \tan\beta \,.
\eeqa
Therefore, all these Higgs bosons have couplings to the top quark
that are of the same order as the Yukawa coupling in the SM.

\Section{Top pair production through SUSY neutral bosons}

The MSSM incorporates 
two Higgs doublets, and after EW
symmetry breaking there are two extra neutral bosons ($H$ and
$A$)  that may be produce through
top-quark loops at hadron colliders. 
The mass of these two fields is not EW (it is SUSY breaking), 
so once produced 
they are {\it naturally} heavy enough to decay into
$t \bar t$. Their mass difference is EW,
between $-2$ and $+10$ GeV.
Moreover, the scalar masses $m_{A,H}$ of interest correspond to the decoupling
regime, where $h$ is basically the SM Higgs and 
\beq
y_{H  t t^c}\approx -{m_t\sqrt{2}\over v}\;{1\over \tan\beta}\approx
-y_{A t t^c}\;.
\eeq

Finally, we will focus on relatively 
low values of $\tan\beta$, where
the decay into bottom quarks is not important and the 
(energy-dependent) widths can be approximated to
\beq
\Gamma_H(s)\approx {3\;y_{H t t^c}^2\; s\;\beta_t^3\over 16\pi\; m_H}\;,
\;\;\;
\Gamma_A(s)\approx {3\;y_{A t t^c}^2\; s\;\beta_t\over 16\pi\; m_A}\;.
\eeq

\begin{figure}
\begin{center}
\begin{tabular}{ccc}
\includegraphics[width=0.45\linewidth]{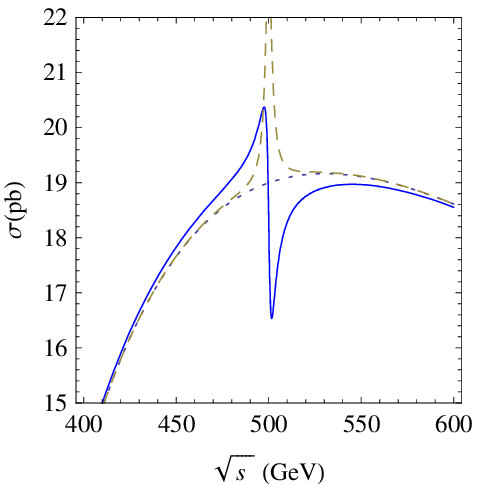} & $\;\;\;$ &
\includegraphics[width=0.45\linewidth]{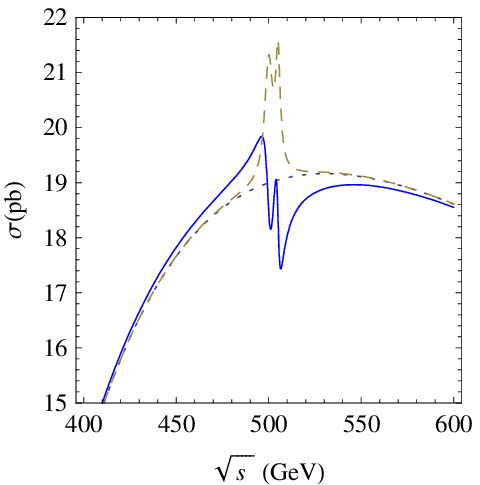} 
\end{tabular}
\end{center}
\caption{$\sigma(gg\rightarrow t\bar t)$ for $\tan\beta=2$
and SUSY bosons of mass $m_A=m_H=500$ GeV (left)
or $m_A=500$ GeV and $m_H=505$ GeV (right). Dots provide the SM cross section and dashes neglect the interference. 
\label{figS3}}
\end{figure}

\begin{figure}
\begin{center}
\includegraphics[width=0.43\linewidth]
{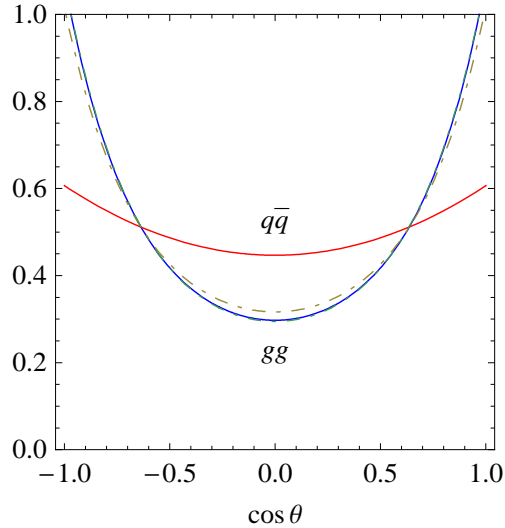} 
\end{center}
\caption{Standard angular distribution for the $t$ quarks
from $q\bar q$ and $g g$ collisions 
at $\sqrt{s}=500$ GeV. We include (dashes) the distribution
from $g g$ at the peak and the dip of Fig.~\ref{figS3}--left.
\label{figS4}}
\end{figure}

In Fig.~\ref{figS3} we plot $\sigma(gg\rightarrow t\bar t)$ at center of
mass energies around $m_A=500$ GeV for  
$m_H=500$ GeV (left) and $m_H=505$ GeV (right). We have
taken $\tan\beta=2$, which implies $\Gamma_H\approx 3.0$ GeV 
and $\Gamma_A\approx 5.3$ GeV.
We observe an average 5.5\% excess and  
8.1\% deficit in the 5 GeV intervals before and after 
$\sqrt{s}=500$ GeV, respectively. We include in dashes
the result ignoring the interference (last term in 
Eqs.~(\ref{csphi}) and (\ref{csA})), which would not be captured if one
uses the narrow-width approximation. It is apparent that the 
interference with the standard amplitude gives the dominant effect. 
In Fig.~\ref{figS3}--left the position of
the peaks and dips caused by
$H$ and $A$ overlap {\it constructively}
(notice, however, that in this $CP$ conserving Higgs sector 
their amplitudes do not interfere). In contrast, in  Fig.~\ref{figS3}--right
their mass difference implies a partial cancellation between 
the dip caused by $A$ and
the peak of $H$. 

The scalar and pseudoscalar couplings with the top quark grow
at smaller values of $\tan\beta$, increasing 
the cross section and the scalar width.
For example, for $\tan\beta=1$ the excess at $\sqrt{s}<500$ GeV
grows to the 6.2\% and 
the deficit to the 9.7\%, whereas for $\tan\beta=5$ the excess 
and deficit
are just a 2.1\% and a 2.6\%, respectively.

The normalized 
angular distribution of the $t$ quark in the center of mass
frame is given in Fig.~\ref{figS4}. We plot the standard distributions for
top-quark production in $g g$ and $q\bar q$ collisions
together with the distribution from $gg$ at the peak and the 
dip obtained in Fig.~\ref{figS3}--left. In the narrow-width 
approximation a scalar resonance gives a flat contribution. However, we
find that the excess or deficit from the scalar interference 
\begin{figure}[h]
\begin{center}
\begin{tabular}{ccc}
\includegraphics[width=0.45\linewidth]{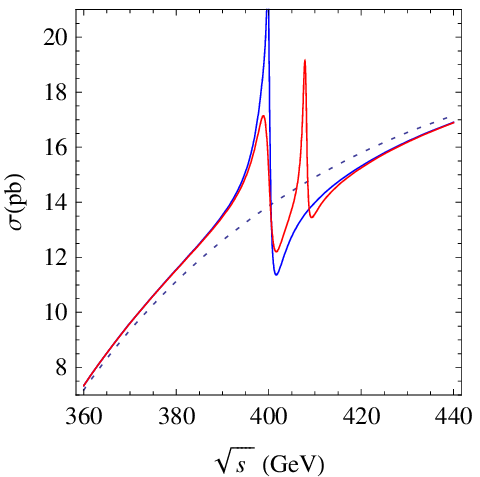}  & $\;\;\;$ &
\includegraphics[width=0.46\linewidth]{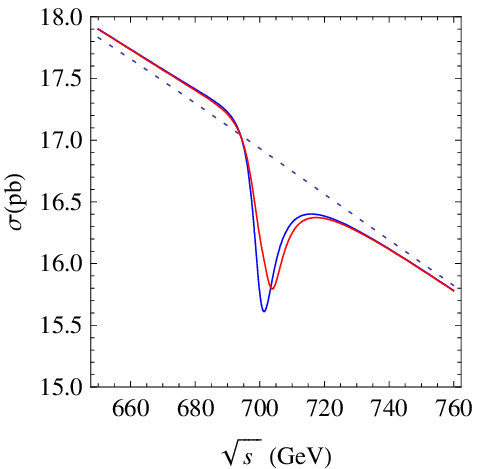} 
\end{tabular}
\end{center}
\caption{$\sigma(gg\rightarrow t\bar t)$ for $m_A=400$ GeV
and $m_A=700$ GeV. We include the cases 
$m_H=400,408$ GeV (left) and $m_H=700,703$ GeV (right). 
\label{figS5}}
\end{figure}
is {\it not} flat and 
does not change significantly the angular distribution. Different
cuts could be applied to reduce the background for $t\bar t$ 
production at the LHC \cite{Barger:2006hm} or even to optimize 
the contribution from $gg$ versus $q\bar q$, but not to enhance
the relative effect of the scalars on $\sigma(gg\rightarrow t\bar t)$.

In Fig.~\ref{figS5} we plot the parton cross section for lower and higher
values of the pseudoscalar mass ($m_A=400, 700$ GeV).
We include the cases where the boson $H$ is degenerate with
$A$ or slightly heavier ($m_H=408$ GeV and $m_H=703$ GeV). We see that 
at lower scalar masses the peak dominates, whereas for
large values of $m_A$ the dip is the dominant effect. This
behaviour, related to the slope of the standard cross section,
reduces in both cases the relevance of the mass difference between the
scalar and the pseudoscalar Higgses.

\Section{Top pair production through LH bosons}

In Chapter 2 we studied in detail LH models. 
There the Higgs appears as a pseudo-GB of a global symmetry broken spontaneously 
at the scale $f>v/\sqrt{2}=174$ GeV. The global 
symmetry introduces an extra $T$ quark of mass 
$\approx y_tf$
that cancels top-quark quadratic corrections 
to the Higgs mass parameter. In addition, the radial
field $r$ associated to the scalar breaking the global
symmetry will have a mass $\approx \sqrt{\lambda} f$
and a large coupling to $T$.
The presence of this vectorlike $T$ quark and of the
massive scalar singlet (the {\it Higgs} 
of the symmetry broken at the scale $f$) 
are then generic features in all these models.

Once the electroweak VEV is included the 
doublet and singlet Higgses (and also the $t$ and $T$
quarks) mix \cite{Barcelo:2007if,Barcelo:2008je}. 
The singlet component $\approx v/(\sqrt{2} f)$
in $h$ will reduce its coupling both to the
top quark and to the gauge bosons and, in turn, 
$r$ will get a doublet
component that couples to these fields.
\begin{figure}
\begin{center}
\begin{tabular}{ccc}
\includegraphics[width=0.33\linewidth]{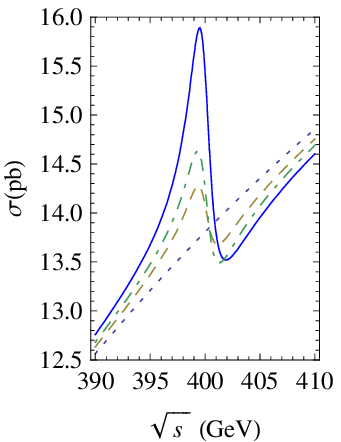} & 
\includegraphics[width=0.32\linewidth]{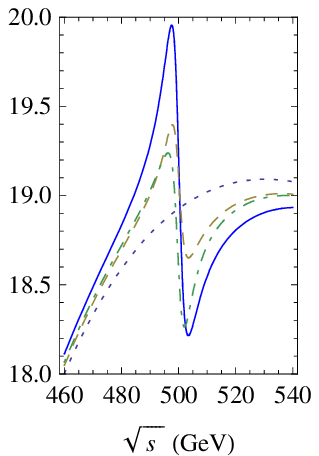} & 
\includegraphics[width=0.32\linewidth]{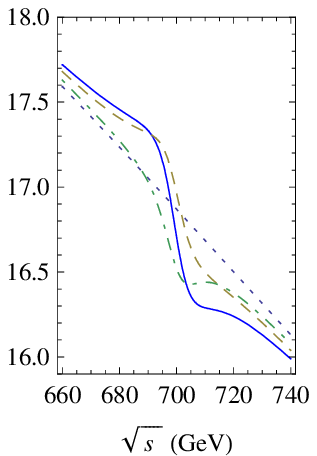} 
\end{tabular}
\end{center}
\caption{$\sigma(gg\rightarrow t\bar t)$ for a LH model with 
$m_r=500$ GeV and $m_T=400,500,700$ GeV. 
Dashes (dot-dashes) correspond to an amplitude with only the 
$T$ ($t$) quark loop.
\label{figS6}}
\end{figure}

The most general\footnote{There could
be an additional mixing, $T\rightarrow c_\beta T + s_\beta t$ 
in the second line of Eq.~(\ref{yt}), but it must be small 
\cite{Aguilar-Saavedra:2002kr} to avoid a too large value of $V_{Tb}$.}
top-quark Yukawa sector with no quadratic corrections at
one loop is 
\beqa
-{\cal L}_t &=& \lambda \; \left( f+{r\over \sqrt{2}}\right)
\sin{u+h\over\sqrt{2}f} \left(c_\alpha t + s_\alpha T \right) t^c
\nonumber \\
&+& \lambda \; \left( f+{r\over \sqrt{2}}\right)
\cos{u+h\over \sqrt{2}f} \; T T^c
+ {\rm h.c.}\;,\label{yt}
\eeqa
where $u$ and $f$ are VEVs satisfying
\beq
f\; \sin {u\over\sqrt{2}f} \equiv f\; s_\theta = {v\over \sqrt{2}}\;.
\eeq
Eq.~(\ref{yt}) becomes 
\beqa
-{\cal L}_t &=& \lambda \; \left( f+{r\over \sqrt{2}}\right)
\left( s_\theta \cos {h\over\sqrt{2}f} +  
c_\theta \sin{h\over\sqrt{2}f} \right)
 \left(c_\alpha t + s_\alpha T \right) t^c +
\nonumber \\
&& \lambda \; \left( f+{r\over \sqrt{2}}\right)
\left( c_\theta \cos {h\over\sqrt{2}f} - 
s_\theta \sin{h\over\sqrt{2}f} \right) T T^c
+ {\rm h.c.}\;.\label{yt2}
\eeqa
Fermion masses, Yukawa couplings and dimension-5 operators
(necessary to check the cancellation of all one-loop 
quadratic corrections) are then obtained by expanding 
\beq
\cos {h\over\sqrt{2}f}\approx 1-{h^2\over 4 f^2}\;,\;\;\;
\sin {h\over\sqrt{2}f}\approx {h\over \sqrt{2} f} \;.
\eeq
The fermion masses and the Yukawas to the heavier scalar 
$r$ have the same structure,
\beq
-{\cal L}_t\supset \lambda \; \left( f+{r\over \sqrt{2}}\right)
\left(\begin{array}{cc} t & T \end{array}\right)
\left(\begin{array}{cc} s_\theta c_\alpha & 0 \\
s_\theta s_\alpha &  c_\theta^2 \end{array}\right)
\left(\begin{array}{c} t^c \\ T^c \end{array}\right)\;.
\eeq
This implies 
\beq 
y_{r t t^c}={m_t\over f}={\sqrt{2}\; s_\theta\; m_t\over v}\;\;\;
{\rm and}\;\;\;
y_{r T T^c}={m_T\over f}\;,
\eeq
where the quarks are mass eigenstates. 
The mass of the heavier $T$ quark is 
$m_T\approx m_t c_\theta/(s_\theta c_\alpha)$, and its mixing
with the doublet 
$V_{Tb}\approx s_\theta^2 s_\alpha c_\alpha/c_\theta^2$.

The extra Higgs $r$ is somehow similar to the heavier scalar 
in a doublet plus singlet model, with the 
doublet component growing with $s_\theta=v/(\sqrt{2} f)$.
If $s_\theta$ is sizable so is its coupling to the top
quark. The coupling to the extra $T$ quark is stronger, 
but if $r$ is lighter
than $2m_T$ its main decay mode will be into $t\bar t$.
Actually, the doublet component in $r$ may also imply large
couplings to the would-be GBs for large values
of $m_r$. More precisely, its decay width $\Gamma_r(s)$ 
at $4m_t^2<s<4m_T^2$ is
\beq
\Gamma_r(s)\approx  
{3\;s_\theta^2\; s\over 8\pi\; v^2} \left[
{m_t^2\;\beta_t^3 \over m_r} 
+{s_\theta^2\; m_r \over 4} \left(
\beta_V^3 +{3\over 4} \beta_V(1-\beta_V^2)^2\right)
\right]\;.
\eeq
Therefore, $r$ is a naturally heavy ($m_r\approx f$) 
but narrow scalar
resonance with large couplings to quarks and an order one
branching ratio to $t \bar t$.

In Fig.~\ref{figS6} we plot the parton-level cross section 
$\sigma(gg\rightarrow t\bar t)$ for $s_\theta=0.5$, $m_T=500$ GeV
and several values of $m_r$. 
We separate the contributions from the top
and the $T$ quark loops (the second one vanishes at $s=m_r^2$).
The plot is similar to the one obtained for SUSY bosons of
the same mass.
At higher values of $m_r$ the decay width 
$\Gamma_{r}$ grows, diluting the 
effect (see Fig.~\ref{figS6}--right). In
contrast, for lower masses the scalar $r$ has a 
narrow width and is strongly coupled 
to quarks, which produces a larger effect (in Fig.~\ref{figS6}--left).
The contribution from the standard $t$-quark loop grows with
$s_\theta$, whereas the contribution from the
extra $T$-quark is 
basically independent of $m_T$.

\Section{Signal at the LHC}

Let us now estimate the invariant mass distribution 
of $t\bar t$ events ($m_{t\bar t}$) in $pp$ collisions at the LHC. 
To evaluate the hadronic cross sections we 
will use the MSTW2008 parton distribution functions (PDFs) \cite{Martin:2009iq}. 
The effect of next-to-leading order (NLO) corrections on the
expressions given in previous sections has been 
studied by several groups (see for example 
\cite{Frederix:2007gi,Frixione:2003ei}).
In particular, the authors in \cite{Frederix:2007gi} analyze 
the dependence of ${\rm d}\sigma/{\rm d} m_{t\bar t}$ on the choice of 
renormalization and factorization scales and of PDFs. They 
show that if the LO cross section 
is normalized to the NLO one at low values of 
$m_{t\bar t}$, 
then the deviations introduced by these scales and by
the uncertainty in the PDFs at $m_{t\bar t}<1$ TeV 
are small (order 10\%). For (scalar and
pseudoscalar) Higgs production in $pp$
collisions and Higgs decay, a complete 
review of NLO results can be found in \cite{Djouadi:2005gj}. 
From the expressions there we obtain that QCD corrections 
enhance the production cross section  
in approximately 
a 20\%, and that the Higgs decay width into $t\bar t$ (for 
$m_{\phi}\gg 2m_t$) is also increased in around a 10\%.
Given these estimates, we have evaluated $pp\rightarrow t\bar t$ 
taking fixed renormalization and factorization 
scales ($\mu_{R,F}=m_t$) and normalizing the LO result 
to the NLO cross section in \cite{Frederix:2007gi} with a 
global factor of 1.3. Our differential cross section 
coincides then with that NLO result at $m_{t\bar t}=500$ GeV.

\begin{figure}
\begin{center}
\begin{tabular}{ccc}
\includegraphics[width=0.33\linewidth]{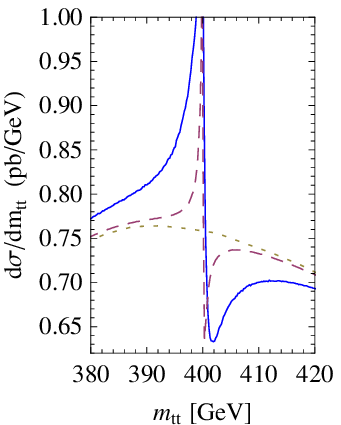} & 
\includegraphics[width=0.318\linewidth]{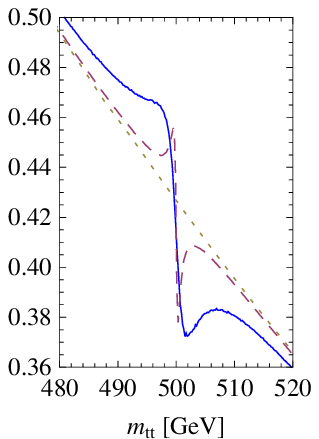} & 
\includegraphics[width=0.328\linewidth]{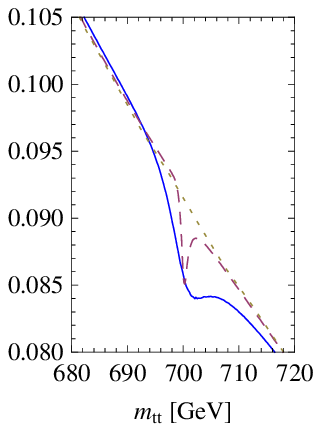} 
\end{tabular}
\end{center}
\caption{
${\rm d}\sigma/{\rm d} m_{t\bar t}$ in SUSY models with
$m_H=m_A=400,500,700$ GeV and $\tan\beta=2$ (solid),
$5$ (dashes). 
\label{figS7}}
\end{figure}

\begin{figure}
\begin{center}
\begin{tabular}{ccc}
\includegraphics[width=0.33\linewidth]{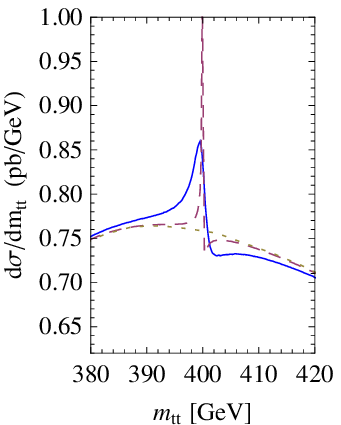} & 
\includegraphics[width=0.318\linewidth]{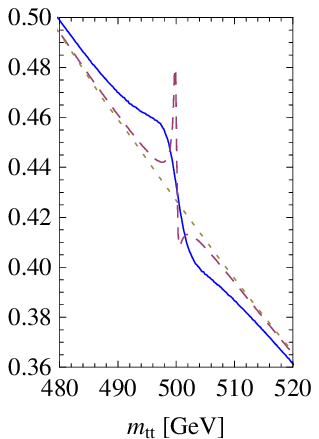}  & 
\includegraphics[width=0.328\linewidth]{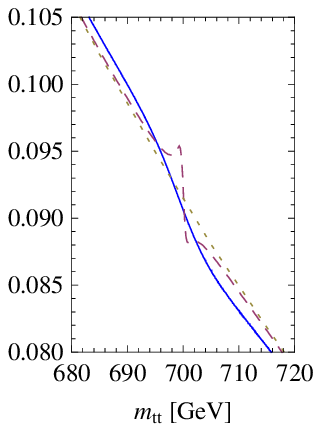} 
\end{tabular}
\end{center}
\caption{
${\rm d}\sigma/{\rm d} m_{t\bar t}$ in LH models with 
$m_r=400,500,700$ GeV and $s_\theta=0.5$ (solid),
$0.2$ (dashes).
\label{figS8}}
\end{figure}

We will take a center of mass energy of 7 TeV. We obtain
that at these energies the cross section $pp\rightarrow t\bar t$
is dominated by $gg$ fusion, with 
$q\bar q\rightarrow t\bar t$ accounting for just 10\% of the
top-quark pairs. In Figs.~\ref{figS7}, \ref{figS8} we plot
${\rm d}\sigma/{\rm d} m_{t\bar t}$ 
for some of the SUSY and LH models described before.
These figures {\it translate} 
the parton-level cross sections in 
Figs.~\ref{figS3}, \ref{figS5}, \ref{figS6} into anomalies 
in the invariant mass distribution in $pp$ collisions.

\begin{figure}
\begin{center}
\includegraphics[width=0.47\linewidth]{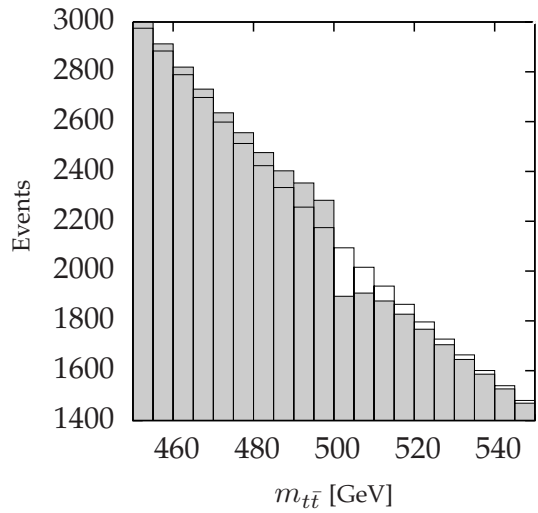} 
\end{center}
\caption{Number of $t\bar t$ events in $pp$ collisions
at 7 TeV and 1 fb$^{-1}$ for
$m_A=m_H=500$ GeV and $\tan\beta=2$ distributed in 5 GeV bins.
\label{figS9}}
\end{figure}

\begin{figure}
\begin{center}
\begin{tabular}{ccc}
\includegraphics[width=0.47\linewidth]{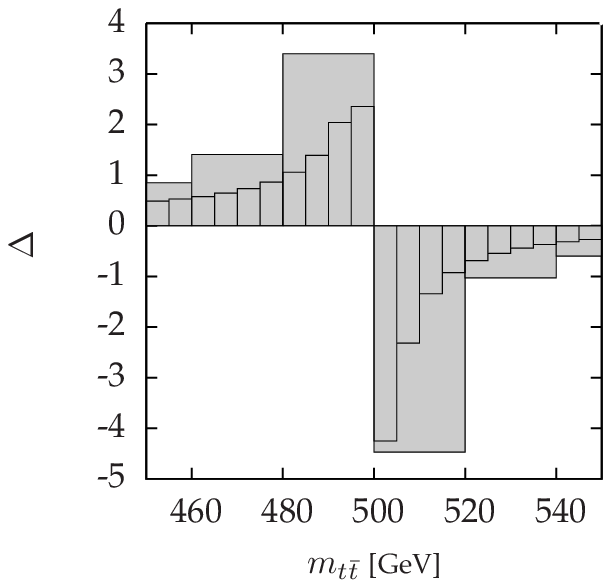} & $\;\;\;$ &
\includegraphics[width=0.47\linewidth]{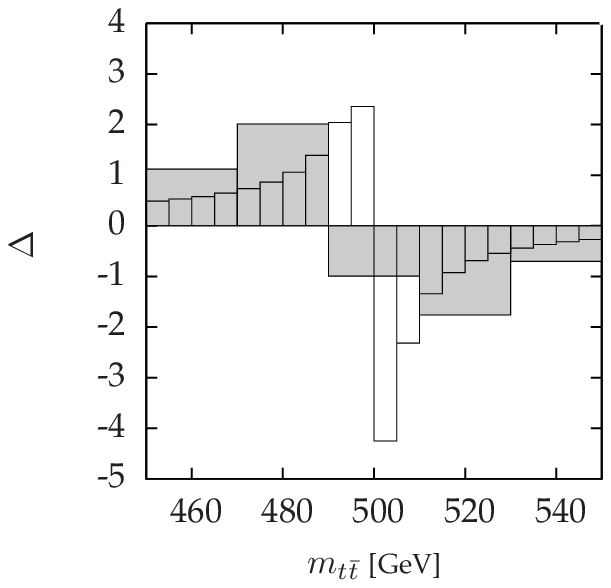} 
\end{tabular}
\end{center}
\caption{Deviation $\Delta=(N-N_{SM})/\sqrt{N_{SM}}$ in the 
number of events 
respect to the standard prediction for two
different binning ($m_A=m_H=500$ GeV and $\tan\beta=2$). 
\label{figS10}}
\end{figure}

To estimate the possible relevance at the LHC of these 
cross sections, 
we will calculate the number of $t\bar t$ events
assuming an integrated 
luminosity of 1 fb$^{-1}$ (we will not apply any cuts). 
In Fig.~\ref{figS9} we plot the number of events per 5 GeV bin of
$m_{t\bar t}$ in
the SUSY model with $m_A=m_H=500$ GeV and $\tan\beta=2$.
We observe a 5\% excess followed by a 9\% deficit, with
smaller deviations as $m_{t\bar t}$ separates from the mass
of the extra Higgs bosons. In Fig.~\ref{figS10} we distribute the 
events in 20 GeV bins and plot the statistical significance 
\beq
\Delta \equiv {N-N_{SM}\over \sqrt{N_{SM}}}
\eeq
of the deviations, where $N$ is the total number
of events in the bin.
The typical signal is an increasing excess in 
a couple of 20 GeV bins that may reach a $+3.4\sigma$ deviation
followed by a deficit of $-4.5\sigma$.
We find that changing the binning is important in order to
optimize the effect. If the same 20 GeV bin includes the peak and
the dip (Fig.~\ref{figS10}, right) then the maximum deviation is just a
$\pm 2\sigma$ effect.

The result is very similar for a LH scalar of
$m_r=500$ GeV with $s_\theta=0.5$. In this LH model
we obtain deviations in consecutive 20 GeV bins 
reaching $+2.5\sigma$ and $-2\sigma$. However, the effect is a bit
more localized, and the cancellation if peak and dip coincide
in a bin is stronger: it may result in three bins 
with just $+1.3\sigma$, $+0.6\sigma$ and $-1.2\sigma$ deviations.

The binning is less important for larger Higgs
masses. For example, in the SUSY case with $m_A=m_H=700$ GeV 
the typical sequence is a couple of 20 GeV bins with a slight
$+0.2\sigma$ excess followed by $-1.2\sigma$, $-0.4\sigma$ and 
$-0.2\sigma$ deficits. In the LH model with $m_r=700$ GeV
the initial excess (caused by the $T$-quark loop)
is a bit more significant, a typical sequence
would consist of two bins with $+0.4\sigma$ excess followed by 
$-0.8\sigma$ and $-0.4\sigma$ deficits.

\begin{figure}
\begin{center}
\begin{tabular}{ccc}
\includegraphics[width=0.47\linewidth]{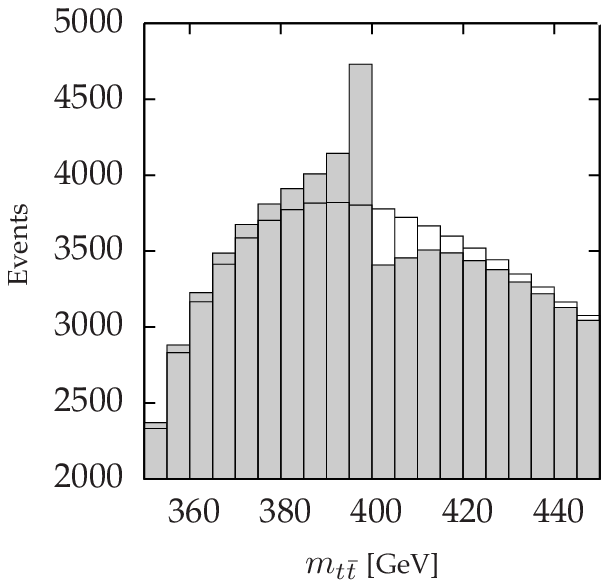} & $\;\;\;$ &
\includegraphics[width=0.47\linewidth]{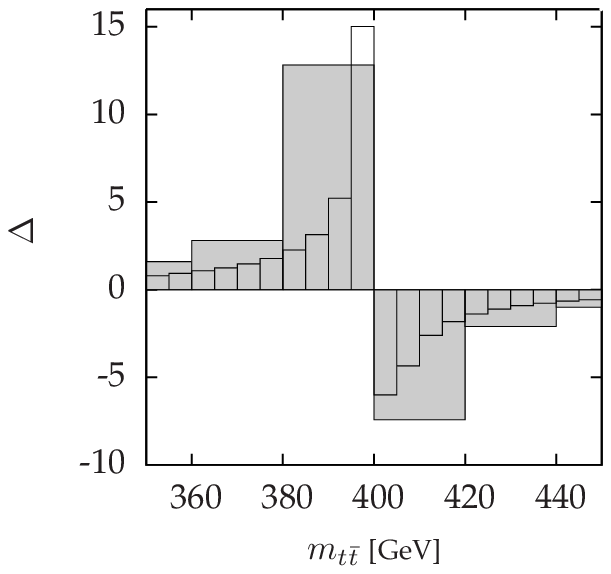} 
\end{tabular}
\end{center}
\caption{Number of $t\bar t$ events in $pp$ collisions (left) 
and deviation $\Delta$ (right)
for $m_A=m_H=400$ GeV and $\tan\beta=2$.
\label{figS11}}
\end{figure}

\begin{figure}
\begin{center}
\begin{tabular}{ccc}
\includegraphics[width=0.47\linewidth]{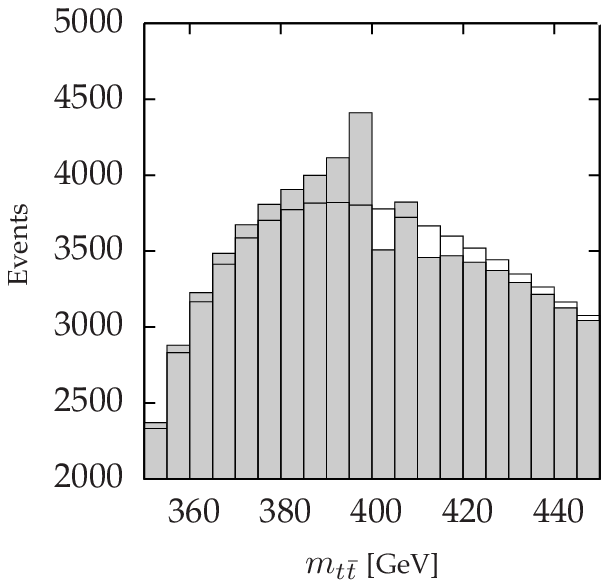} & $\;\;\;$ &
\includegraphics[width=0.47\linewidth]{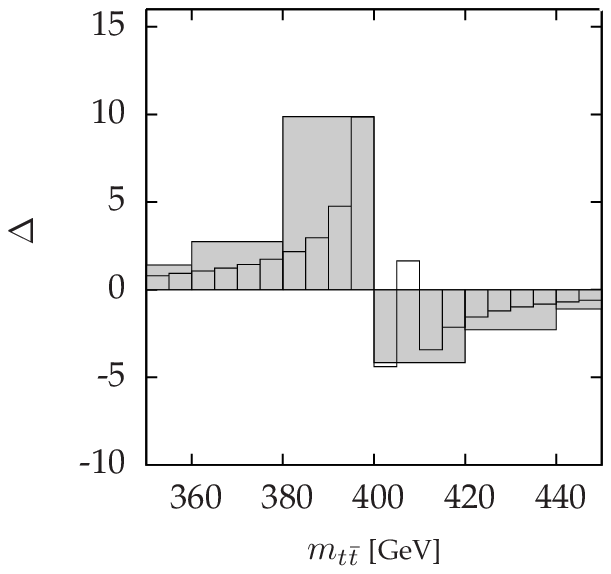} 
\end{tabular}
\end{center}
\caption{Number of $t\bar t$ events in $pp$ collisions (left) 
and deviation $\Delta$ (right)
for $m_A=400$ GeV, $m_H=408$ GeV and $\tan\beta=2$.
\label{figS12}}
\end{figure}

Let us finally focus on lighter Higgses, as they provide the
most promising signal. In Fig.~\ref{figS11} we plot the event distribution
(left) and the statistical significance (right) 
for $\tan\beta=2$ and $m_A=m_H=400$ GeV,
whereas Fig.~\ref{figS12} corresponds to a mass difference of 8 GeV
($m_A=400$ GeV and $m_H=408$ GeV). The sequence of deviations in
both cases would be seen as a clear anomaly, reaching an excess
of up to $13\sigma$ (for $m_H-m_A=-2$ GeV) in a 20 GeV bin. 
The LH case
is analogous but, again, more localized. We obtain an excess 
of $+3.4\sigma$ in a 20 GeV bin followed by a $-1.7\sigma$ deficit.

Since no anomaly has been found in the $m_{t \bar{t}}$ at the LHC, the scenario
with a low $\tan \beta$ ($\leq 2 $) and light extra Higsses ($m_{A,H} \sim 400 - 500$ GeV) seems excluded. A more precise
statement would require a more elaborate simulation using the experimental cuts and event selection and reconstruction.

Finally, we would like also to comment on the possibility that
these processes introduce anomalies at the Tevatron, which achieved
almost 10 fb$^{-1}$ at $\sqrt{s}=1.96$ TeV. The main difference with
the LHC is that at the Tevatron 90\% of the top-quark pairs
were produced through $q\bar q$ interactions. Since the signal 
that we have explored is caused by interference in the 
$gg\rightarrow t\bar t$ channel, for the same integrated
luminosity the deviations there would be 
9 times weaker than at the LHC (where gluon fusion provides
90\% of the top pairs). We find, however, that 1$\sigma$ 
deviations could be obtained at the Tevatron 
for low masses of the heavy Higgs bosons.
This signal could be enhanced by {\it separating} the $t\bar t$
events in two or three sets according to the $\cos \theta$
of the final $t$ quark. As we see in Fig.~\ref{figS4}, the $gg$ and
$q \bar q$ contributions at 
$m_{t\bar t}\approx m_{\phi} \approx 500$ GeV 
have different angular distributions
(this difference, however, vanishes at lower invariant 
masses). One could separate, for example, the events with 
$\left| \cos\theta \right|$
larger or smaller than 0.6. Then the
anomalies in $d\sigma/ d m_{t\bar t}$ that we have discussed
should increase in the $\left| \cos\theta \right| > 0.6$
interval. Unfortunately, after researchers from Tevatron became 
interested in our work, we found that the signal is too small to be
observable in that collider.

%% file: Ch4.tex

\def \lsim{\mathrel{\vcenter
     {\hbox{$<$}\nointerlineskip\hbox{$\sim$}}}}
\def \gsim{\mathrel{\vcenter
     {\hbox{$>$}\nointerlineskip\hbox{$\sim$}}}}

{\Chapter{The $t\bar t$ 
FB Asymmetry, Massive Gluons and Heavy Quarks}\label{Ch4}}

The 1.96 TeV Tevatron has recorded 
over 10 fb$^{-1}$ of data, while the 7 TeV LHC is already above
4 fb$^{-1}$ of integrated luminosity.
This means that the physics beyond the
SM is currently being searched with an important
degree of detail. Until now no discovery has been
reported by any experimental collaboration, although the 
Tevatron has observed a {\it persistent} anomaly in $t\bar t$
physics that is an intriguing departure from the SM 
predictions and that will be the main object of this chapter.

Generically, the absence of experimental anomalies 
in $pp$ collisions at the LHC puts bounds on 
extensions of the SM that may reach the TeV scale and 
sometimes higher. As a consistent alternative
with the Tevatron $t\bar t$ data, however, 
these results may just imply that the experimental signature of the
new physics is peculiar and easy to miss despite being at relatively 
low scales. After all, the presence of new physics below the 
TeV scale in the top-quark sector is a clear result 
from naturalness arguments. 
We will take this approach and will study the $t\bar{t}$
forward-backward (FB) asymmetry at the Tevatron~\cite{AFBTEV,AFB2,AFB3}
and its implications at the LHC.

We have explained in previous chapters that 
the large coupling of the top quark to the
EW symmetry breaking sector implies that the 
new physics stabilizing the
latter could also appear in top-quark observables. This argument makes the \mbox{2--3$\sigma$} deviation (see below) versus
the standard value in the Tevatron FB asymmetry specially interesting.
Even if it is not
statistically significant at the level of discovery, 
the consistency among different CDF and D$\emptyset$ measurements 
strengthens the case for new physics.
However, any candidate responsible for the
asymmetry has to be carefully disguised, as its large contribution there
should not translate into any significant departure from the SM in
other related observables. In particular, 
the $t\bar{t}$ total cross section, its 
invariant-mass distribution, dijet production, same-sign top-pair
production, or the $t\bar{t}$ charge asymmetry at the LHC are
observables where correlated anomalies could be expected 
\cite{Relatedobservables1,Relatedobservables2,Relatedobservables3,Relatedobservables4,Relatedobservables5,Relatedobservables6}.

We will start reviewing the status of the Tevatron asymmetry.
Then we will motivate the framework that we propose to explain the asymmetry: 
a \mbox{700--900 GeV} gluon of very large width caused by new decay channels. 
In particular, heavy quarks strongly coupled to the gluon 
will be introduced in the model. We will make a 
complete analysis of the model in order to reveal the
best strategy for its observation at the LHC. 
This study will include implementing the model in
MADGRAPH/MADEVENT v4~\cite{Alwall:2007st}, simulating the hadronization and showering effects with PYTHIA~\cite{pythia}, 
simulating the detectors with PGS~\cite{PGS4} and DELPHES~\cite{Ovyn:2009tx}
and reconstructing the final state using computer algorithms
as experimentalists do. Such a {\it sophisticated} analysis
is necessary to decide about the consistency of the model with 
current observations and its observability in future searches.
The results in this chapter have been published 
in \cite{Barcelo:2011fw,Barcelo:2011vk,Barcelo:2011wu}.

\vspace{0.7cm}

\Section{The top-quark FB Asymmetry}

The top quark was discovered in 1995 at the $p\bar p$ 
Tevatron collider at the Fermilab by the CDF and D$\emptyset$
collaborations. It is the heaviest elementary 
particle known so far, and its mass and total inclusive cross section
in pair production are currently known with a precision of 
about 1.1\% \cite{TeVprecision} and 10\% \cite{Abazov:2011cq}, respectively. 
Nevertheless, the measurements of other top-quark 
properties are still statistically limited, so the
question to be answered soon by the LHC is whether the SM 
successfully predicts all of these properties.

\begin{figure}[]
\begin{center}
\includegraphics[width=0.6\linewidth]{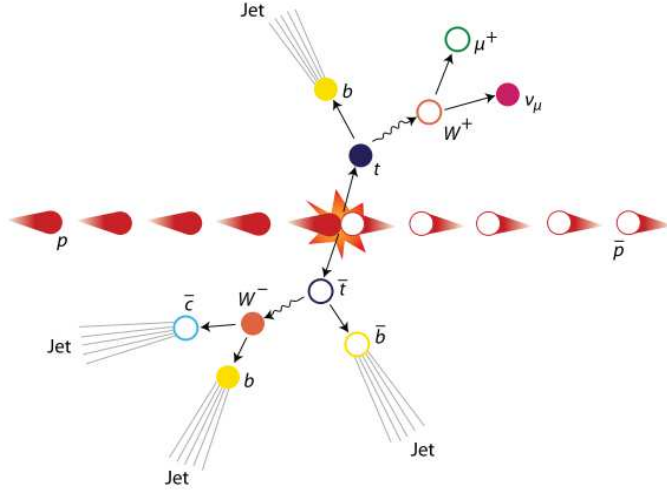} 
\caption{Typical $p \bar{p} \rightarrow t \bar{t}$ at the Tevatron.}
\label{collision}
\end{center}\end{figure}

At the Tevatron, with a c.o.m.~energy 
$\sqrt{s} = 1.96$ TeV, most top quarks are pair-produced via
strong interactions. In particular, quark annihilation  
contributes an 90\% while gluon-gluon
fusion provides a 10\% of the $t\bar t$ pairs. Due to 
the large value of its mass, top-quark production is an ideal
testing ground to study perturbative QCD effects. 
The $t\bar t$ FB asymmetry is one of such examples. 
It appears at NLO 
in $q\bar{q} \rightarrow t\bar{t}X$
reactions, and it translates into an asymmetry in $p\bar{p}$ 
collisions (Fig.~\ref{collision}) due to the higher content of  quarks in $p$ and
antiquarks in $\bar p$. In $pp$ collisions at the LHC the asymmetry
vanishes, although  one can use the fact that quarks tend
to carry more momentum than antiquarks in a proton to 
define a similar anomaly in events at intermediate rapidities \cite{LHCasy}.

\begin{figure}[t]
\begin{center}
\begin{tabular}{ccccc}
\includegraphics[width=0.2\linewidth]{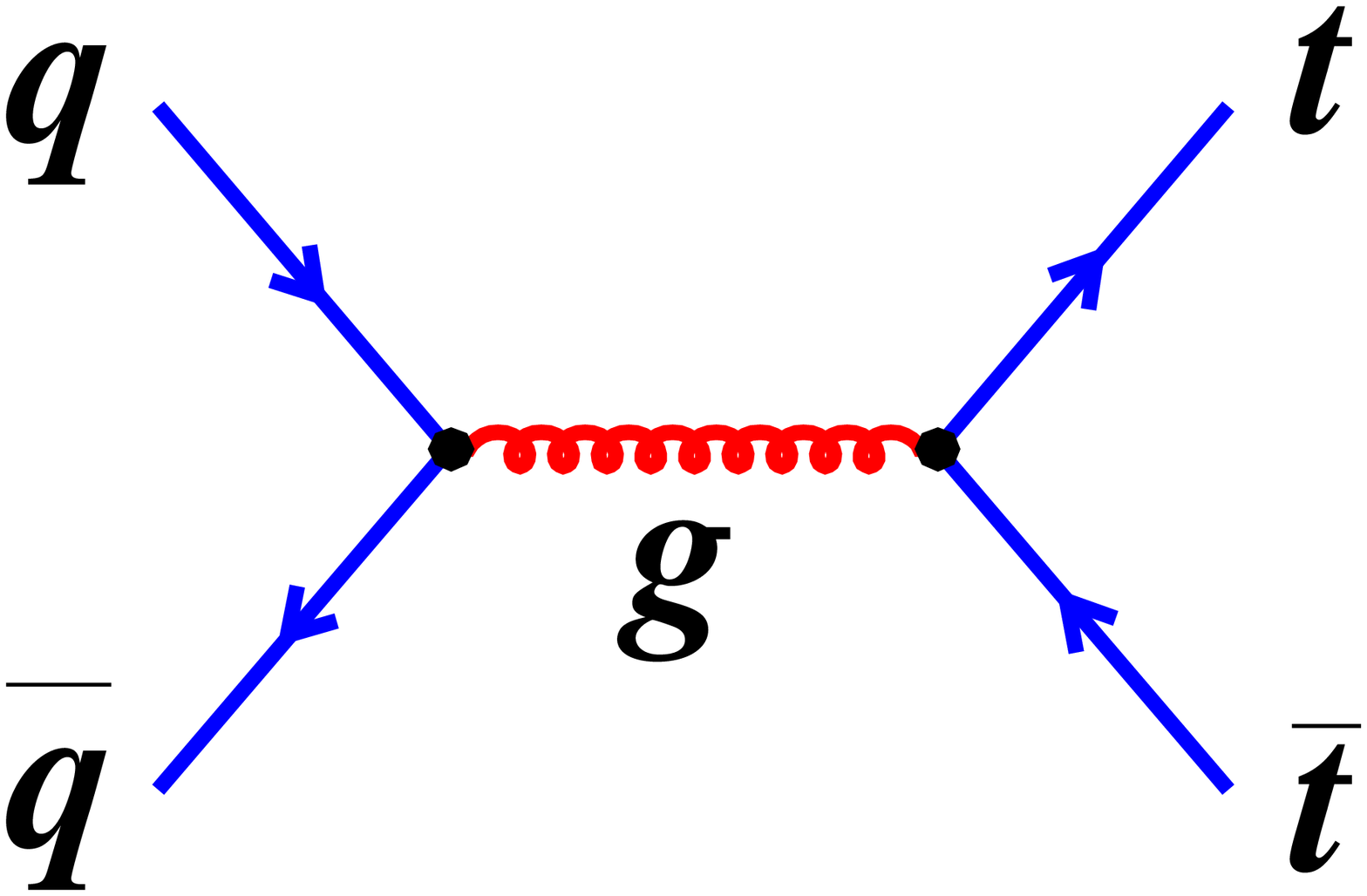} &
\includegraphics[width=0.2\linewidth]{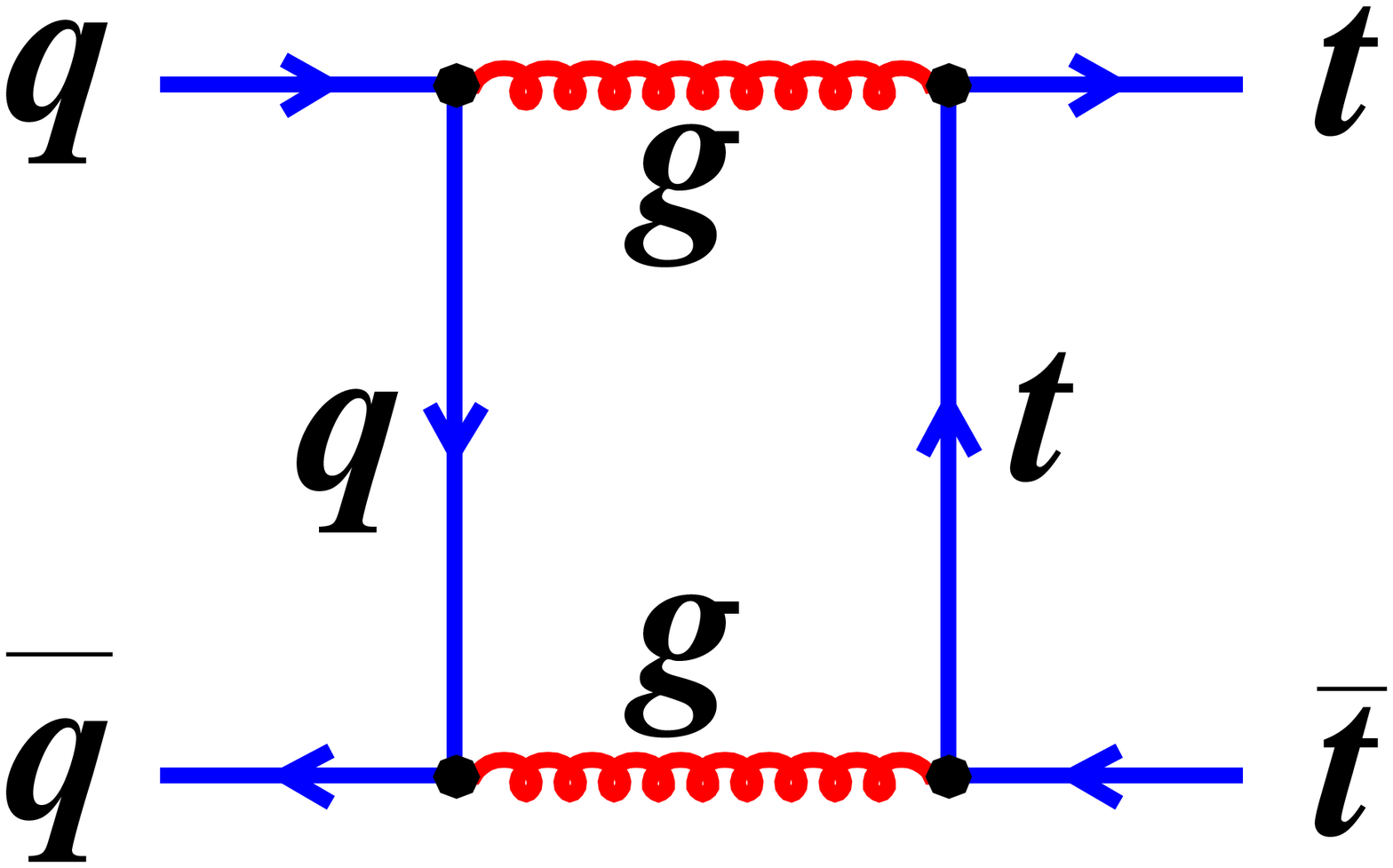} & \;\;\;\;\;\;\;\; &
\includegraphics[width=0.2\linewidth]{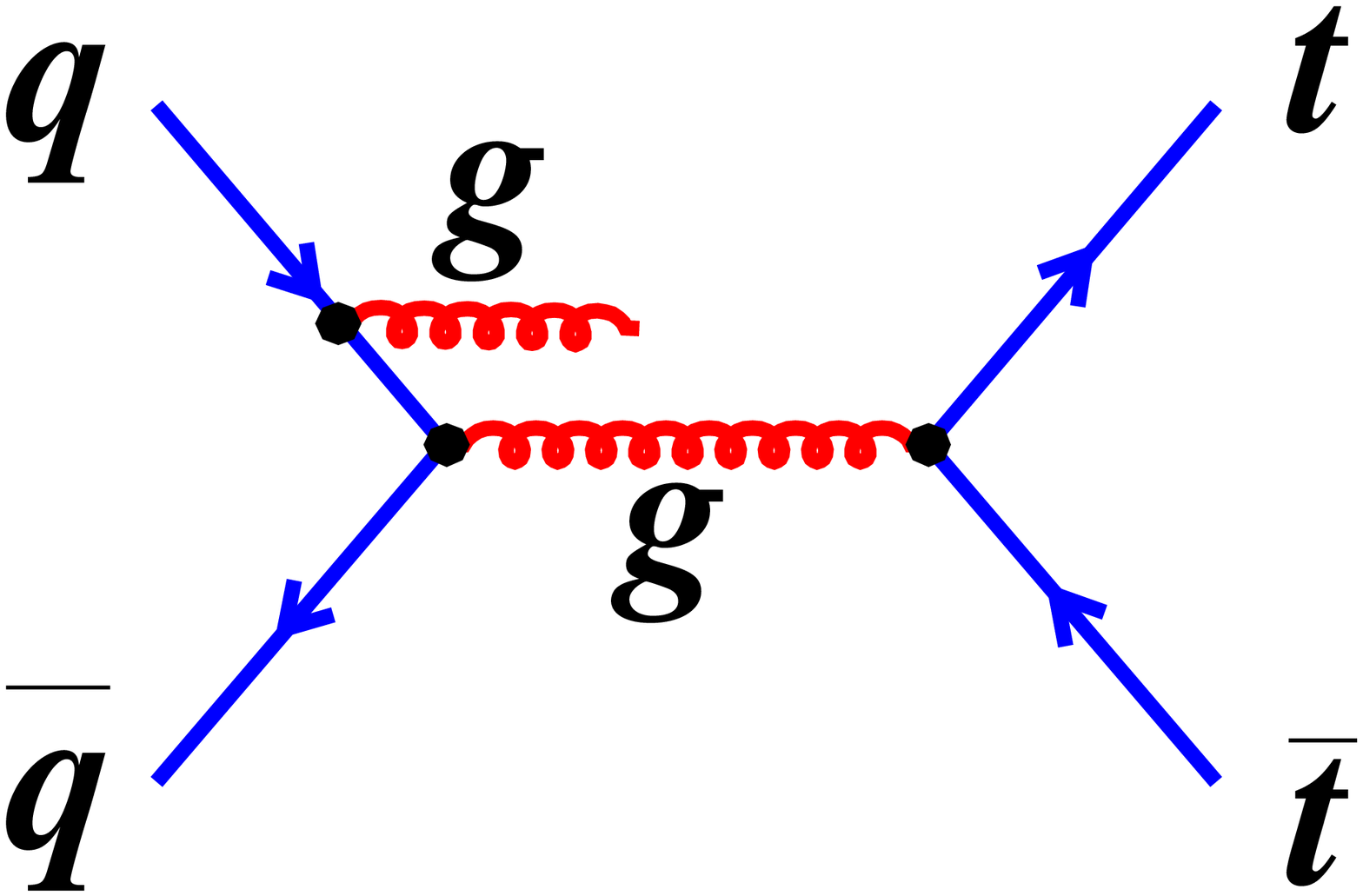} &
\includegraphics[width=0.2\linewidth]{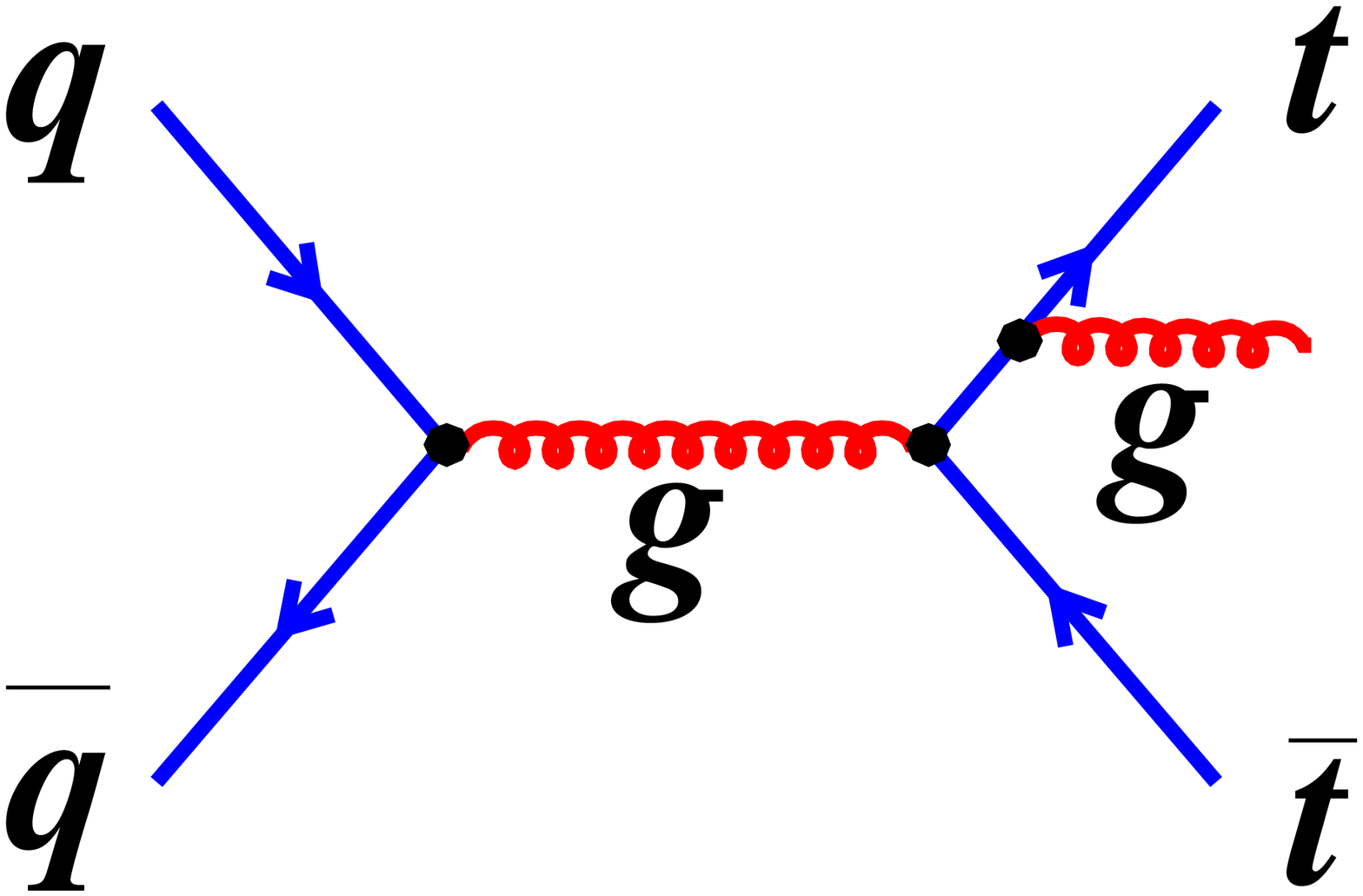} 
\end{tabular}
\end{center}
\caption{Leading order and box diagram (left) and ISR and FSR diagram (right).}
\label{box}
\end{figure}

Diagrammatically the asymmetry arises 
from the interference between initial and final state gluon
radiation on the one hand (Fig.~\ref{box}--right), and the interference 
of the Born and the box diagram on
the other hand (Fig.~\ref{box}--left) \cite{Kuhn:1998jr}. These are the only processes that contribute, as 
heavy flavor excitation $qg \rightarrow qt\bar{t}$ is negligible 
and gluon fusion symmetric.
The asymmetry is usually expressed  in terms of the variable 
$\cos \alpha$, 
where $\alpha$ is the angle of the top quark in the rest frame of the incoming
partons:
\begin{equation}
A(\alpha)= \frac{N_t(\cos \alpha) - N_{\bar{t}}(\cos \alpha)} {N_t(\cos \alpha) + N_{\bar{t}}(\cos \alpha)} = 
\frac{N_t(\cos \alpha) - N_t(-\cos \alpha)} {N_t(\cos \alpha) + N_t(-\cos \alpha)}\;.
\end{equation}
Since charge conjugation is a symmetry of the strong interactions, 
$\sigma_{q\bar{q} \rightarrow t\bar{t}} (\alpha) = 
\sigma_{\bar{q}q  \rightarrow\bar{t}t} (180◦ - \alpha)$ and this can be interpreted as a FB asymmetry:
\begin{equation}
A_{FB} = \frac{N_t(\cos \alpha \ge 0) - N_t(\cos \alpha < 0)} {N_t(\cos \alpha \ge 0) + N_t(\cos \alpha < 0)}\;.
\end{equation}

In $p\bar{p} \rightarrow t\bar{t}$ reactions at the Tevatron energies 
the SM predicts a (6--7)\% asymmetry\footnote{ EW radiative corrections would increase this prediction in a $1.4\%$
 \cite{Kuhn:1998kw,Hollik2011ps,Kuhn:2011ri}.} in the $t\bar{t}$ c.o.m.~frame. 
The interference of the $q\bar q$ Born and box amplitudes leads to 
a positive contribution,
while the interference between the initial and final state radiation 
amplitudes yields a smaller negative value. In different invariant-mass
intervals the SM gives
\begin{equation}
A_{SM}^{t\bar t} \approx \left\{
\begin{array}{l l} 
\displaystyle \phantom{-} 0.040 \pm 0.006, \quad
& m_{t\bar t}<450\;{\rm GeV}\,; \\
\phantom{-}0.088\pm 0.013,
& m_{t\bar t}>450\;{\rm GeV}\,,
\end{array} \right. 
\end{equation} 
where the values refer to the asymmetry measured
in the $t\bar{t}$ c.o.m.~frame. 
Both CDF and D$\emptyset$ have observed a significant deviation.
In particular, the CDF measurement with 5.3 fb$^{-1}$ gives
\begin{equation}
A_{Exp}^{t\bar t} \approx \left\{
\begin{array}{l l} 
\displaystyle -0.116 \pm 0.153, \quad
& m_{t\bar t}<450\;{\rm GeV}\,; \\
\phantom{-}0.475\pm 0.114,
& m_{t\bar t}>450\;{\rm GeV}\,,
\end{array} \right. 
\end{equation} 
implying a 3$\sigma$ deviation at large values of 
$m_{t\bar t}$ \cite{AFB3}.
The D$\emptyset$ measurement \cite{AFBTEV} is consistent with this one although with a weaker energy dependence.
In particular, D$\emptyset$ has also measured
the asymmetry ($A_{FB}^l$) given by the charged lepton from $t  \rightarrow W^+ b  \rightarrow l^+ \nu b$ and
$\bar{t}  \rightarrow W^- \bar{b}  \rightarrow l^- \bar{\nu} \bar{b}$.
Recent calculations \cite{Bernreuther:2010ny} predict $A_{FB}^l = (3.5 \pm 1)\%$ within the SM
while D$\emptyset$ finds at 5.4 fb$^{-1}$
$A_{FB}^l = (12.7 \pm 5.5)\%$ for events where the lepton
charge is positive and $A_{FB}^l = (15.6 \pm 5.0)\%$ for events
where the lepton charge is negative \cite{Abazov:2011rq} (all uncertainties are
statistical). This $2 \sigma$ deviation, again, goes in the same direction as the CDF data.

If caused by new physics, these unexpected results would be 
an order-one departure from the standard quark physics  
at 450--800 GeV, and similar
anomalies could be expected in other observables at
the Tevatron and the early LHC. However, the asymmetry 
has not been {\it supported} by current data on 
the total $t\bar t$ cross section or the invariant-mass 
distributions of top-quark pairs and dijets. As a consequence,
possible new particles proposed to explain it are typically 
pushed above 1--2 TeV, 
out of reach both from Tevatron energies and from the
current LHC luminosity. Such high values, in turn, become 
ineffective to produce the large asymmetry or should be apparent
with a slightly increased LHC luminosity, as general effective
Lagrangian studies indicate~\cite{Relatedobservables1,Relatedobservables2}.

We will show that a heavy gluon with mass
$700$--$900$ GeV and small and mostly axial-vector couplings to the light quarks and 
relatively large couplings to the right-handed top quark 
can still explain the observed asymmetry with no
conflict with current data. The mechanism that could hide 
it relies on a very large width
caused by new decay channels opening at $\sqrt{\hat s}\lsim 600$
GeV.

\Section{Vector and axial-vector gluons at the Tevatron}

Before considering a more motivated model, let us review 
the impact that a massive  gluon $G$ may have on $t\bar t$ physics in
the simplest cases (for a related discussion
see~\cite{Ferrario:2008wm1,Ferrario:2008wm2,Moreau1,Moreau2,Frampton:2009rk,Chivukula,Burdman:2010gr,Haisch,Aguilar:2011ci,Aguilar:2006gw,Aguilar:2011cp,Falkowski:2011zr}).  
We will focus on the $m_{t\bar t}$ distribution and $A_{FB}^{t\bar t}$, two observables that have been measured at the Tevatron. We will consider two
different options according to the coupling of $G$ to the light quarks.
\begin{itemize}
\item Coupling to vector currents only\footnote{The vector and axial couplings are related to the
left and right handed couplings: $g_{V,A}=(g_R \pm g_L)/2$.} ({\it case V}):
\beq
g_V^q=g^q_R=g^q_L\,;\;\;\;g_A^q=0\;.
\label{g1}
\eeq
\item Coupling to axial currents only ({\it case A}):
\beq
g_A^q=g^q_R=-g^q_L\,;\;\;\;g_V^q=0\;.
\label{g2}
\eeq
\end{itemize}
For the top quark we will simply assume
\beq
g^t_R\ge g^t_L>0\;.
\label{gt}
\eeq

\begin{figure}[h]
\begin{center}
\begin{tabular}{cc}
\includegraphics[width=0.5\linewidth]{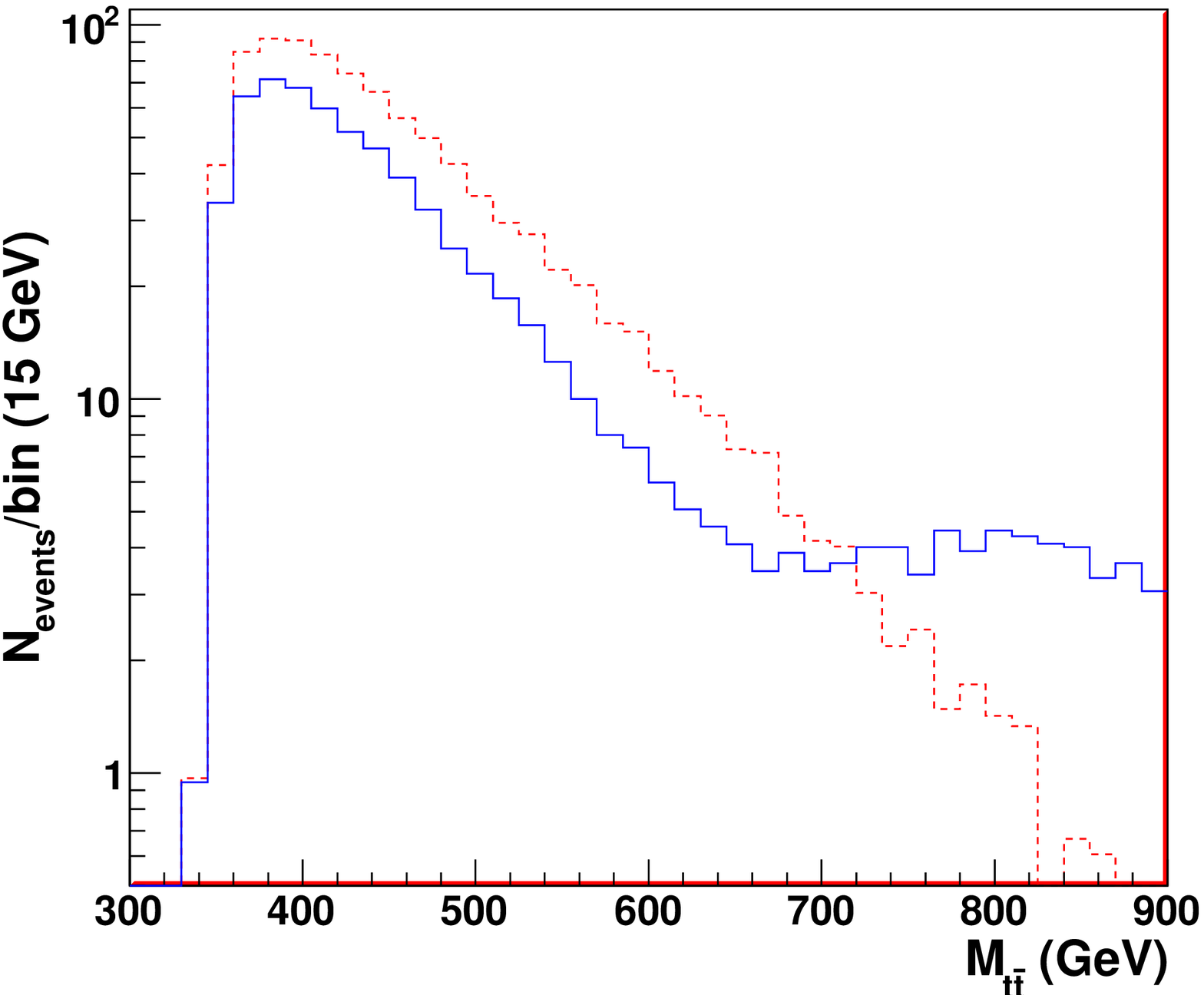} & 
\includegraphics[width=0.5\linewidth]{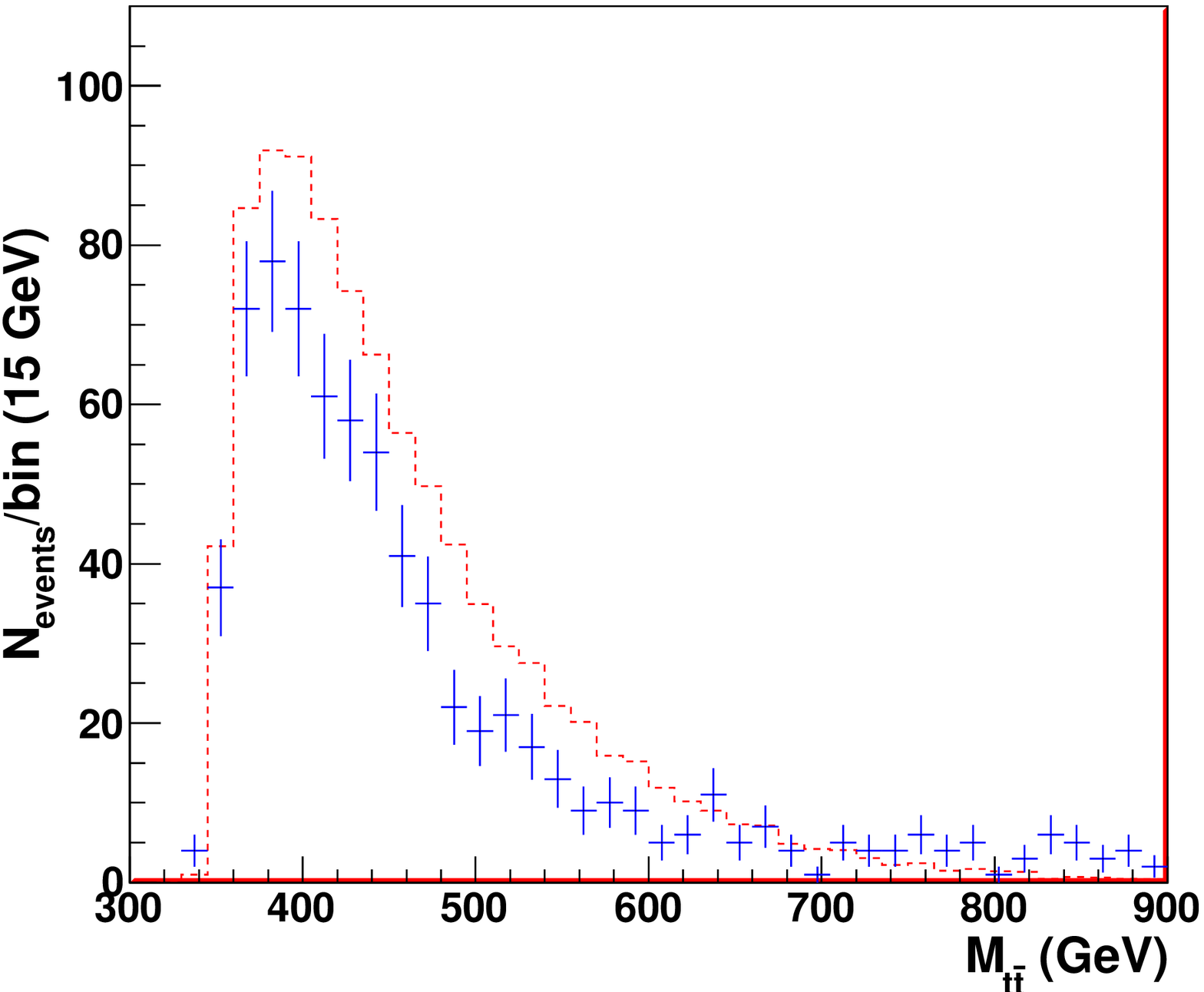} 
\end{tabular}
\end{center}
\caption{$m_{t\bar t}$ distribution at the Tevatron 
in the SM (dashes) and in model $V$ (solid) 
for a luminosity of 5.3 fb$^{-1}$ and $g_V^q=+ 0.2 g_s$.
On the left we plot the average number of events expected in each 
case, and on the right a particular Montecarlo simulation. 
The errors shown are statistical only.
}
\label{fig1FB}
\end{figure}

We have implemented the model in MADGRAPH/MADEVENT, used PYTHIA for hadronization
and showering and PGS4 for detector simulation. 
In Fig.~\ref{fig1FB} we plot $m_{t\bar t}$ distribution for {\it case V}
with\footnote{$g_s$ is the strong gauge coupling.} $(g_V^q=0.2g_s,\;g_A^q=0)$,  
$(g_R^t=6g_s,\;g_L^t=0.2g_s)$ and a mass $\textrm{M}_\textrm{G}=850$ GeV. 
For these couplings the gluon width is
$\Gamma_G\approx 0.32 \textrm{M}_\textrm{G}$. 
We have taken an integrated luminosity of 5.3 fb$^{-1}$
and the cuts/acceptances described in~\cite{:2007dia} (we have normalized
our samples so that our SM prediction 
agrees with the background-subtracted data there).
The event selection strategy is the following. 
We select $t \bar t$ candidate events in the 
lepton+jets topology, where one top decays semileptonically 
($t \rightarrow l \nu b$) and the other hadronically 
($t \rightarrow q \bar{q}' b$).
More precisely, we select events with an isolated electron or muon
in the central portion of the detector with high transverse momentum 
($p_T > 20$ GeV and $| \eta | < 1.0$) and a large amount of 
missing transverse energy $E^{\textrm{miss}}_T \geq 20$
GeV, consistent with the presence of an undetected 
neutrino. We require four
or more hadronic jets with $| \eta | < 2.0$,  
three of them must have $E_T > 15$ GeV and a fourth
$E_T > 8$ GeV. The jets must be clustered in fixed cones 
of radius $\Delta R= \sqrt{ (\Delta \eta)^2 + (\Delta \phi)^2} \leq 0.4$ 
and at least one
of them is required to be b-tagged.
In this qualitative analysis we do not
reconstruct the events, we just implement these cuts into
the parton-level information provided by MADGRAPH/MADEVENT. 

The $682$ semileptonic $t\bar t$ pairs given by model $V$ 
(see Fig.~\ref{fig1FB}) result from the destructive 
interference of the standard [$\approx g_s^2/\hat s$]
and the massive-gluon [$\approx 0.2g_s\cdot 6g_s/(-M^2)$] amplitudes.
We obtain a 30\% reduction for 
$\textrm{m}_{t\bar t}<\textrm{M}_\textrm{G}-\Gamma_G$ and an excess
at higher invariant masses with respect to the SM. The distribution does not
show a clear peak, but the change in the {\it slope} at 
$m_{t\bar t}\approx 650$ GeV
would have been apparent in the data. Taking the opposite sign
for the light-quark vector coupling $(g_V^q=-0.2g_s,g_A^q=0)$ the 
situation is similar, although the interference is now constructive
at low values of $m_{t\bar t}$.

In these models with only vector couplings of the light quarks to the massive gluon the FB asymmetry will 
appear only at NLO, since 
$A_G^{t\bar t}\propto -g_A^qg_A^t=0$ (see for
example~\cite{Ferrario:2009bz}).  
In particular, 
the  interference of 
the tree-level and  the one-loop box amplitudes will provide the
standard contribution, of order $A_{NLO}^{t\bar t} \approx 0.09$
at high invariant masses as estimated in~\cite{AFB3} using the
Monte Carlo for FeMtobarn processes software
(MCFM)~\cite{Campbell:1999ah}.  
An analogous interference between the massive gluon 
and the box diagrams will also contribute to 
the asymmetry.
At $m_{t\bar t}\ll M_G$ we estimate (see also~\cite{Bauer:2010iq})
\beq
A_{V-NLO}^{t\bar t} \approx A_{NLO}^{t\bar t} \times {m_{t\bar t}^2\over -M_G^2}
{g_V^q g_V^t\over g^2}\,,
\eeq
implying an additional contribution
of order $A_{V-NLO}^{t\bar t} \approx \mp 0.04$ for $g^q_V=\pm 0.2g_s$.
Therefore, 
the total value seems in this {\it case V}
very far (over 3$\sigma$) from 
the asymmetry observed at the Tevatron.

\begin{figure}
\begin{center}
\begin{tabular}{cc}
\includegraphics[width=0.5\linewidth]{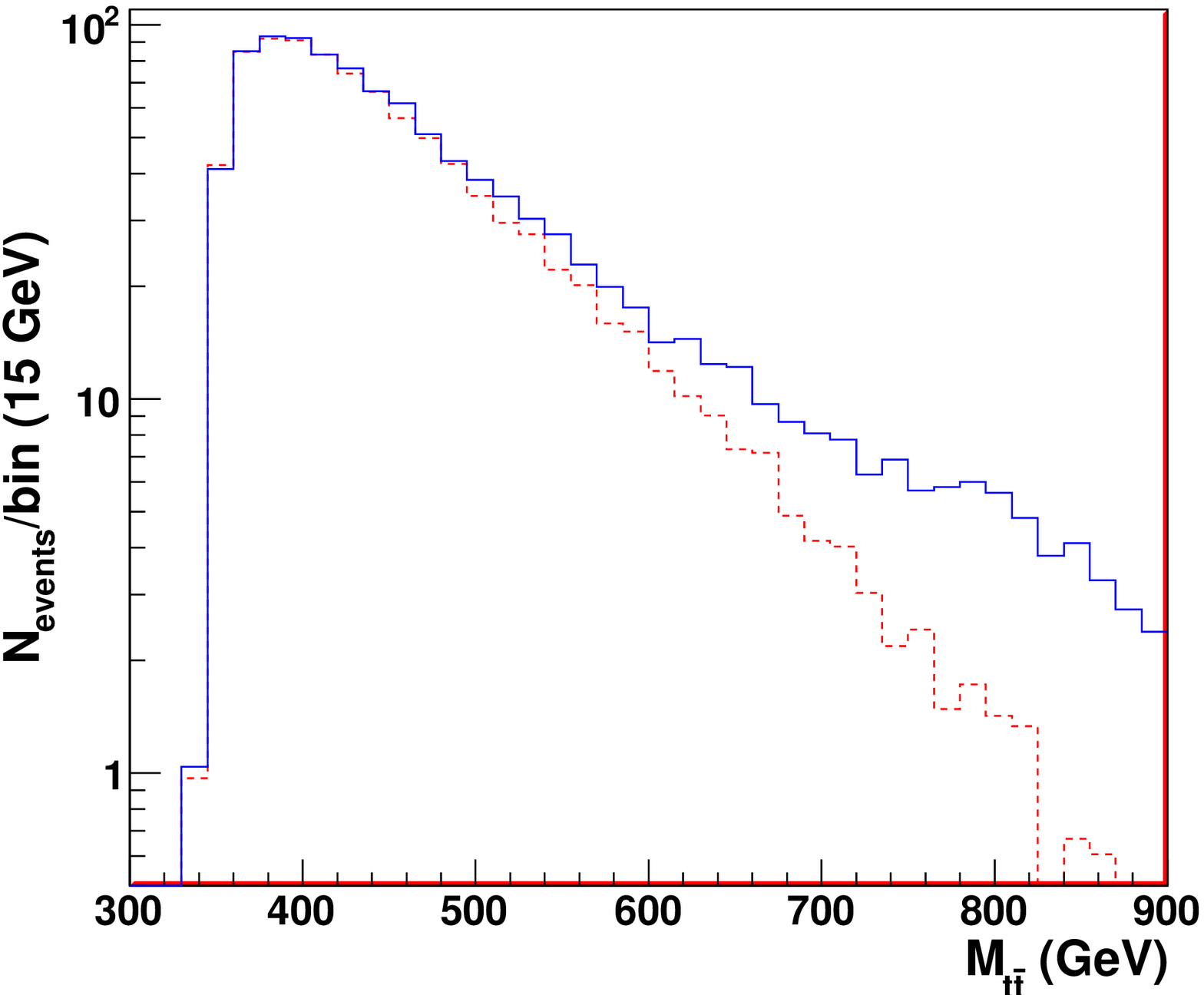} & 
\includegraphics[width=0.5\linewidth]{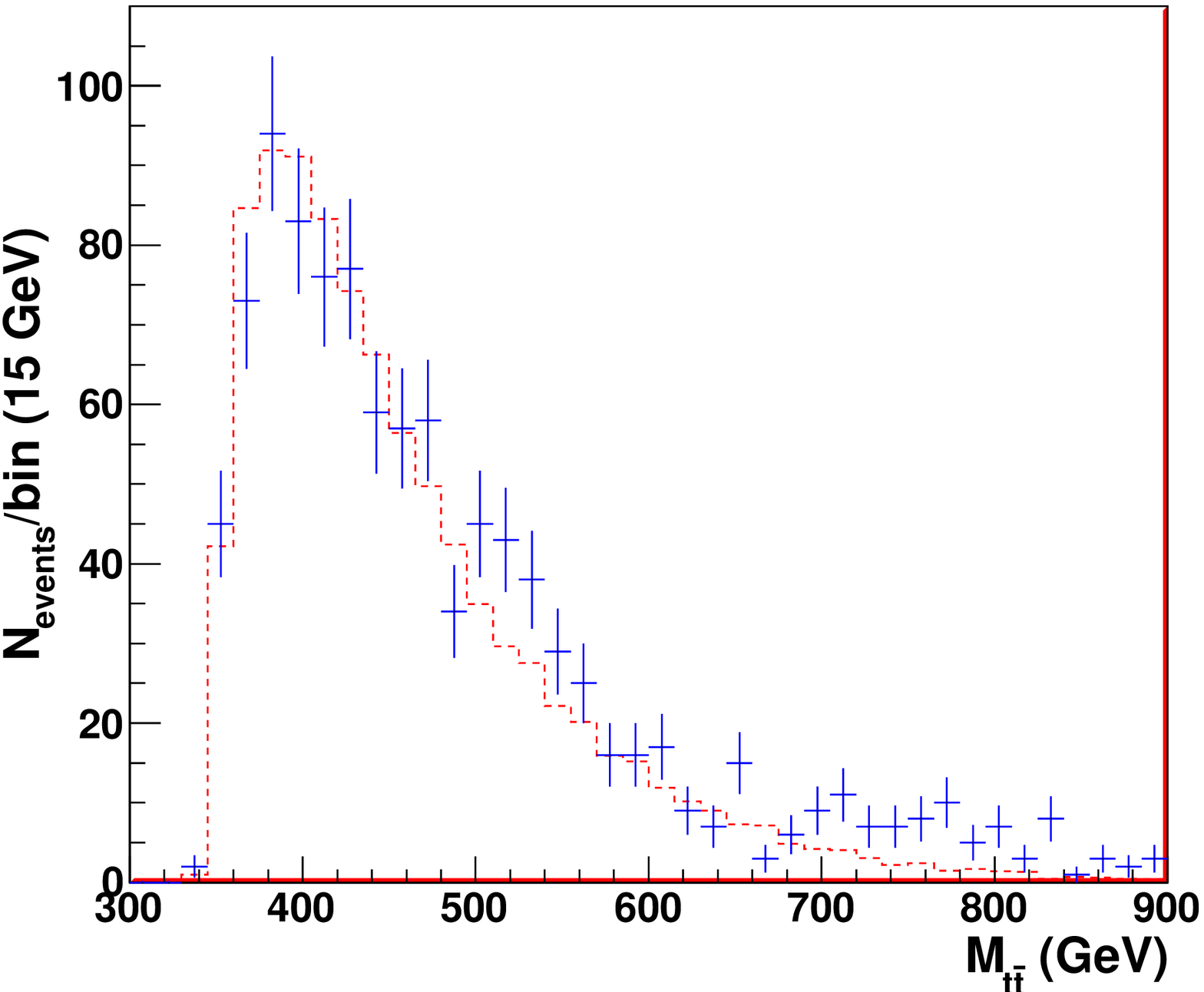} 
\end{tabular}
\end{center}
\caption{$m_{t\bar t}$ distribution at the Tevatron 
in the SM (dashes) and in model $A$ (solid) 
for a luminosity of 5.3 fb$^{-1}$ and $g_A=-0.2 g_s$.
On the left we plot the average number of events expected in each 
case, and on the right a particular Montecarlo simulation.
The errors shown are statistical only.}
\label{fig2FB}
\end{figure}
{\it Case A}, with a purely axial-vector coupling to the light
quarks, is completely
different. Both $q_L\bar q_L\to t\bar t$ and $q_R\bar q_R\to t\bar t$
parton-level cross sections will 
have large contributions from the interference between the amplitudes with the massless and the massive gluons. 
However, since their couplings
are opposite ($g_L^q=-g_R^q$), it will be constructive
in the first process and destructive in the 
second one, and both effects tend to cancel each other.
Up to invariant masses $m_{t\bar t}\approx M_G - \Gamma_G$ where the
resonant contribution becomes important, the number of $t\bar t$
events and their $m_{t\bar t}$ distribution will be very close to
the one in the SM. Note that the top-quark couplings do not need
to be purely axial for this to happen (alternatively, it would be
enough to have just the top couplings to be purely axial for the same
cancellation). The region around the peak 
will be hidden by the low statistics if $M_G$ is large
enough. In Fig.~\ref{fig2FB} we plot {\it case A} with
$(g_A^q=-0.2g_s,\;g_V^q=0)$,  
$(g_R^t=6g_s,\;g_L^t=0.2g_s)$, $M_G=850$ GeV and  
$\Gamma_G=0.32 \textrm{M}_\textrm{G}$ GeV. After cuts we obtain
1042 $t\bar t$ pairs, a number only 12\% higher than
the one expected in the SM. At
$m_{t\bar t} \approx 600$ GeV the distribution exhibits
a change in the 
slope, but the region where the
differences are important (around 750 GeV) is of 
little statistical significance (see a 
particular Montecarlo simulation in Fig.~\ref{fig2FB}--right).

In contrast to the case with vector couplings to the light quarks, 
$A_{G}^{t\bar t}$ is in {\it case A} large: the total
number of events does not change, but there is a large
forward excess that coincides with the backward deficit. 
In the $t\bar t$ rest frame we obtain
\beq
A_{G}^{t\bar t} \approx \left\{
\begin{array}{l l} 
\displaystyle 0.07
& m_{t\bar t}<450\;{\rm GeV}\,; \\
0.20
& m_{t\bar t}>450\;{\rm GeV}
\;.
\end{array} \right. 
\eeq

Therefore, {\it case A} provides a promising framework for model
building.  However, although in this model the peak 
at $m_{t\bar t} = 850\pm 272$ GeV is
practically non-existent, the tails in the invariant-mass distribution
could be accessible to the current experiments. 

In any case the results above suggest a way to go in order to find a `working' gluon:

\begin{itemize}

\item The large value of the asymmetry at $m_{t\bar t} \sim 500 $ GeV implies
that the gluon must be relatively light (700 GeV $\lesssim M_G \lesssim 900$ GeV) and
with strong couplings to the standard quarks ($g^q \cdot g^t \approx g_s^2$).

\item The absence of anomalies at $m_{t\bar t} \sim 500 $ GeV and the value of
$A_{FB}^{t\bar t}$ forces small and almost axial-vector couplings with the light
quarks ($g_L^q=-g_R^q$). This, and a large coupling to the right-handed top quark (e.g., $g_R^t \sim 4g_s$),
are natural features obtained in holographic models, where the massive gluon and the third quark family
live toward the TeV brane~\cite{Barcelo:2011fw}. 

\item It would be necessary to reduce the `tail' that the model implies at
$m_{t\bar t} > 600 $ GeV. This can be achieved increasing the gluon width.
However we do not want to reduce the gluon effect on the asymmetry. This
prevents an increase of the width through decays into light particles and suggest
new decay channels that open at $m_{t\bar t} \geq 600 $ GeV. The two-point 
gluon function would not include the contribution to its imaginary part from these
decay modes at $m_{t\bar t} < 600 $ GeV, leaving the $A_{FB}^{t\bar t}$ value
unchanged. However, it would reduce $t\bar t$ production through the massive gluon at higher energies,
hiding a possible peak in the $t\bar t$ or the dijet distributions.
Remarkably enough holographic models provide such a framework with decays
$G \rightarrow qQ$ where $q$ is a standard quark and $Q$ a massive excitation. Then the model requires $m_Q + m_q \sim 600$ GeV.

\end{itemize}

It is apparent that the model will demand a proper treatment of the gluon width and,
in particular, of its energy dependence. A Breit-Wigner
with constant width would offer a poor description of the gluon-mediated
amplitude. Instead, when a new channel 
\beq
q\bar q\rightarrow G\rightarrow q \bar Q ,\, \bar q  Q\;
\label{gt}
\eeq
opens at $\sqrt{\hat s}= m_Q+m_q$ 
it contributes to  $\Gamma_G(s)$
\beqa
\Gamma_G^{Q q}(\hat s)&=\theta\left[\hat s-(m_q+m_Q)^2\right]\;
 \displaystyle {g^2 \over 12 \pi} {\hat s \over M_G}\, 
\displaystyle \left(1 - {(m_q+m_Q)^2\over \hat s} \right)^{1\over 2} 
\left(1 - {(m_q-m_Q)^2\over \hat s} \right)^{1\over 2}\times
\nonumber \\ 
&
\displaystyle \left[ \left( 1 - {m_q^2+m_Q^2+6 m_q m_Q\over 2 \hat s } -
{(m_Q^2 - m_q^2)^2\over 2 \hat s^2 } \right) \right. g_A^{Qq\,2} +
\nonumber \\
&
\displaystyle \left. \left( 1 - {m_q^2+m_Q^2-6 m_q m_Q\over 2 \hat s } -
{(m_Q^2 - m_q^2)^2\over 2 \hat s^2 } \right)g_V^{Qq\,2}  \right] 
\;, 
\label{GtT}
\eeqa
where $g_{V,A}^{Qq}$ are the vector and axial 
coupling of the massive gluon
to $Q$ and $q$, respectively.

\begin{figure}[t]
\begin{center}
\includegraphics[width=.5\linewidth]{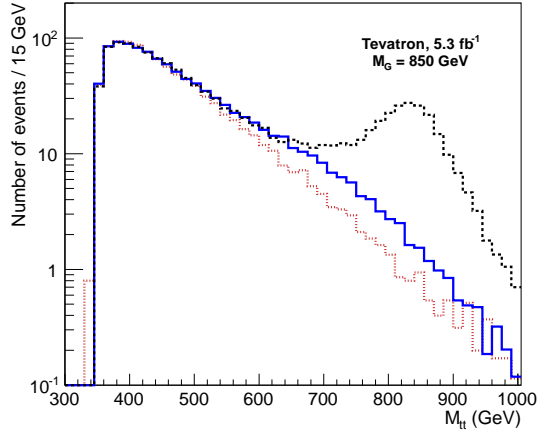}
\caption{\label{toymodel} Prediction for the $m_{t\bar{t}}$ distribution at the
Tevatron with a luminosity of 5.3 fb$^{-1}$ for the SM (dotted), 
and for a model similar to $case$ $A$ with (solid) and without (dashed) the new decay channels $G \rightarrow q Q$
opening at $m_Q + m_q \sim 600$ GeV. The coupling of the heavy gluon to $qQ$ are fixed in such a way
that the total width at the gluon mass is $\Gamma_G=0.7M_G$.}
\end{center}
\end{figure} 

Fig.~\ref{toymodel} illustrates the effect of new decay channels $G \rightarrow q Q$ opening at $m_Q + m_q \sim 600$ GeV enlarging the gluon width
in a model similar to $case$ $A$. We have chosen the coupling of the heavy gluon to $qQ$ in such a way
that the total width at the gluon mass is $\Gamma_G=0.7M_G$.
The figure shows that without these new channels 
the peak is clearly visible. Once they
are included the large width makes the gluon completely invisible.
Moreover, implementing the energy dependence of the width\footnote{We modified the matrix element in MADGRAPH/MADEVENT
in order to implement the energy-dependence width.} leaves the asymmetry unchanged.
Including the SM contribution, the FB asymmetry in the large
$m_{t\bar t}$ region goes from $A_G^{t\bar t}=0.30$ with
no extra $Q$ quarks to $A_G^{t\bar t}=0.33$ in this 
model, just $1.2 \sigma$ away from the central value of the CDF
measurement \cite{AFB3}. It is then clear that our \textit{stealth}-gluon model (as we name it) can reproduce 
the Tevatron data on the FB asymmetry and the $m_{t\bar{t}}$ distribution.   

The low masses of the gluon and the new quarks,
together with the sizable couplings required to generate the large width,
make the production of single new quarks mediated by the massive gluon a
very attractive channel at the LHC. 
Indeed, the signal there will depend strongly on the nature of the vectorlike
quark involved. In the next section we classify all the 
possibilities and introduce a benchmark model. We show
that current analysis could easily miss such model, whereas specific
searches would very likely reveal the 
mechanism responsible for the Tevatron asymmetry. 

\Section{A benchmark model\label{benchmark}}

Let us now introduce a benchmark model that successfully
reproduces the Tevatron FB asymmetry with no conflict with other
experimental tests\footnote{Models with warped extra
dimensions naturally fulfill all the necessary ingredients to realize this scenario
\cite{Barcelo:2011fw}.}. It contains simultaneously all 
possible decay channels: $G \rightarrow t T, b B, q Q$ with $q$ denoting a light flavor.
Therefore, it will allow us to perform
a comprehensive study of the stealth-gluon scenario. The model 
admits variations where one or several channels are suppressed 
while the others are enhanced in such a way that the total 
gluon width does not change significantly. 
We take $\textrm{M}_\textrm{G}=850$ GeV, although similar setups
can be obtained for gluon masses as low as 700 GeV. 
We fix the couplings to $G$ of the SM
quarks to 
\beqa
g_L^q&=&0.3\,g_s \;, \quad  g_R^q =g^b_R=-0.3\,g_s \;, 
\quad
g_R^t=+4\,g_s \;, \quad 
g_L^t=g_L^b=0\;.
\label{gencoup}
\eeqa
We assume the presence
of six fields for the vectorlike quarks, corresponding to the excitations of  
$t_R$, $b_R$ and the four light flavors $q_L$. We fix their masses to
\beqa
M_T &=& 450\;{\rm GeV}\;,\quad M_B = M_Q= 600\;{\rm GeV} \;,
\eeqa
and their flavor-changing couplings to the heavy gluon to
\beqa
g_R^{Tt} &=& 4\,g_s\;,\quad
g_R^{Bb} = 3.5\,g_s\;,\quad
g_L^{Qq} = 3.5\,g_s\;.
\eeqa
With these values the total width is $\Gamma_G\approx 0.7\,\textrm{M}_\textrm{G}$ 
while the 
decay branching fractions are 
\begin{eqnarray}
&&{\rm BR}(G\to t\bar{t})\approx 0.2 \;,\qquad\quad
\;\;{\rm BR}(G \to T\bar{t},t\bar{T})\approx 0.24\;,\nonumber \\
&&{\rm BR}(G\to B\bar{b},b\bar{B})\approx 0.11\;,\quad
{\rm BR}(G\to Q\bar{q},q\bar{Q})\approx 0.44\;.
\end{eqnarray}
The extreme cases where all
the decay modes of $G$ are absent except one are defined with:
\begin{eqnarray}
\mbox{Extreme $T$ model:} 
&& g_R^{Tt} = 7.28\,g_s\;,
\quad g_R^{Bb} = g_L^{Qq} = 0\;,
\label{ext:T}\\
\mbox{Extreme $B$ model:} 
&& g_R^{Bb} = 9.36\,g_s\;,
\quad g_R^{Tt} = g_L^{Qq} = 0\;,
\label{ext:B}\\
\mbox{Extreme $Q$ model:} 
&& g_L^{Qq} = 4.68\,g_s\;,
\quad g_R^{Tt}=g_R^{Bb} = 0\;,
\label{ext:Q}
\end{eqnarray}
and all the other couplings unchanged. In these cases 
the heavy gluon has a $20 \%$ 
branching ratio into $t\bar{t}$ and $80 \%$ into the new
channel. Note that in some of these models the required coupling is
unrealistically large. We just take them as limiting examples to get
clear idea of the LHC reach for these signatures (realistic models 
should lie somewhere in between the benchmark and the extreme cases).

The new heavy quarks will then be produced through
$G$ in the $s$--channel as $Q\bar Q$ pairs or as a
single particle together with a standard quark, $Q\bar q$. Pair
production will also receive the standard QCD contribution (in fact,
due to the axial nature of the $G$ coupling to light quarks, the
interference terms cancel and away from the resonance pair production
is like in the SM).
Single heavy-quark 
production, on the other hand, 
is unsuppressed and opens kinematically at lower
energies ($\sqrt{\hat{s}}=m_q+m_Q\ll 2m_Q$),
appearing as a very promising mechanism 
unexplored in previous literature.
The vectorlike quarks will then decay in a 
model-dependent way, according to their EW quantum numbers
and their mixing with the SM quarks. Assuming weak couplings, their
width will be narrow, and a simple scaling allows to go from one
model to another. To be definite we will take the  
branching ratios obtained in the large-mass 
limit of the usual Higgsless models, 
\begin{equation}
\mathrm{BR}(Q\to W q^\prime)
=\frac{2}{3}\;,\quad
\mathrm{BR}(Q\to Z q)
=\frac{1}{3}\;.
\end{equation}
Higgs decays can potentially lead to interesting
signatures~\cite{Aguilar:2006gw} but we will
not consider it here. 

With these assumptions the final states produced in $q\bar q$ collisions 
will be the following\footnote{The conjugated processes are not explicitly shown but
are included in our analyses.}:

\noindent {\it (i)} $W^+W^-b\bar b\,$, from 
\beq
q\bar q\to G\to T \bar t\to  (W^+ b) W^-\bar b
\eeq
and 
\beq
q\bar q\to G\to B \bar b\to (W^-t)\bar b\to (W^-W^+b)\bar b\;.
\eeq
Notice that the final state in these two channels coincides with the one in $t \bar t$ production.\\

\noindent {\it (ii)} $Zb\bar b\,$\;, from 
\beq
q\bar q\to G\to B \bar b\to (Zb)\bar b\;.
\eeq

\noindent {\it (iii)} $Zt\bar t\,$, from 
\beq
q\bar q\to G\to T \bar t\to (Zt)\bar t\to (Z W^+ b) W^-\bar b\,.
\eeq

\noindent {\it (iv)} $W\!+\!\mathrm{jets}\,$\;, from 
\beq
q\bar q\to G\to Q \bar q\to (Wq')\bar q\;.
\eeq

\noindent {\it (v)} $Z\!+\!\mathrm{jets}\,$\;, from 
\beq
q\bar q\to G\to Q \bar q\to (Zq)\bar q\;.
\eeq

\begin{figure}[t]
\begin{center}
\includegraphics[width=.5\linewidth]{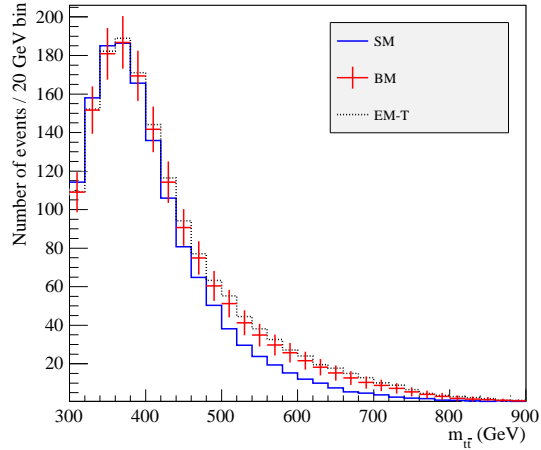}
\caption{\label{mtt_Tevatron} $m_{t\bar{t}}$ distribution at the
Tevatron for 5.3 fb$^{-1}$ in the SM (solid/blue), 
the benchmark model (points with error bars) and the extreme $T$ case
(dotted/black). We include the contribution from
$T\bar{t},t\bar{T}$ and $B\bar{b},b\bar{B}$ when present.}
\end{center}
\end{figure}

Through the following sections we show that these signals do not introduce
observable anomalies in current LHC analyses, 
but that simple modifications
in the reconstruction of the final state could very likely provide a
signal. The impact of this scenario on top-quark physics at the Tevatron
has been discussed in \cite{Barcelo:2011vk}, where we name it
as the stealth-gluon model due to its ability to explain the 
FB asymmetry without introducing anomalies (peaks or tails)
in the $t\bar t$
invariant-mass distribution. In particular, it implies 

\beq
A_{G}^{t\bar t} \approx \left\{
\begin{array}{l l} 
\displaystyle 0.12
& m_{t\bar t}<450\;{\rm GeV}\,; \\
0.33
& m_{t\bar t}>450\;{\rm GeV}
\;,
\end{array} \right. 
\eeq
values that are 
compatible with the D$\emptyset$ and CDF observations~\cite{AFBTEV,AFB2,AFB3}. 
The $m_{t\bar t}$ distribution at the Tevatron
is given in Fig.~\ref{mtt_Tevatron}, where we compare the reconstruction
as $t\bar t$ pairs of all the events giving $W^+W^-b\bar{b}$ in the
benchmark model with 
the SM prediction. In our simulation we have followed the analysis
in~\cite{:2007dia} (the same as in the previous section). Here we 
have included in the analysis the event reconstruction (see App.~\ref{AppA}).

The figure includes the prediction 
in the extreme $T$ model (the prediction in the extreme $B$
model is similar, whereas in the extreme $Q$ model it is below
the benchmark one). The deviations are never larger than
2.5$\sigma$ (assuming statistical errors only), and below 2$\sigma$
in all the bins for the benchmark and the extreme-$Q$ models.

\Section{Single $T$ and $B$ quark production at the LHC\label{TB}}

\subsubsection*{$W^+ W^- b\bar{b}$ channel}

\begin{figure}[t]
\begin{center}
\includegraphics[width=.5\linewidth]{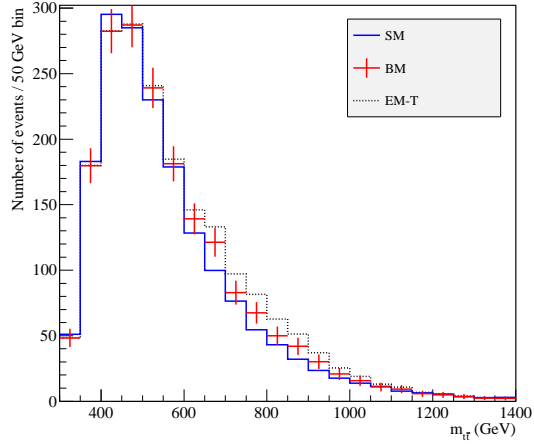}
\caption{\label{LHC_ttbar} $m_{t\bar{t}}$ distribution at the LHC for 
0.2 fb$^{-1}$ in the SM (solid/blue), the benchmark model
(points with error bars) and the extreme $T$ model (dotted/black). 
We include the contribution from 
$T\bar{t},t\bar{T}$ and $B\bar{b},b\bar{B}$ when present.}
\end{center}
\end{figure}

As described in the previous section, the processes $q\bar q \to 
T\bar t, B\bar b$ followed by the charged-current decay of the
heavy quark will result in the same
$W^+W^-b\bar{b}$ final state as $t\bar{t}$ production. 
Thus, this signal would add to the one from 
top-quark pairs produced
through the massive gluon plus the standard contribution, and it is then necessary
to check that these processes do not imply any observable
excess in current analyses of $t\bar{t}$ production (see for example \cite{LHCttbar})
or fourth generation $T\bar{T}$ searches \cite{T4thgen}. 
In particular, we have simulated the analysis of ATLAS \cite{LHCttbarATLAS}\footnote{App.~\ref{AppB} shows a similar analysis done by the CMS collaboration.}
to study the effect of the channels
\begin{equation}
pp\to T\bar{t},\,t\bar{T},\,B\bar{b},\,b\bar{B}
\end{equation}
together with all the contributions to $t\bar{t}$ production. We show the result in Fig.~\ref{LHC_ttbar}.
We have assumed a 
$10\%$ uncertainty in the $t\bar{t}$ prediction and allowed a
normalization factor (within this $10\%$) to correctly reproduce the
three bins around the peak at $m_{t\bar{t}}\approx 500$ 
GeV. We show the SM, the benchmark model (with statistical error bars)
and the extreme $T$ model. 
The deviation in the extreme $B$ case
is similar to the one in the extreme $T$ model, whereas 
the extreme $Q$ case is closer than the benchmark to the SM.
The $\approx 20\%$ excess at $m_{t\bar t}=600$--$900$ GeV in
the  extreme $T$ and $B$
models seems in the limit of being probed with the current LHC data.
Increasing the luminosity to 
4 fb$^{-1}$ we find 8 consecutive
bins with differences above 3$\sigma$ for the DELPHES simulation and
7 consecutive ones for the PGS4 simulation in the case of the extreme T
model. The benchmark and extreme Q models are not that
clear. For instance, using PGS4~\cite{PGS4} we find 3 and 2 consecutive bins with departures
larger than 3$\sigma$ in these cases for a luminosity of 4 fb$^{-1}$ 
(in all our estimates we only include statistical
errors). In summary, in our model one could expect a 10\% excess relative 
to the SM prediction in all the $m_{t\bar t}$ 
bins below 1 TeV. These events are just $t\bar t$ pairs mediated by
the heavy gluon $G$. In addition, the bins between $600$--$900$ GeV
could be increased an extra $15\%$ with 
$T\bar t$ and/or $B\bar b$ events that are reconstructed as $t\bar t$ 
pairs.

\begin{figure}[t]
\begin{center}
\includegraphics[width=.46\linewidth]{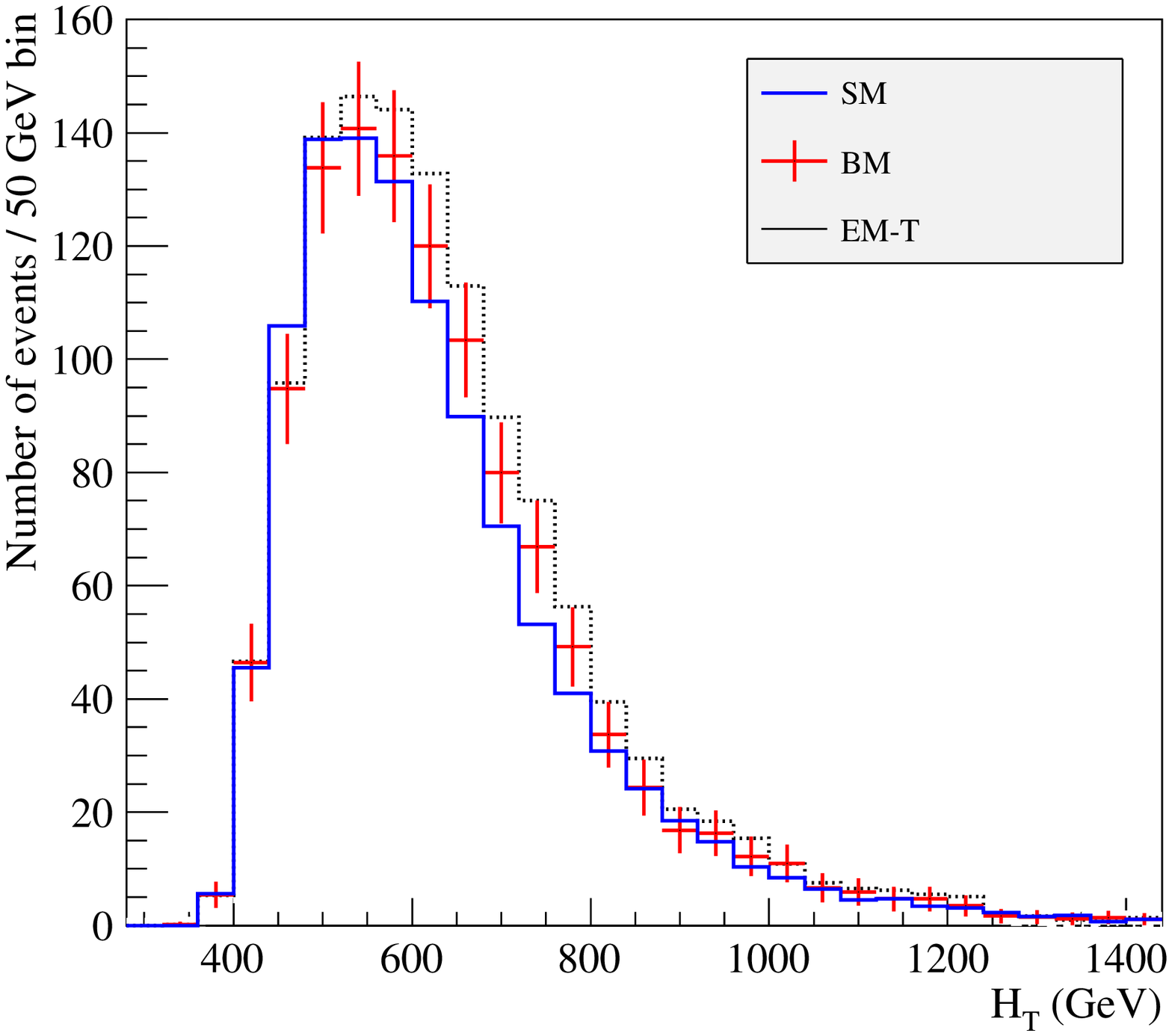}
\includegraphics[width=.46\linewidth]{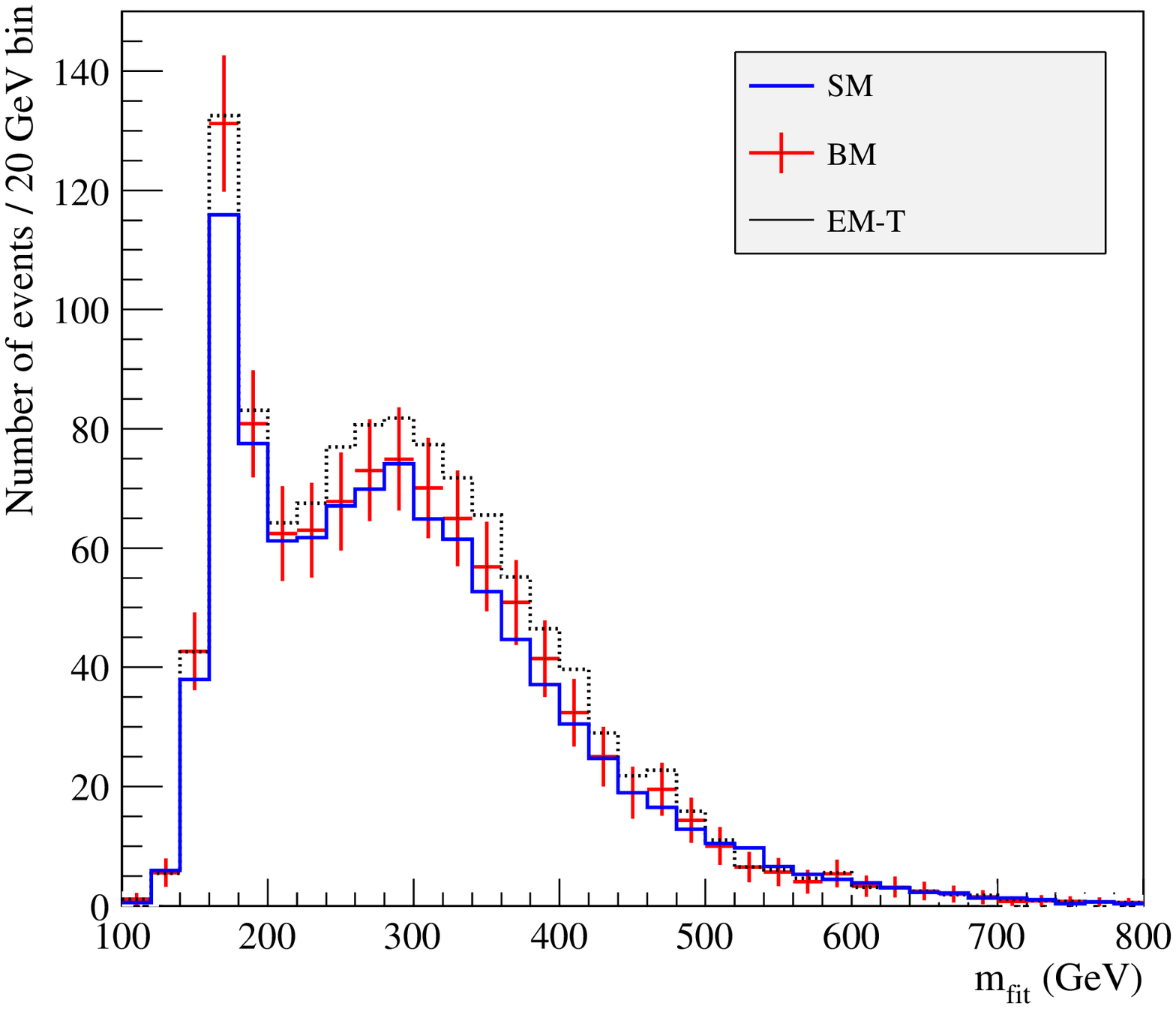}
\caption{\label{LHC_TTbar} $T\bar{T}$ search at the LHC for 0.821
  fb$^{-1}$. Left panel: $H_T$ distribution. Right panel:
  $m_{\mathrm{fit}}$ distribution. In both cases we show the 
predictions in the SM (solid/blue), in the benchmark model
(data points with statistical errors) and 
in the extreme $T$ case (dotted/black). 
We include the contribution from 
$T\bar{t},t\bar{T}$ and $B\bar{b},b\bar{B}$ when present.}
\label{TTbarra}
\end{center}
\end{figure}

Another LHC study sensitive to our model is the search for a fourth 
generation of quarks 
produced as $T\bar{T}$ pairs~\cite{T4thgen}. We have reproduced the
corresponding CMS analysis there (see App.~\ref{AppC}) and plot our results in Fig.~\ref{LHC_TTbar}
for the muon channel with the published luminosity of 0.821 fb$^{-1}$.
We plot the SM, the benchmark and the extreme $T$ cases in
solid/blue, data points (with error bars), and dotted/black, respectively. 
The left panel shows the
$H_T$ distribution (defined in this case as the scalar sum of the $p_T$
of the jets, the charged lepton and the missing $E_T$), 
and the right panel gives the $T$
reconstructed mass in the events generated with 
our model(s) and  with the SM. In both plots the number of
standard events has been normalized by the same factor.
Our results are similar to the ones obtained for 
$t\bar{t}$ production. The benchmark and the 
extreme $Q$ models are not visible, whereas the extreme
$T$ and $B$ models are starting to be probed by the data.
We have also checked that pair production of $T$ quarks gives 
in our model a negligible
contribution, compatible with the bound obtained in~\cite{T4thgen}. 
Similarly, the recent search for pair
production of vectorlike $T$ quarks decaying to $Z t$~\cite{TTtoZ} does
not imply any restriction to our model. 

\begin{figure}[t]
\begin{center}
\begin{tabular}{cc}
\includegraphics[width=0.5\linewidth]{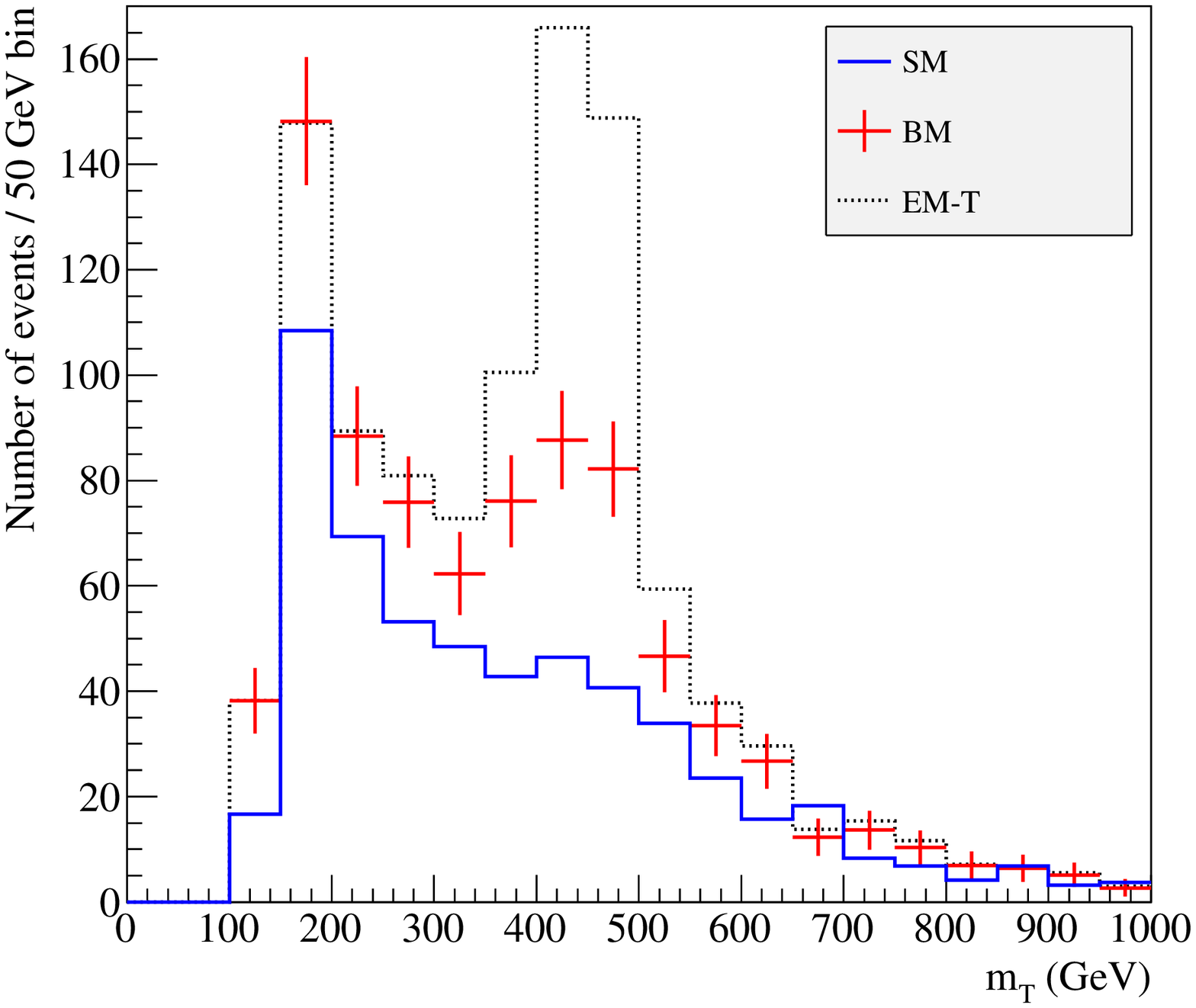} &
\includegraphics[width=0.5\linewidth]{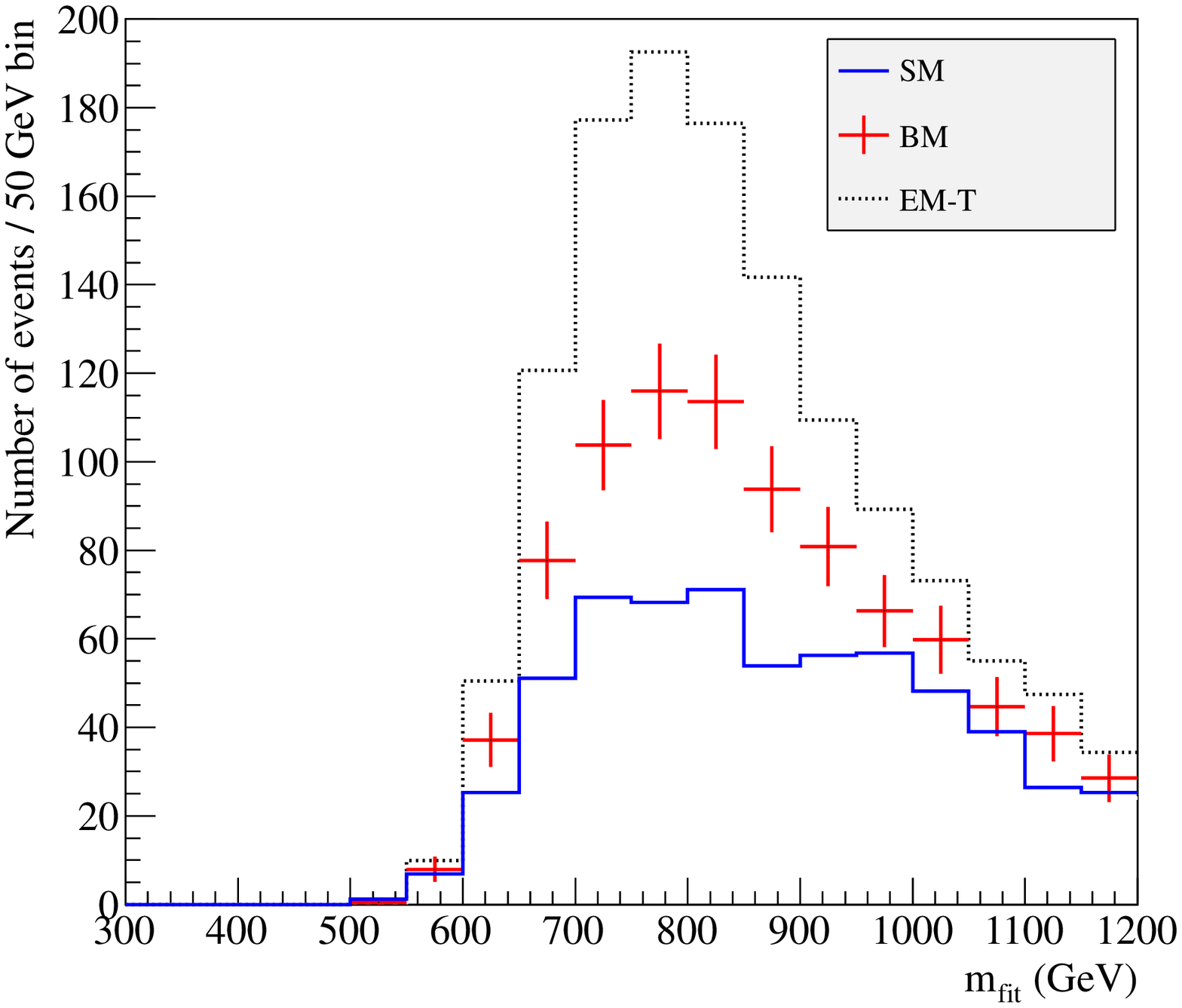}  
\end{tabular}
\end{center}
\caption{\label{T:discovery} Reconstruction of $m_T$ (left) and
$\textrm{M}_\textrm{G}$ (right) at the LHC. In both cases we
have normalized the distributions to 4 fb$^{-1}$ data 
and represent the results for the SM
(solid/blue), the benchmark model (data points  
with statistical errors) and the extreme $T$ case (dotted/black). 
Details of the reconstruction method can be found in the text.}
\end{figure}

Our results indicate that the model, proposed
to explain the large FB Tevatron asymmetry, is almost invisible in
$t\bar t\to W^+ W^- b\bar{b}$ searches. 
The reason for that is twofold. First, the
large gluon width suppresses the number of $t\bar t$ 
events in the region $m_{t\bar{t}} = 600$--$900$ GeV, while its axial
couplings to the light quarks does the same at lower and higher invariant
masses. Second, $T\bar t$ or $B\bar b$ events 
are reconstructed as $t\bar t$ or $T\bar T$ pairs, resulting into 
a poorer fit and a wider spread.
The key to isolate events of type $T\bar t$ would
be to reconstruct them not like 
two objects with the same mass, 
but like a $t$ quark plus a $T$ quark of arbitrary mass. These
events will only occur at large invariant masses, $m_{T\bar{t}}>m_T+m_t$,
a region already accessible at the LHC with the current
luminosity. Therefore, we 
can use the more stringent cuts used in the $T\bar{T}$ analysis
of~\cite{T4thgen} (we use the muon channel). 
Actually, we will require the hardest jet to have
$p_T\geq 200$ GeV instead of the 120 GeV.
We will then identify just 
one 173 GeV $t$ quark 
(using a $\chi^2$ similar to the one in
\cite{LHCttbar} and requiring $\chi^2\leq 10$, see App.~\ref{AppB} and~\ref{AppC}) 
and will 
plot the mass of the second one in events of invariant
mass above 600 GeV (Fig.~\ref{T:discovery}--left) 
for SM and extreme $T$ model simulations. 
We have normalized the plots to the
recorded luminosity of 4 fb$^{-1}$. 
As it is apparent in the plot, 
we find three consecutive bins around $m_T=450$ GeV 
departing more than 3$\sigma$ from the SM
prediction even in the benchmark model. 
Counting the total excess $S$ of events versus the standard background
$B$ on the peak (three bins between 350 and
500 GeV) we get
\begin{equation}
\frac{S}{\sqrt{B}}\approx \left\{ 
\begin{array}{ll}
8,\; \quad \mbox{benchmark},\\
21, \quad \mbox{extreme T}.
\end{array}
\right.
\end{equation}

Thus, the extreme $T$ case would 
imply a stunning deviation in this
kind of searches, and even the benchmark model 
could show evidence for new
physics. With the large excess in the extreme $T$ model one can
also try to reconstruct the massive gluon peak. In order to do that, we
remove the total invariant mass 
cut and compute the total invariant mass $m_{T\bar t}$
for the events with a 
reconstructed $T$ mass above 350 GeV. The result is shown in
Fig.~\ref{T:discovery}--right. Although the SM and the new
physics model peak in the same region, the factor of $\sim 3(2)$
excess in the extreme T (benchmark) model is quite evident. 

\begin{figure}[ht]
\begin{center}
\begin{tabular}{cc}
\includegraphics[width=0.5\linewidth]{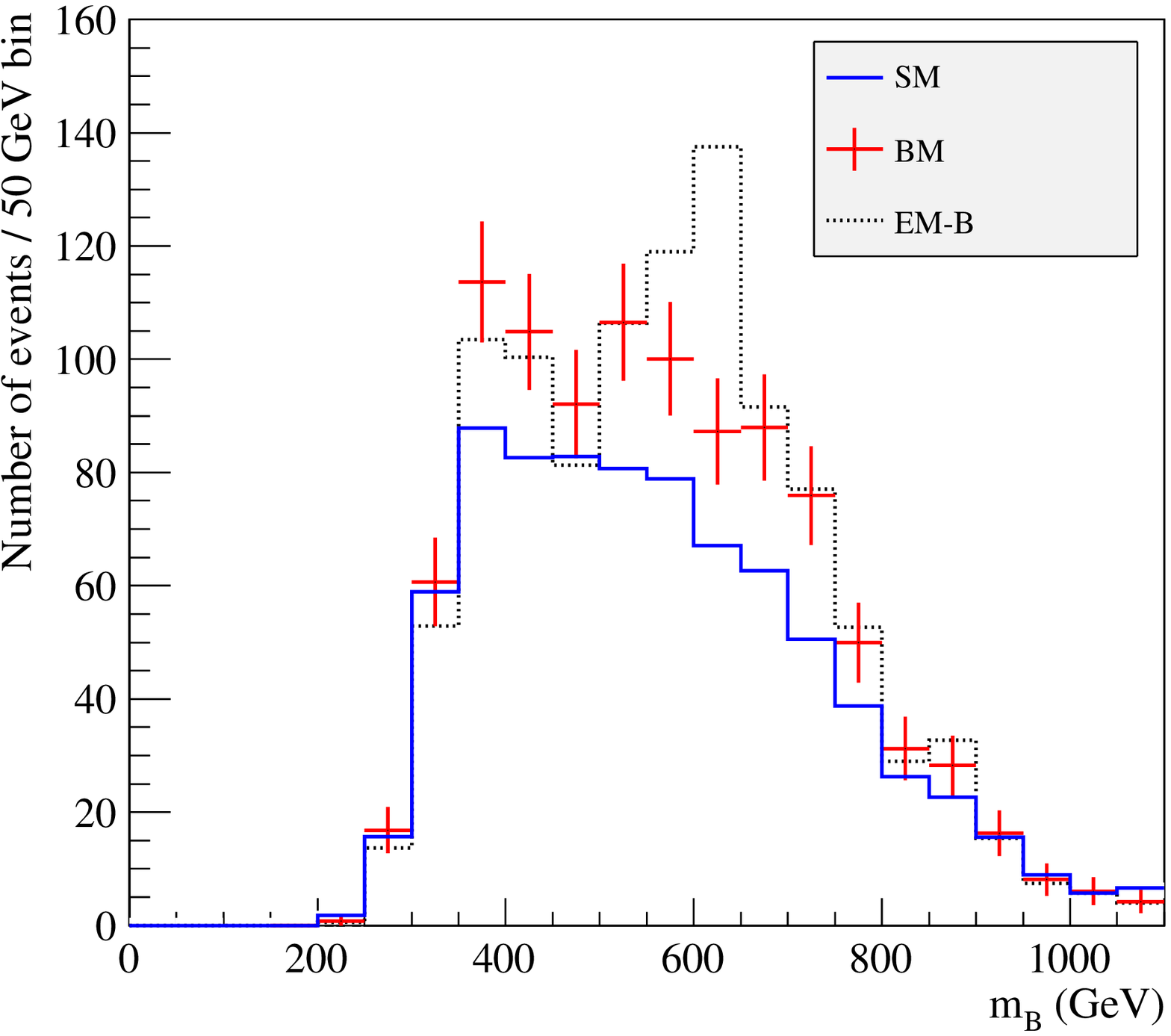} & 
\includegraphics[width=0.5\linewidth]{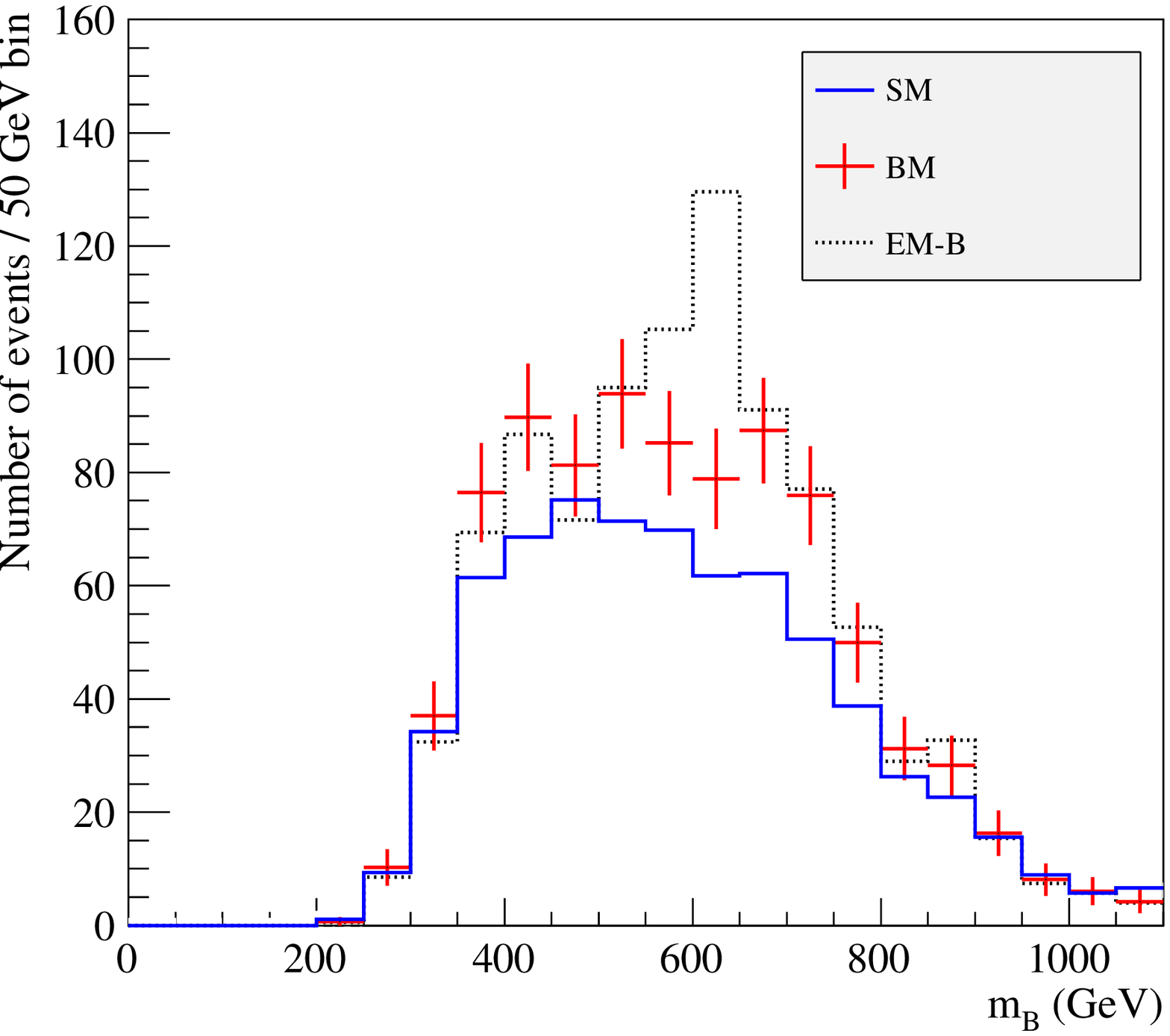} 
\end{tabular}
\end{center}
\caption{\label{B:discovery} Reconstruction of $m_B$ at the LHC for 
4 fb$^{-1}$ in the SM (solid/blue), the benchmark model (data points
with errors)
and the $B$ case (dotted/black). We consider the cuts $m_{B\bar{b}}>600$ GeV
(left) and $m_{B\bar{b}}>700$ GeV (right).
Details of the reconstruction method can be found in the text.
}
\end{figure}

The $B\bar b,\,b\bar B\to W^+ W^- b\bar{b}$ 
channel is slightly different. Instead of
producing two top-like objects, the heavy bottom decays into a $W$
plus a top that subsequently decays into another $W$ (with opposite
charge) and a $b$. 
 
\begin{figure}[h]
\begin{center}
\begin{tabular}{c}
\includegraphics[width=0.45\linewidth]{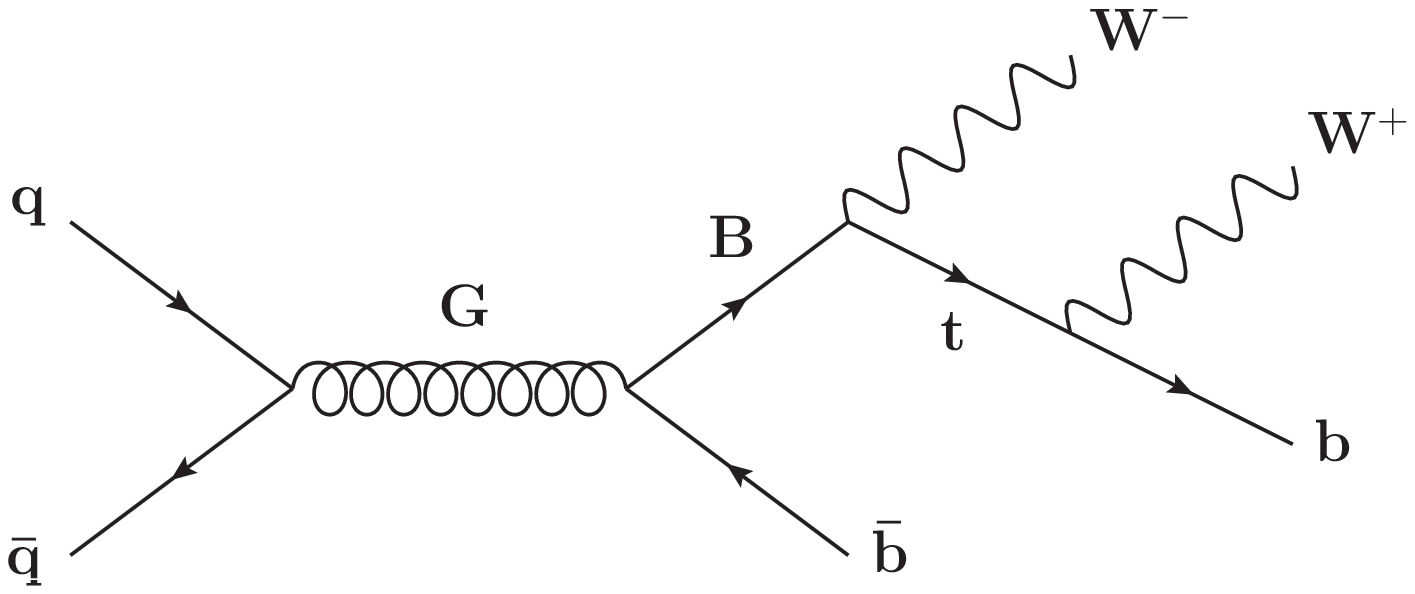}
\end{tabular}
\end{center}
\end{figure}
\vspace{-0.5cm}
\hspace{-0.83cm} We will still follow the selection procedure in
our previous analysis, with the cuts in~\cite{T4thgen} 
(muon channel) except for the cut on the $p_T$ of the hardest jet,
that is moved from 120 GeV to 200 GeV and a $\chi^2\leq 10$ 
(again we use the $\chi^2$ used in~\cite{LHCttbar}, App.~\ref{AppA}) choosing
the best configuration reconstructing a $173$ GeV top quark, 
and will plot the invariant mass of this $t$ quark plus the
extra $W$. The result is shown in Fig.~\ref{B:discovery} for the
benchmark and the extreme $B$ models with two different cuts in the
total invariant-mass distribution and 4
fb$^{-1}$. In this case, our reconstruction of the $B$ quark is not as
clear as the one of the $T$ quark, and 
more sophisticated analyses should be used to dig out
the signal from the background. Nevertheless, we will see in the
next section that the extreme $B$ model can be probed much more
efficiently using the neutral decay of the $B$ quark.

\subsubsection*{$Z b\bar{b}$ channel}

\begin{figure}[t]
\begin{center}
\includegraphics[width=0.45\linewidth]{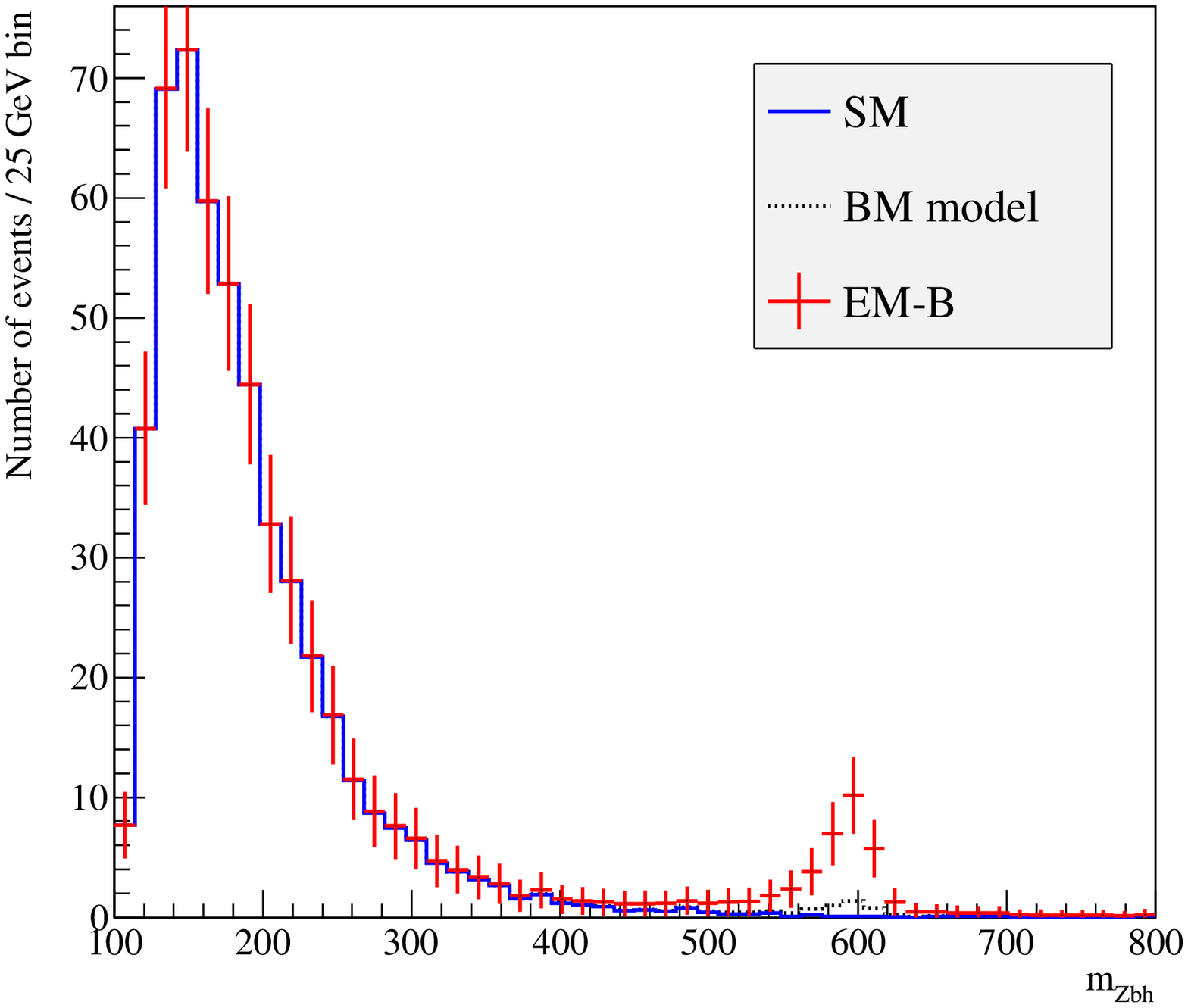} 
\includegraphics[width=0.45\linewidth]{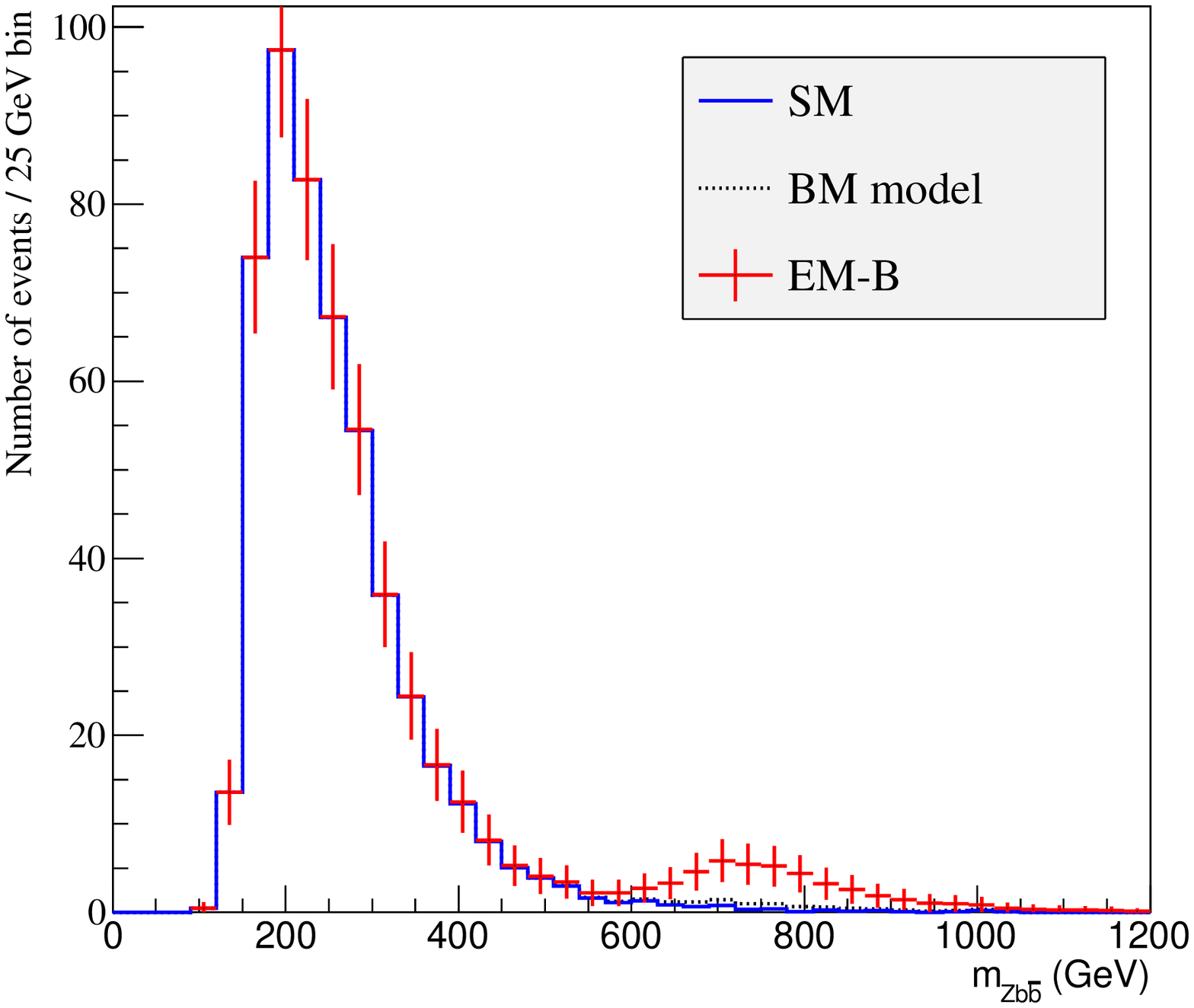} 
\end{center}
\caption{\label{discovery} Left panel: 
reconstruction of $m_{Zb_h}$ at the LHC.
Right panel: reconstruction of $m_{zb\bar{b}}$ to
show the heavy gluon mass.
In both cases we have normalized the distributions to 
4 fb$^{-1}$ of data and have represented the SM with thick solid blue
line, the benchmark model with thin solid red line
and the extreme $B$ case (data points with statistical errors). 
}
\end{figure}

Let us now turn to the neutral decays of the heavy $T$ and $B$
quarks, starting with the $B\bar{b},b\bar{B}$ channel into a
$Z b\bar{b}$ final state. The SM irreducible background to
this process is small ($\sigma(Zb\bar{b})$ with a leptonic $Z$
decay is around $2$ pb), whereas
the background from 
final states with larger cross sections like $Z$+jets and 
$t\bar t$ can be reduced with a very simple set of 
cuts.\footnote{We have also checked that our model does not conflict
with current searches of $H\to ZZ\to Z b\bar b$ \cite{Aad:2011ec} or
measurements of $Z+b$ cross-section~\cite{zb:atlas}.}
To isolate the
signal we will require two same-flavor opposite-sign 
leptons with $p_T\geq
25$ GeV and $|m_{l^+l^-}-m_Z|\leq 25$ GeV, and two $b$-tagged jets with
$p_T\geq 20$ GeV and $|\eta\leq 2.8|$. We will also impose 
a veto on missing energy
$E_T\leq 40$ GeV, to reduce the $t\bar{t}$ background. With this
selection we compute the invariant mass of the $Z$ and the hardest of
the two $b$-jets (denoted by $b_h$), since the $b$ quark from the
decay of the heavy $B$  
is typically
the hardest one. We plot the result in Fig.~\ref{discovery}. In
the left panel we show the $m_{Zb_h}$ invariant-mass distribution in
the SM, the benchmark model and the extreme $B$ case. It is clear that
the distributions in the SM and the new model peak in very different
regions. The benchmark model leads to a too small cross
section and would require higher luminosity for discovery. The extreme
$B$ model, however, shows a clear peak with a total number of $\approx
40$ events at $m_{Zb_h} \approx m_B= 600$ GeV, 
versus $\approx 3$ background events,
implying a statistical significance of  
\begin{equation}
\frac{S}{\sqrt{B}}\approx 21,\qquad (Zb\bar{b}\mbox{ for
extreme $B$}).
\end{equation}
Given the
presence of a distinct peak we can attempt to reconstruct the mass
of the heavy gluon. In the right panel of Fig.~\ref{discovery} we
show the total invariant-mass of the three objects $Zb\bar{b}$ for the
events passing the cuts. Due to the large width of the heavy gluon
(the kinematical threshold prevents the full width to be apparent
at energies below $\sim 600$ GeV) the number of events peaks
slightly below $\textrm{M}_\textrm{G}=850$ GeV, but the effect is clearly observable. 
The approximate statistical significance of the excess above
$600$ GeV is 
\begin{equation}
\frac{S}{\sqrt{B}}\approx \frac{38}{\sqrt{5}}=17,
\qquad (\mbox{$\textrm{M}_\textrm{G}$ peak in }Zb\bar{b}\mbox{ for
  extreme B}).
\end{equation} 

The $Zb\bar{b}$ channel appears then as 
very promising even with the very simple cuts that we have used. 
In the extreme case the reconstruction of
the $B$ quark and of the massive gluon at the  4 fb$^{-1}$ LHC could
be correlated with the $t\bar{t}$ anomalies discussed before, 
disentangling the origin of the Tevatron FB asymmetry. 

\subsubsection*{$Z t \bar{t}$ channel}

\begin{figure}
\begin{center}
\includegraphics[width=0.5\linewidth]{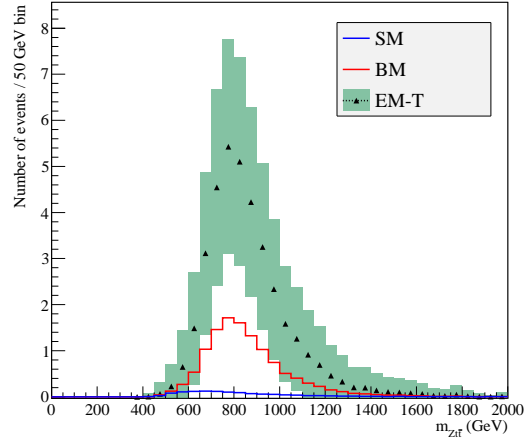} 
\end{center}
\caption{\label{Zttdiscovery} Total invariant-mass reconstruction for
  the $Zt\bar{t}$ channel in the SM (solid/blue almost flat),
  benchmark (solid/red) and extreme $T$ (data with statistical
  errors shown as a band) models for the $Zt\bar{t}$ analysis
  described in the text for the LHC with
4 fb$^{-1}$.}
\end{figure}

The $Z t\bar{t}$ production channel resulting into a $Z W^+ W^- b \bar{b}$
final state has also a very small SM background, but it 
is harder to reconstruct due to its
large multiplicity. Instead of trying to
reconstruct the $T$ mass, it is simpler to reconstruct the total final
state in the search for the massive gluon. We do that requiring
{\it (i)} three charged leptons with
$p_T\geq 25$ GeV, and at least two of them with the 
same flavor and opposite sign
reconstructing the $Z$ within 25 GeV; {\it (ii)}
at least two $b$--tagged and at
least two non--$b$--tagged jets with $p_T>20$ and $|\eta|< 2.8$. We
reconstruct the neutrino momentum using the on-shellness condition for
a $W$ and take the two hardest jets and $b$-jets if there are
more of them. The result is shown in
Fig.~\ref{Zttdiscovery}. The extreme $T$ model shows a clear peak
with $\approx 36$ events with no expected background events (the
benchmark gives a weaker deviation). 
A more detailed analysis, trying to 
reconstruct both top quarks, would certainly help in the reconstruction of
the heavy $T$ mass. Since the extreme $T$ model would also 
show up in the charged decay channel, a hint on 
the $T$ mass could be used in the reconstruction of this channel. 

\Section{Light flavor excitations: $Wq'\bar q$ and $Zq\bar q$ \label{Q}}

\begin{figure}[t]
\begin{center}
\includegraphics[width=0.45\linewidth]{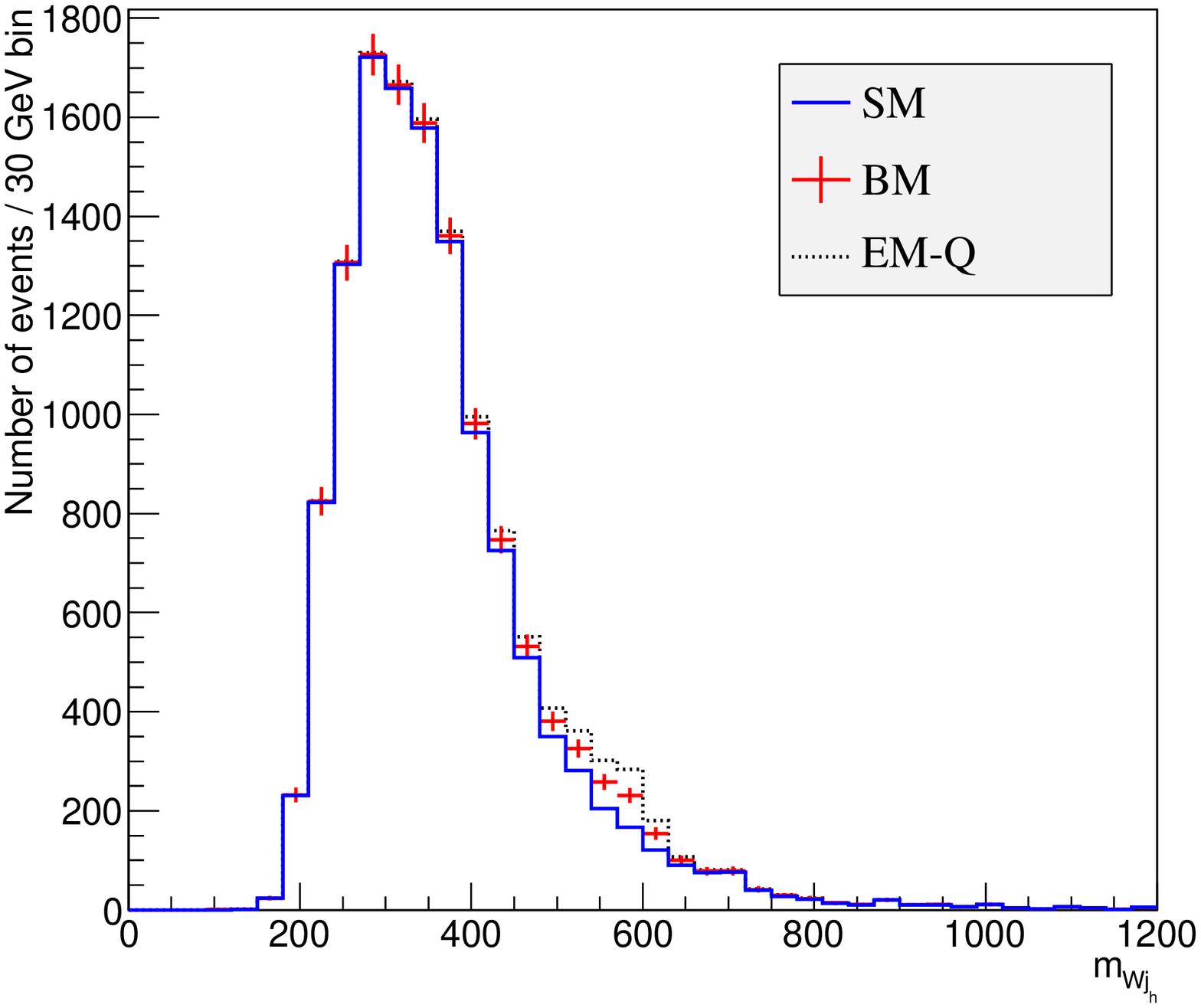} 
\includegraphics[width=0.45\linewidth]{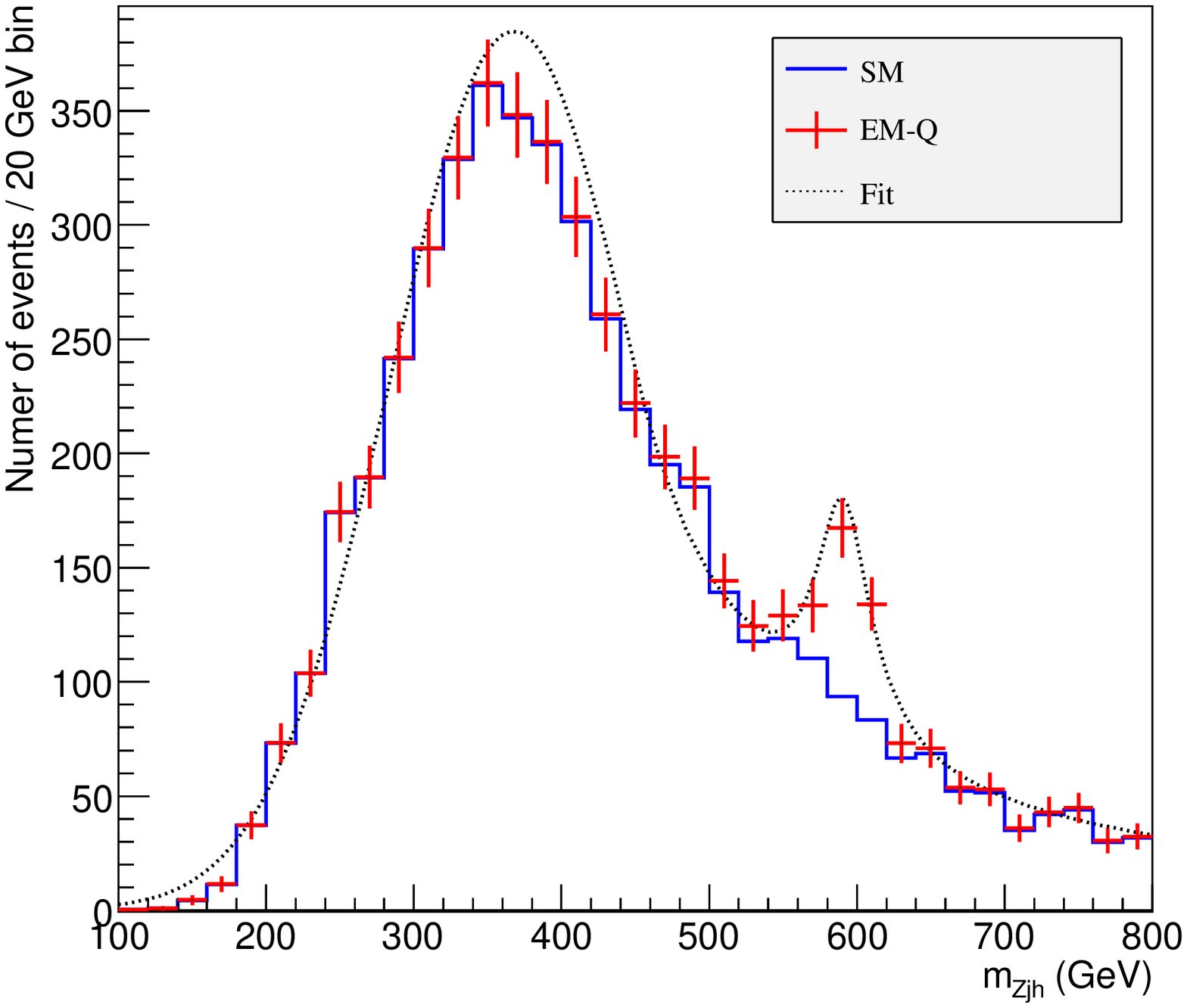} 
\end{center}
\caption{\label{WZj:discovery} Left panel: transverse mass for the $W
  j_h$ system in the $Wjj$ analysis described in the text for the SM
  (solid/blue), benchmark model (data points with errors) and extreme
  Q model (dotted/black). Right
  panel: Result of the fit of the $m_{Zj_h}$ distribution for the
  $Zjj$ analysis described in 
  the text for the SM (solid blue), extreme Q model (data points with
  statistical errors) and the fit to both distributions (dotted/black). 
Both plots are for the 7 TeV LHC with 4 fb$^{-1}$.
}
\end{figure}

We have seen in previous sections that the production of single 
$T$ or $B$ quarks tend to introduce anomalies in current searches 
and could be seen 
if the reconstruction algorithms are slightly modified. 
However, $Q\bar q$ production is less apparent in these searches,
being the best example of stealth new
physics~\cite{Barcelo:2011vk}. 
We discuss in this section the
best strategy to observe the extreme
Q model at the LHC.  
In the benchmark (extreme $Q$) 
model the production of heavy excitations $Q$
of the light flavors has a total cross
section of $2.9$ ($5.4$) pb  at the 7 TeV LHC, resulting
with a 2:1 ratio the final states $Wq^\prime q$ and $Zqq$. 
The SM irreducible background is 17 nb for $W$ plus $\ge 1$ jets 
and 6 nb for $Z$ plus $\ge 1$ jets. 
Therefore, we need to impose cuts to
disentangle our signal from these large backgrounds.
First of all, 
these extra $Q\bar q$ events will only appear at invariant-masses 
above $m_Q=600$ GeV, with the maximum at $\approx 700$ GeV.
In addition, the jet from the decay of the heavy
quark, with a $p_T\sim
m_Q/2$, will be typically harder than the second jet.
We should then impose an stringent cut on the hardest jet in order
to reduce the SM backgrounds. In particular, requiring a
hardest jet with $p_T\geq 150$ GeV on top of the cuts defined in
\cite{Wjets:ref} reduces the $W$+jets background to
manageable levels. We show in
Fig.~\ref{WZj:discovery}--left the transverse mass distribution
of the $W$ and the hardest jet. The signal does not seem significant in the
benchmark model but may be observable in the extreme $Q$ case, 
with 6 bins departing more than 3$\sigma$ from the expected background. 

The neutral case is even more promising. Requiring two same-flavor,
opposite-charge leptons with $p_T\geq 25$ GeV that reconstruct the $Z$
mass within 25 GeV, and two or more jets, with $p_T\geq 150$ GeV for
one of them, and computing the invariant mass of the $Z$ and the
hardest jet, we obtain the distribution in
Fig.~\ref{WZj:discovery}--right. Although the
benchmark model is still unobservable, there is a clear peak for the
extreme model. We have fitted the signal plus background histogram to
a Crystal Ball plus gaussian shape and obtained an excess of 170 events
over the expected 540 background events in the region of two standard
deviations around the center of the gaussian. This leads to a
statistical significance of $7\sigma$ deviation and a best fit 
of $m_Q^{\mathrm{fit}}=590$ GeV, very close to the actual heavy
quark mass.
This analysis is interesting as it gives a very clean signal
for a model that is otherwise very difficult to find.

%% file: Contents/Conclusions/Conclusions.tex

\Conclusions


The Standard Model of particle physics has proved to be an 
extraordinarily successful
theory to describe the physics at colliders below the TeV scale. 
Nevertheless, one of 
its basic pieces, the mechanism responsible for the electroweak 
symmetry breaking, has not been fully confirmed yet. 
There are also formal 
arguments indicating that the model must be completed at 
energies accessible to the LHC. In
particular, the hierarchy problem has been the main motivation 
for model building during the past 30 years.
On the other hand, there have been a series of experimental 
anomalies that 
almost always have vanished with an increase in the 
amount of available data. In any case, these anomalies have been
a motivation to explore possible scenarios for new physics. 
The main results of this Thesis are related 
to phenomenological implications of some of these 
models beyond the standard one.
We have focused on the interplay between the Higgs and 
the top-quark sectors because of  two basics reasons.
First, we expect that the top quark plays a fundamental role 
in the mechanism of gauge symmetry breaking, since it is responsible 
for the largest radiative corrections
destabilizing the electroweak scale. Second, Tevatron has 
observed an asymmetry in $t \bar t$ production that can not
be explained within the Standard Model. 
We can summarize our results as follows.

\begin{itemize}

\item Little Higgs models could be a {\it bridge} connecting the 
Standard Model and a more general theory like 
Supersymmetry or Technicolor, which would 
define a framework consistent 
up to the Planck scale. 
In the second chapter we have studied models
where the Higgs appears as a pseudo-Goldstone boson of a 
global symmetry broken spontaneously at a scale $f$ slightly higher
than the electroweak one. These models incorporate an extra $T$ 
quark that must be light in order to cancel top-quark
quadratic corrections 
to the Higgs mass parameter. On the other hand, the extra 
gauge bosons also present in these models have masses of the same 
order, introducing a 
conflict with electroweak precision data if $f < 3$ TeV. 
We have studied and solved this tension in the original 
simplest model using 
VEVs ($f_1 \approx 0.1 f_2$) and couplings 
($\lambda_2, \lambda'_{1,2} \lesssim 0.1 \lambda_1$) suppressed 
by an approximate symmetry in the top-quark sector.
Thus, we have changed the collective symmetry breaking 
principle ($\lambda_1 \approx \lambda_2$ and 
$\lambda'_{1,2} = 0$) for an approximate symmetry breaking one.

We have also shown that after electroweak symmetry breaking
the physical Higgs in these models has 
both doublet and singlet $SU(2)_L$ components.
As a consequence, its gauge and Yukawa couplings
are reduced, and Higgs production through
gluon and $W$ boson fusion is weaker than in the Standard Model.
This, together with the new decay channel 
$h \rightarrow \eta \eta$, where $\eta$
is a light pseudoscalar singlet, could relax LEP
bounds on the Higgs mass.

Finally, we have also studied the one-loop effective
potential of this model. We have shown that 
it provides the observed electroweak symmetry 
breaking and an acceptable
Higgs mass. To do that we have worked at all 
order in $v^2 / f^2$, writing 
the results in terms of sines and cosines, as the 
first-order expansion
breaks down when $f_1$ is small. The usual case with 
collective breaking requires the  addition of new
terms breaking the symmetry in the scalar potential  
in order to get a Higgs mass above LEP bounds.

The observation of a $T$ quark and of anomalous gauge 
and Yukawa couplings for the Higgs boson 
at the LHC would be hints pointing 
towards this class of models.
Their discovery could be taken as an invitation to 
build a more powerful
accelerator, since these models define a framework just 
valid up to energies $4\pi f\approx 10$ TeV.

\item In the third chapter we have analyzed Higgs boson effects 
on $t \bar t$ production at the Tevatron
and the LHC. We have initially shown that a standard Higgs 
heavy enough to decay into top-quark pairs would
couple very strongly to itself. As 
a consequence, its decay width grows and dilutes the effect on
the $t \bar t$ invariant  mass distribution.
On the other hand, Supersymmetry and Little Higgs models 
provide heavy Higgses with no need for large scalar self-couplings, 
as their mass is not electroweak. We have seen 
that interference effects are important and 
the narrow-width approximation gives a poor (even a
misleading) estimate.
We find peak-dip structures in $m_{t\bar t}$
that require optimized strategies 
at hadron colliders. In 
supersimetric models we obtain that
the mass difference between the neutral gauge bosons $H$ 
and $A$ is a crucial parameter  that could amplify 
or make disappear the effects
(something that would not happen by adding peaks).  
Choosing the right binning would be important in order to make 
these effects observable.

We have also found that the top-quark angular distribution  
in the invariant mass region $m_{t\bar t}\approx m_{H,A}$ 
does not correspond to the spin of the intermediate particle, 
{\it i.e.}, the excess caused by the interference 
does not have a flat distribution 
in the center of mass frame (as one would obtain  
in the narrow-width approximation).

Finally, we have found that these effects
would not be observable at the Tevatron 
even for the 10 fb$^{-1}$ finally achieved. 
The main difference with the LHC is that at the Tevatron 
90$\%$ of the top-quark pairs
are produced through $q \bar q$ interactions. Since the 
signal we have explored is caused by interference in the 
$gg \rightarrow t \bar t$
channel, for the same integrated luminosity the 
deviations there would be 9 times weaker than at the LHC, 
where gluon fusion provides 90$\%$
of the top pairs. After researchers from Tevatron became 
interested in our work, we obtained that the signal would be too small to be
observable in that collider.

\item The 2--3$\sigma$ deviation in the forward-backward 
asymmetry respect to the Standard Model seems consistent, but it has not been
supported by anomalies in any other observables at the 
Tevatron nor the LHC. As a consequence, the new physics proposed 
to explain it is 
typically pushed above 1 TeV. In Chapter 4 we 
have proposed a scenario with an experimental signature that is 
peculiar and 
easy to miss in current searches despite being at low energy.
It would be defined by a relatively light gluon with 
a very large width produced by new decay channels  $G\to qQ$.
We have shown that the model reproduces both the asymmetry 
and the $t\bar t$ invariant mass distribution observed 
at hadron colliders. 

We have analyzed the phenomenological implications 
of the heavy quarks $Q$ at the LHC.
In particular, we have studied their consistency with current
studies of 
$t \bar t$ production and $T \bar T$ searches. 
We have found that $T \bar t$ production 
gives the same $WWb\bar b$ signal as $t \bar t$, and that it
could be observed by  
slightly changing the reconstruction criteria used 
currently for $t \bar t$ at ATLAS and CMS.
The situation is similar for the $B \bar b$ channel.

We have also studied other signals that, if 
analyzed, could reveal single heavy quark
production through the $qQ$ channel 
at the LHC. In particular, $Zq \bar q$ where the $Q$ 
quark is reconstructed with the
$Z$ boson and the highest-energy jet looks promising. Other 
signals, like $Zb \bar b$ or $Zt \bar t$, are
predicted here and could offer 
a signal above  
the Standard Model backgrounds after applying 
optimal cuts. With a slight increase 
in the integrated luminosity, the scenario could be confirmed 
or excluded at the LHC (we are currently collaborating with 
researchers from ATLAS to implement
our model in their simulations).

We have focused our study on the region motivated by the
Tevatron asymmetry, but our analyses can be also applied to 
a wider range of couplings and
quark and gluon masses.

\end{itemize}

To conclude, in this Thesis we have studied the Higgs 
boson and different extensions of the Standard Model that it
motivates. The 
study of the top quark
in hadronic colliders should be essential to understand how to 
complete the Standard Model. In this sense, the forward-backward 
asymmetry in $t \bar t$ appears as a
promising hint of new physics. Its explanation could be linked
to physics slightly {\it different} from the one we are used to, 
with massive 
quarks that are not pair produced and appear 
together with a light quark or with massive gluons of
very large width.
The analysis of the new data should tell us sooner than later
whether we are 
looking for in the right direction. Certainly, we are living 
exciting times in particle physics.

%% file: Contents/Appendices/Appendix_A/AppA.tex

{\FirstAppendix{Event Reconstruction}\label{AppA}}

The reconstruction of $m_{t \bar t}$ is done in three steps: first a leptonically decaying $W$ boson is 
reconstructed, then the jets are associated to partons in the $t \bar t$ decay chain, and finally a kinematic fit
is performed.

It is assumed that one W boson from a top decay has decayed into the observed lepton and an
undetected neutrino. The $E^{\textrm{miss}}_T$ is taken as a measurement of the transverse momentum of the
neutrino, but its longitudinal momentum (i.e. parallel to the beam direction) is unmeasured.
Imposing the condition that the invariant mass of the lepton and neutrino is the mass of the
$W$ boson (80.4 GeV) allows the construction of a quadratic equation for the longitudinal
momentum of the neutrino. If there are two real solutions, both are retained. If the equation
only has imaginary solutions, the components of $E^{\textrm{miss}}_T$ are modified by the minimal amount in
$| \Delta E^{\textrm{miss}}_{T \;\;  x} | + | \Delta E^{\textrm{miss}}_{T \;\; y} |$ to give one real solution.
The association of the jets to the hadronic $W$ decay and to the two bottom quarks is done by 
calculating a $\chi^2$ for each possible combination (including the two neutrino solutions if they are both
physical and only allowing b-tagged jets to be associated to a bottom quark) and choosing the
combination with the smallest value. The $\chi^2$ is a sum over several terms

\beq
\chi^2 = \sum \chi^2_i = \sum {( x_i - x_i^{\textrm{ref}})^2 \over  \sigma^2_i} \;,
\eeq

where $x_i$ is a reconstructed quantity, $x_i^{\textrm{ref}}$ is a reference value for this quantity and $\sigma^2_i$ is a
resolution parameter. The reconstructed quantities and the central values and widths used are
listed in Table B.1. The appropriate reference values and resolutions for the masses are obtained
from the distributions of these quantities in a Montecarlo simulation. Since no flavor-specific
jet energy scale corrections are applied, deviations from the generated values of the top
quark and $W$ boson masses are expected. The association of jets to the $W$ boson and the bottom
quarks is found to be correct in approximately $80\%$ of events.

\hspace{-1cm}
\begin{table}
\begin{center}
\begin{tabular}{|c|c|c|}
\hline
\textbf{Quantity} & \textbf{Reference Value} & \textbf{Resolution} \\
\hline \hline
Leptonic top Mass & 169.0 GeV & 16.3 GeV \\
Hadronic top Mass & 174.7 GeV & 14.6 GeV\\
Hadronic $W$ Mass  & 83 GeV & 10.9 GeV \\
$p_T$ of $t \bar t$ System & 0 GeV & 50 GeV \\
$H_T$ Fraction & 1. & 0.15 \\
\hline
\end{tabular} 
\label{chisquaredchapter5}
\caption{Quantities, with their reference values and resolutions, used in the definition of the $\chi^2$
for jet-parton association. The `$H_T$ Fraction' is the scalar sum of the transverse energy in the
selected jets divided by the scalar sum of the transverse energy in all jets.}
\end{center}
\end{table}

%% file: Contents/Appendices/Appendix_B/AppB.tex

{\Appendix{Event Selection for $t \bar{t}$ Production}\label{AppB}}

The $t \bar t$ candidate events are selected\footnote{This selection of events is similar to the one done by the CMS collaboration in \cite{LHCttbar}.} 
in the lepton+jets topology, where one top quark decays semileptonically 
($t \rightarrow l \nu b$) and the other hadronically ($t \rightarrow q \bar{q}' b$).
In particular, we select events with an isolated muon or electron
in the central portion of the detector with high transverse momentum 
and a large amount of missing transverse energy, consistent with the presence of an undetected 
neutrino. Moreover, we select events with three or more jets. 

Muons(Electrons) are required to have $p_T > 20$ GeV and $| \eta | < 2.1$($p_T > 30$ GeV and $| \eta | < 2.5$).
To select leptonic $W$ boson decays, events are required to contain either one isolated muon or
electron. The isolation requirement is based on the ratio of the total transverse
energy observed from all hadrons and photons in a cone of size $\Delta R = \sqrt{(\Delta \eta)^2 + (\Delta \phi)^2} < 0.3$
around the lepton direction to the transverse momentum of the lepton itself, known as relative
isolation, where $\Delta \phi$ and $\Delta \eta$ are the
azimuthal angle and pseudorapidity differences between the electron and the jet. This quantity must be less than 10\%.

In order to reduce the background from Drell-Yan production and $t \bar t$ production in which both
$W$ bosons decay leptonically, events in which two lepton candidates are identified are vetoed.
To increase the rejection, the second lepton may be allowed to satisfy looser requirements: an
additional muon must have $p_T > 10$ GeV, $| \eta | < 2.5$ and relative isolation $< 0.2$; an electron
in an event with a muon candidate must have $p_T > 15$ GeV, $| \eta | <$ 2.5 and relative isolation
$< 0.2$.

Events are additionally required to contain three or more jets with
$p_T > 30$ GeV and $| \eta | <$ 2.4, and must not overlap with
any lepton candidate within $ \Delta R < 0.4$. To enhance the rejection of background from $W$ boson
and Drell-Yan production in association with relatively low $p_T$ jets, the leading jet is required
to have $p_T > 70$ GeV and the second leading jet to have $p_T > 50$ GeV.

The negative of the vector sum of the momenta of all reconstructed jets and leptons in the plane
transverse to the beam is the missing transverse energy $E_T^{\textrm{miss}}$ vector. QCD background is
suppressed further by requiring $E_T^{\textrm{miss}} > 20$ GeV.

%% file: Contents/Appendices/Appendix_C/AppC.tex

{\Appendix{Event Selection for $T \bar{T}$ Production}\label{AppC}}

The selection is chosen to maximize $S / \sqrt{B}$ in the accepted signal sample for a $T$ mass of
$400$ GeV, where $S$ is the number of $T \bar{T}$ signal events and $B$ the number of events expected
from $t \bar{t}$ production and all other EW background processes. 

The search is performed for the strong pair production of a $T$ quark and its antiparticle, followed
by each of their decays to a $W$ boson and a $b$ or $\bar{b}$ quark. Lepton+jets events are selected
with a single charged lepton, missing transverse momentum, and at least four jets of high
transverse momenta, indicative of events in which one of the $W$ bosons decays to leptons (electrons or muons) 
and the other $W$ boson decays to quarks. \\

\begin{flushleft}
In the $e$+jets channel, we require
\end{flushleft}

\begin{itemize}

\item an isolated electron with $p_T > 30 - 45$ GeV and $| \eta | < 2.5$;

\item missing $p_T > 20$ GeV;

\item at least four jets with $p_T > 120, 90, 35, 35$ GeV and $| \eta | < 2.4$. Jets that satisfy
$\Delta R = \sqrt{(\Delta \eta)^2 + (\Delta \phi)^2} < 0.3$ of the electron are rejected;

\item at least one jet must be b-tagged.

\end{itemize}
In the $\mu$+jets channel, we require

\begin{itemize}

\item an isolated muon with $p_T > 35$ GeV and $| \eta | < 2.1$; 

\item missing $p_T > 20$ GeV;

\item at least four jets with $p_T > 120, 90, 35, 35 GeV$, and $| \eta | < 2.4$. Jets that satisfy
$\Delta R < 0.3$ of the muon are rejected;

\item at least one jet must be b-tagged.

\end{itemize}

%% file: PhDThesis.bbl
\begin{thebibliography}{99}
\fancyhead[LE]{\thepage~~~~Bibliography}
\fancyhead[RO]{Bibliography~~~~~\thepage}




\bibitem{Barcelo:2007if}
  R.~Barcelo, M.~Masip and M.~Moreno-Torres,
  Nucl.\ Phys.\  B {\bf 782} (2007) 159
  [arXiv:hep-ph/0701040].
  
\bibitem{Barcelo:2008je}
  R.~Barcelo and M.~Masip,
  Phys.\ Rev.\  D {\bf 78} (2008) 095012
  [arXiv:0809.3124 [hep-ph]].
  
\bibitem{Barcelo:2009uy} 
  R.~Barcelo, M.~Masip and I.~Mastromatteo,
  JCAP {\bf 0906}, 027 (2009)
  [arXiv:0903.5247 [hep-ph]].

  
\bibitem{Barcelo:2010bm}
  R.~Barcelo and M.~Masip,
  Phys.\ Rev.\  D {\bf 81} (2010) 075019
  [arXiv:1001.5456 [hep-ph]].
  
\bibitem{Barcelo:2011fw} 
  R.~Barcelo, A.~Carmona, M.~Masip and J.~Santiago,
  Phys.\ Rev.\ D {\bf 84}, 014024 (2011)
  [arXiv:1105.3333 [hep-ph]].

\bibitem{Barcelo:2011vk} 
  R.~Barcelo, A.~Carmona, M.~Masip and J.~Santiago,
  Phys.\ Lett.\ B {\bf 707}, 88 (2012)
  [arXiv:1106.4054 [hep-ph]].

\bibitem{Barcelo:2011wu} 
  R.~Barcelo, A.~Carmona, M.~Chala, M.~Masip and J.~Santiago,
  Nucl.\ Phys.\ B {\bf 857}, 172 (2012)
  [arXiv:1110.5914 [hep-ph]].
  
\bibitem{Frascati:2010}{R.~Barcelo, Frascati Physics Series Vol. LI (2010) 7-12.}






\bibitem{GlashowSM}{S. L. Glashow, Nucl. Phys. {\bf 22} (1961) 579.}

\bibitem{WeinbergSM}{S. W. Weinberg, Phys. Rev. Lett. {\bf 19} (1967) 1264.}

\bibitem{SalamSM}{A. Salam, {\it Originally printed in *Svartholm: Elementary Particle Theory, Proc. of the Nobel Symposium Held 1968 at Lerum, Sweden*, Stockolm 1968, 367-377}.}

\bibitem{Higgs}{P. W. Higgs, Phys. Lett. {\bf 12} (1964) 132.}

\bibitem{Novaes}{S. F. Novaes, arXiv:hep-ph/0001283.}

\bibitem{Pokorski}{P. Pokorski, arXiv:hep-ph/0502132.}

\bibitem{Hollik}{W. Hollik, J. Phys. Conf. Ser. {\bf 53} (2006) 7-43.}

\bibitem{Pich}{A. Pich, arXiv:0705.4264[hep-ph].}

\bibitem{Langacker1}{P. Langacker, arXiv:0901.0241[hep-ph].}

\bibitem{Langacker2}{P. Langacker, Boca Raton, USA: CRC Pr. (2010) 663 p.}

\bibitem{Amsler}{C. Amsler {\it et al.} [Particle Data Group], Phys. Lett. B {\bf 667}, 1 (2008) and 2009 partial update for the 2010 edition.}

\bibitem{DjouadiSM}{A. Djouadi,
  Phys.\ Rept.\  {\bf 457} (2008) 1
  [arXiv:hep-ph/0503172].}
  
\bibitem{IllanaSM}{`J. I. Illana, http://www.ugr.es/$\sim$jillana/curso.pdf'.}


\bibitem{Englert}{F. Englert and R. Brout, Phys. Rev. Lett. {\bf 13} (1964) 321.}

\bibitem{Gural}{G. S. Guralnik, C. R. Hagen and T. W. B. Kibble, Phys. Rev. Lett. {\bf 13} (1964) 585.}

\bibitem{Kibble}{T. W. B. Kibble, Phys. Rev. {\bf 155} (1967) 1554.}

\bibitem{Anderson}{P. W. Anderson, Phys. Rev. {\bf 130} (1963) 439.}

\bibitem{SuperNambu}{Y. Nambu, Phys. Rev. {\bf 117} (1960) 648.}


\bibitem{Nambu}{Y. Nambu, Phys. Rev. Lett. {\bf 4} (1960) 380.}

\bibitem{Goldstone}{J. Goldstone, Nuovo Cim {\bf 19} (1961) 154.}

\bibitem{Agrawal:1997gf}
V.~Agrawal, S.~M.~Barr, J.~F.~Donoghue and D.~Seckel,
Phys.\ Rev.\ D {\bf 57} (1998) 5480.

\bibitem{Arkani-Hamed:2004fb}
N.~Arkani-Hamed and S.~Dimopoulos,
JHEP {\bf 0506} (2005) 073.

\bibitem{LEPEWWG}
The LEP Collaboration (ALEPH, DELPHI, L3 and OPAL), the LEP Electroweak
Working Group. `http://lepewwg.web.cern.ch/LEPEWWG'.

\bibitem{Moroi:1995yh} 
  T.~Moroi,
  Phys.\ Rev.\ D {\bf 53}, 6565 (1996)
  [Erratum-ibid.\ D {\bf 56}, 4424 (1997)]
  [hep-ph/9512396].
  
\bibitem{Carena:1996qa} 
  M.~S.~Carena, G.~F.~Giudice and C.~E.~M.~Wagner,
  Phys.\ Lett.\ B {\bf 390}, 234 (1997)
  [hep-ph/9610233].

\bibitem{Malkawi:1996fs} 
  E.~Malkawi, T.~M.~P.~Tait and C.~P.~Yuan,
  Phys.\ Lett.\ B {\bf 385}, 304 (1996)
  [hep-ph/9603349].
  
\bibitem{Djouadi:2006rk} 
  A.~Djouadi, G.~Moreau and F.~Richard,
  Nucl.\ Phys.\ B {\bf 773}, 43 (2007)
  [hep-ph/0610173].


\bibitem{Kado}{M. M. Kado and C. G. Tully, Annu. Rev. Nucl. Part. Sci. B {\bf 52} (2002) 65.}

\bibitem{Egli:1989vu}{S.~Egli {\it et al.}  [SINDRUM Collaboration], Phys.\ Lett.\  B {\bf 222} (1989) 533.}

\bibitem{Barr:1989pv}{G.~D.~Barr {\it et al.}  [NA31 Collaboration], Phys.\ Lett.\  B {\bf 235} (1990) 356.}

\bibitem{CLEO1989}{Alam MS, {\it et al.} [CLEO Collab.], Phys. Rev. Lett. {\bf 40} (1989) 712.}

\bibitem{CUSB1982}{Sievertz M, {\it et al.} [CUSB Collab.], Phys. Rev. Lett. {\bf 26} (1982) 717.}

\bibitem{CUSB1988}{Lee-Franzini J, {\it et al.} [CUSB Collab.], Proc. Int. Conf. High Energy Phys., XXIVth, Munich, Aug. 4 - 10, 1988, p. 891, (1989).}

\bibitem{LEP11}{ALEPH Collaboration (D. Buskulic {\it et al.}), Phys. Lett. B {\bf 384} (1996) 427.}

\bibitem{LEP12}{DELPHI Collaboration (P. Abreu {\it et al.}), Nucl. Phys. B {\bf 421} (1994) 3.}

\bibitem{LEP13}{L3 Collaboration (M. Acciarri {\it et al.}), Phys. Lett. B {\bf 385} (1996) 454.}

\bibitem{LEP14}{OPAL Collaboration (G. Alexander {\it et al.}), Z. Phys. C {\bf 73} (1997) 189.}

\bibitem{LEP2}{The LEP Collaboration (ALEPH, DELPHI, L3 and OPAL), Phys. Lett. B {\bf 565} (2003) 61.}

\bibitem{Tevatron}{The CDF and D0 Collaborations, the {\it Tevatron New Phenomena and the Higgs} Working Group, arXiv:1107.5518 [hep-ex].}

\bibitem{LHCCMS}{The CMS Collaboration, arXiv:1202.1488v1 [hep-ex].}

\bibitem{LHCATLAS}{The ATLAS Collaboration, arXiv:1202.1408v2 [hep-ex].}




\bibitem{Erler:2004nh}
J.~Erler and P.~Langacker,
``Electroweak model and constraints on new physics,''
arXiv:hep-ph/0407097,
in S.~Eidelman {\it et al.}  [Particle Data Group],
Phys.\ Lett.\ B {\bf 592} (2004) 1.

\bibitem{Roy:2005hg}
  T.~Roy and M.~Schmaltz,
  JHEP {\bf 0601} (2006) 149.

\bibitem{Berezhiani:2005pb}
  Z.~Berezhiani, P.~H.~Chankowski, A.~Falkowski and S.~Pokorski,
  Phys.\ Rev.\ Lett.\  {\bf 96} (2006) 031801.

\bibitem{Csaki:2005fc}
C.~Csaki, G.~Marandella, Y.~Shirman and A.~Strumia,
Phys.\ Rev.\ D {\bf 73} (2006) 035006.

\bibitem{Bai:2007tv}
  Y.~Bai, J.~Fan and Z.~Han,
  Phys.\ Rev.\  D {\bf 76} (2007) 065003.

\bibitem{Contino:2003ve1}
R.~Contino, Y.~Nomura and A.~Pomarol,
Nucl.\ Phys.\ B {\bf 671} (2003) 148.

\bibitem{Contino:2003ve2}
K.~Agashe, R.~Contino and A.~Pomarol,
Nucl.\ Phys.\ B {\bf 719} (2005) 165.

\bibitem{Contino:2003ve3}
K.~Agashe and R.~Contino,
Nucl.\ Phys.\ B {\bf 742} (2006) 59.

\bibitem{Giudice:2007fh}
  G.~F.~Giudice, C.~Grojean, A.~Pomarol and R.~Rattazzi,
  JHEP {\bf 0706} (2007) 045.
  
\bibitem{Perelstein:2005ka}
M.~Perelstein,
Prog.\ Part.\ Nucl.\ Phys.\  {\bf 58} (2007) 247.

\bibitem{Schmaltz:2005ky}
M.~Schmaltz and D.~Tucker-Smith,
Ann.\ Rev.\ Nucl.\ Part.\ Sci.\  {\bf 55} (2005) 229.

\bibitem{Arkani-Hamed:2001nc}
N.~Arkani-Hamed, A.~G.~Cohen and H.~Georgi,
Phys.\ Lett.\ B {\bf 513} (2001) 232.

\bibitem{delAguila:2008zu} 
  F.~del Aguila, J.~I.~Illana and M.~D.~Jenkins,
  JHEP {\bf 0901}, 080 (2009)
  [arXiv:0811.2891 [hep-ph]].
  
\bibitem{delAguila:2010nv} 
  F.~del Aguila, J.~I.~Illana and M.~D.~Jenkins,
  JHEP {\bf 1009}, 040 (2010)
  [arXiv:1006.5914 [hep-ph]].

\bibitem{Schmaltz:2004de}
M.~Schmaltz,
JHEP {\bf 0408} (2004) 056.

\bibitem{delAguila:2011wk} 
  F.~del Aguila, J.~I.~Illana and M.~D.~Jenkins,
  JHEP {\bf 1103}, 080 (2011)
  [arXiv:1101.2936 [hep-ph]].

\bibitem{Casas:2005ev}
  J.~A.~Casas, J.~R.~Espinosa and I.~Hidalgo,
  JHEP {\bf 0503} (2005) 038.

\bibitem{Han:2005dz}
  Z.~Han and W.~Skiba,
  Phys.\ Rev.\  D {\bf 72} (2005) 035005.

\bibitem{Marandella:2005wd}
  G.~Marandella, C.~Schappacher and A.~Strumia,
  Phys.\ Rev.\  D {\bf 72} (2005) 035014.

\bibitem{Aguilar-Saavedra:2002kr}
J.~A.~Aguilar-Saavedra,
Phys.\ Rev.\ D {\bf 67} (2003) 035003
[Erratum-ibid.\ D {\bf 69} (2004) 099901].

\bibitem{Georgi}
H.~M.~Georgi, S.~L.~Glashow, M.~E.~Machacek and D.~V.~Nanopoulos,
Phys.\ Rev.\ Lett.\  {\bf 40} (1978) 692.

\bibitem{Rizzo:1979mf}
T.~G.~Rizzo,
Phys.\ Rev.\ D {\bf 22} (1980) 178
[Addendum-ibid.\ D {\bf 22} (1980) 1824].

\bibitem{O'Connell:2006wi}
D.~O'Connell, M.~J.~Ramsey-Musolf and M.~B.~Wise,
``Minimal extension of the standard model scalar sector,''
arXiv:hep-ph/0611014.

\bibitem{Bahat-Treidel:2006kx}
O.~Bahat-Treidel, Y.~Grossman and Y.~Rozen,
``Hiding the Higgs at the LHC,''
arXiv:hep-ph/0611162.

\bibitem{Chen:2006cs}
C.~R.~Chen, K.~Tobe and C.~P.~Yuan,
Phys.\ Lett.\ B {\bf 640} (2006) 263.

\bibitem{Han:2005ru}
T.~Han, H.~E.~Logan and L.~T.~Wang,
JHEP {\bf 0601} (2006) 099.

\bibitem{Aguilar-Saavedra:2006gw1}
J.~A.~Aguilar-Saavedra,
Phys.\ Lett.\ B {\bf 625} (2005) 234
[Erratum-ibid.\ B {\bf 633} (2006) 792]. 


\bibitem{Aguilar-Saavedra:2006gw2}
J.~A.~Aguilar-Saavedra,
`Light Higgs boson discovery from fermion mixing',
arXiv:hep-ph/0603200.

\bibitem{Kilian:2004pp1}
W.~Kilian, D.~Rainwater and J.~Reuter,
Phys.\ Rev.\ D {\bf 71} (2005) 015008.

\bibitem{Kilian:2004pp2}
W.~Kilian, D.~Rainwater and J.~Reuter,
Phys.\ Rev.\ D {\bf 74} (2006) 095003
[Erratum-ibid.\ D {\bf 74} (2006) 099905].

\bibitem{Cheung:2006nk}
K.~Cheung and J.~Song,
``Light pseudoscalar $\eta$ and $H\rightarrow \eta \eta$ 
decay in the simplest little Higgs
model,''
arXiv:hep-ph/0611294.

\bibitem{Coleman:1973jx} 
  S.~R.~Coleman and E.~J.~Weinberg,
  Phys.\ Rev.\ D {\bf 7}, 1888 (1973).

\bibitem{delAguila:2005yi}
  F.~del Aguila, M.~Masip and J.~L.~Padilla,
  Phys.\ Lett.\  B {\bf 627} (2005) 131
  [arXiv:hep-ph/0506063].





\bibitem{Martin:1997ns} 
  S.~P.~Martin,
  In *Kane, G.L. (ed.): Perspectives on supersymmetry II* 1-153
  [hep-ph/9709356].


\bibitem{Djouadi:2005gj}
  A.~Djouadi,
  Phys.\ Rept.\  {\bf 459} (2008) 1
  [arXiv:hep-ph/0503173].
  
\bibitem{Gaemers:1984sj}
  K.~J.~F.~Gaemers and F.~Hoogeveen,
  Phys.\ Lett.\  B {\bf 146} (1984) 347.

\bibitem{Dicus:1994bm}
  D.~Dicus, A.~Stange and S.~Willenbrock,
  Phys.\ Lett.\  B {\bf 333} (1994) 126
  [arXiv:hep-ph/9404359].
  
\bibitem{Frederix:2007gi}
  R.~Frederix and F.~Maltoni,
  JHEP {\bf 0901} (2009) 047
  [arXiv:0712.2355 [hep-ph]].

\bibitem{Barger:2006hm}
  V.~Barger, T.~Han and D.~G.~E.~Walker,
  Phys.\ Rev.\ Lett.\  {\bf 100} (2008) 031801
  [arXiv:hep-ph/0612016].

\bibitem{:2007dia}
  T.~Aaltonen {\it et al.}  [CDF Collaboration],
  Phys.\ Rev.\  D {\bf 77} (2008) 051102
  [arXiv:0710.5335 [hep-ex]].

\bibitem{Abazov:2008ny}
  V.~M.~Abazov {\it et al.}  [D0 Collaboration],
  Phys.\ Lett.\  B {\bf 668} (2008) 98
  [arXiv:0804.3664 [hep-ex]].

\bibitem{Cabrera:2009zza}
  S.~Cabrera  [ATLAS Collaboration],
  J.\ Phys.\ Conf.\ Ser.\  {\bf 171} (2009) 012085.

\bibitem{Baur:2007ck}
  U.~Baur and L.~H.~Orr,
  Phys.\ Rev.\  D {\bf 76} (2007) 094012
  [arXiv:0707.2066 [hep-ph]].

\bibitem{Hioki:2009hm}
  Z.~Hioki and K.~Ohkuma,
  Eur.\ Phys.\ J.\  C {\bf 65} (2010) 127
  [arXiv:0910.3049 [Unknown]].

\bibitem{Kumar:2009vs}
  K.~Kumar, T.~M.~P.~Tait and R.~Vega-Morales,
  JHEP {\bf 0905} (2009) 022
  [arXiv:0901.3808 [hep-ph]].
  
\bibitem{Berdine:2007uv}
  D.~Berdine, N.~Kauer and D.~Rainwater,
  Phys.\ Rev.\ Lett.\  {\bf 99} (2007) 111601
  [arXiv:hep-ph/0703058].
  
\bibitem{Golfand:1971iw} 
  Y.~.A.~Golfand and E.~P.~Likhtman,
  JETP Lett.\  {\bf 13}, 323 (1971)
  [Pisma Zh.\ Eksp.\ Teor.\ Fiz.\  {\bf 13}, 452 (1971)].
  
\bibitem{Wess:1973kz} 
  J.~Wess and B.~Zumino,
  Phys.\ Lett.\ B {\bf 49}, 52 (1974).
  
\bibitem{Wess:1974jb} 
  J.~Wess and B.~Zumino,
  Nucl.\ Phys.\ B {\bf 78}, 1 (1974).

\bibitem{Wess:1974tw} 
  J.~Wess and B.~Zumino,
  Nucl.\ Phys.\ B {\bf 70}, 39 (1974).
  
\bibitem{Haber:1984rc} 
  H.~E.~Haber and G.~L.~Kane,
  Phys.\ Rept.\  {\bf 117}, 75 (1985).
  
\bibitem{Errata}{J. F. Gunion and H. E. Haber, Nucl. Phys. B {\bf 272} (1986) 1 [Erratum-ibid. B {\bf 402}
(1993) 567].}

\bibitem{GunionHaber}{J. F. Gunion and H. E. Haber, Nucl. Phys. B {\bf 278} (1986) 449.}

\bibitem{Ibanez:1982fr} 
  L.~E.~Ibanez and G.~G.~Ross,
  Phys.\ Lett.\ B {\bf 110}, 215 (1982).


\bibitem{deBoer:1994he}
  W.~de Boer, R.~Ehret and D.~I.~Kazakov,
  Z.\ Phys.\  C {\bf 67} (1995) 647
  [arXiv:hep-ph/9405342].

\bibitem{Martin:2009iq}
  A.~D.~Martin, W.~J.~Stirling, R.~S.~Thorne and G.~Watt,
  Eur.\ Phys.\ J.\  C {\bf 63} (2009) 189
  [arXiv:0901.0002 [hep-ph]].

\bibitem{Frixione:2003ei}
  S.~Frixione, P.~Nason and B.~R.~Webber,
  JHEP {\bf 0308} (2003) 007
  [arXiv:hep-ph/0305252].




\bibitem{AFBTEV}
  V.~M.~Abazov {\it et al.} [ D0 Collaboration ],
  Phys.\ Rev.\ Lett.\  {\bf 100 } (2008)  142002.
  [arXiv:0712.0851 [hep-ex]].
  
\bibitem{AFB2}
  T.~Aaltonen {\it et al.} [ CDF Collaboration ],
  Phys.\ Rev.\ Lett.\  {\bf 101 } (2008)  202001.
  [arXiv:0806.2472 [hep-ex]].
  
\bibitem{AFB3}
  T.~Aaltonen {\it et al.} [ CDF Collaboration ], [arXiv:1101.0034 [hep-ex]].

\bibitem{Relatedobservables1}
  C.~Degrande, J.~-M.~Gerard, C.~Grojean, F.~Maltoni and G.~Servant,
  JHEP {\bf 1103} 125 (2011)
  [arXiv:1010.6304 [hep-ph]].
  
\bibitem{Relatedobservables2}
  C.~Degrande, J.~-M.~Gerard, C.~Grojean, F.~Maltoni and G.~Servant,
  Phys.\ Lett.\ B {\bf 703}, 306 (2011)
  [arXiv:1104.1798 [hep-ph]].

\bibitem{Relatedobservables3}
  J.~A.~Aguilar-Saavedra and M.~Perez-Victoria, JHEP {\bf 1105 } (2011) 034 [arXiv:1103.2765 [hep-ph]].
  
\bibitem{Relatedobservables4}
  J.~A.~Aguilar-Saavedra and M.~Perez-Victoria,
  Phys.\ Lett.\ B {\bf 701}, 93 (2011)
  [arXiv:1104.1385 [hep-ph]].
  
\bibitem{Relatedobservables5}
  J.~A.~Aguilar-Saavedra and M.~Perez-Victoria,
  Phys.\ Rev.\ D {\bf 84}, 115013 (2011)
  [arXiv:1105.4606 [hep-ph]].

\bibitem{Relatedobservables6}
  J.~A.~Aguilar-Saavedra and M.~Perez-Victoria,
  JHEP {\bf 1109}, 097 (2011)
  [arXiv:1107.0841 [hep-ph]].
  
\bibitem{Alwall:2007st}
  J.~Alwall {\it et al.},
  JHEP {\bf 0709} (2007) 028
  [arXiv:0706.2334 [hep-ph]].
  
\bibitem{pythia}
  T.~Sjostrand, S.~Mrenna and P.~Z.~Skands,
  JHEP {\bf 0605} (2006) 026
  [arXiv:hep-ph/0603175].

\bibitem{PGS4} PGS4,
`http://www.physics.ucdavis.edu/$\sim$
conway/research/software/pgs/pgs4-general.htm'.

\bibitem{Ovyn:2009tx}
  S.~Ovyn, X.~Rouby, V.~Lemaitre,
    [arXiv:0903.2225 [hep-ph]].
  
\bibitem{TeVprecision}
T.E.W. Group (Tevatron Electroweak Working Group) (2007), [arXiv:0703034 [hep-ph]].

\bibitem{Abazov:2011cq}
  V.~M.~Abazov {\it et al.}  [D0 Collaboration],
  Phys.\ Lett.\  B {\bf 704} (2011) 403
  [arXiv:1105.5384 [hep-ex]].
  
\bibitem{LHCasy}
CMS Collaboration, note CMS PAS TOP-11-030.

\bibitem{Kuhn:1998jr} 
  J.~H.~Kuhn and G.~Rodrigo,
  Phys.\ Rev.\ Lett.\  {\bf 81}, 49 (1998)
  [hep-ph/9802268].

\bibitem{Kuhn:1998kw} 
  J.~H.~Kuhn and G.~Rodrigo,
  Phys.\ Rev.\ D {\bf 59}, 054017 (1999)
  [hep-ph/9807420].

  
 \bibitem{Hollik2011ps}
W.~Hollik and D.~Pagani, Phys.\ Rev.\ D {\bf 84}, 093003 (2011) [arXiv:1107.2606 [hep-ph]].
  
\bibitem{Kuhn:2011ri} 
  J.~H.~Kuhn and G.~Rodrigo,
  JHEP {\bf 1201}, 063 (2012)
  [arXiv:1109.6830 [hep-ph]].
  
\bibitem{Bernreuther:2010ny} 
  W.~Bernreuther and Z.~-G.~Si,
  Nucl.\ Phys.\ B {\bf 837}, 90 (2010)
  [arXiv:1003.3926 [hep-ph]].
  
\bibitem{Abazov:2011rq} 
  V.~M.~Abazov {\it et al.}  [D0 Collaboration],
  Phys.\ Rev.\ D {\bf 84}, 112005 (2011)
  [arXiv:1107.4995 [hep-ex]].

\bibitem{Ferrario:2008wm1}
  P.~Ferrario, G.~Rodrigo, Phys.\ Rev.\  {\bf D78}, 094018 (2008) arXiv:0809.3354 [hep-ph].
  
\bibitem{Ferrario:2008wm2}
  P.~Ferrario and G.~Rodrigo,
  Phys.\ Rev.\ D {\bf 80}, 051701 (2009)
  [arXiv:0906.5541 [hep-ph]].
  
\bibitem{Moreau1} 
  A.~Djouadi, G.~Moreau, F.~Richard and R.~K.~Singh,
  Phys.\ Rev.\ D {\bf 82}, 071702 (2010)
  [arXiv:0906.0604 [hep-ph]].
 
\bibitem{Moreau2} 
  A.~Djouadi, G.~Moreau and F.~Richard,
  Phys.\ Lett.\ B {\bf 701}, 458 (2011)
  [arXiv:1105.3158 [hep-ph]].
 
\bibitem{Frampton:2009rk}
  P.~H.~Frampton, J.~Shu, K.~Wang,
  Phys.\ Lett.\  {\bf B683}, 294-297 (2010)
  [arXiv:0911.2955 [hep-ph]].

\bibitem{Chivukula}
  R.~S.~Chivukula, E.~H.~Simmons, C.~-P.~Yuan,
  Phys.\ Rev.\  {\bf D82}, 094009 (2010)
  [arXiv:1007.0260 [hep-ph]]

\bibitem{Burdman:2010gr}
  G.~Burdman, L.~de Lima, R.~D.~Matheus,
  Phys.\ Rev.\  {\bf D83}, 035012 (2011)
  [arXiv:1011.6380 [hep-ph]].

\bibitem{Haisch}
  U.~Haisch, S.~Westhoff,
  JHEP {\bf 1108}, 088 (2011).
  [arXiv:1106.0529 [hep-ph]];

\bibitem{Aguilar:2011ci}
  J.~A.~Aguilar-Saavedra and M.~Perez-Victoria,
  Phys.\ Lett.\ B {\bf 705}, 228 (2011)
  [arXiv:1107.2120 [hep-ph]].

\bibitem{Aguilar:2006gw}
  J.~A.~Aguilar-Saavedra,
  JHEP {\bf 0612}, 033 (2006).
  [hep-ph/0603200].

\bibitem{Aguilar:2011cp}
  J.~A.~Aguilar-Saavedra, A.~Juste and F.~Rubbo,
  Phys.\ Lett.\ B {\bf 707}, 92 (2012)
  [arXiv:1109.3710 [hep-ph]].

\bibitem{Falkowski:2011zr}
  A.~Falkowski, G.~Perez, M.~Schmaltz,
  [arXiv:1110.3796 [hep-ph]].


  
\bibitem{Ferrario:2009bz} 
  P.~Ferrario and G.~Rodrigo,
  Phys.\ Rev.\ D {\bf 80}, 051701 (2009)
  [arXiv:0906.5541 [hep-ph]].
  
\bibitem{Campbell:1999ah} 
  J.~M.~Campbell and R.~K.~Ellis,
  Phys.\ Rev.\ D {\bf 60}, 113006 (1999)
  [hep-ph/9905386].
  
\bibitem{Bauer:2010iq} 
  M.~Bauer, F.~Goertz, U.~Haisch, T.~Pfoh and S.~Westhoff,
  JHEP {\bf 1011}, 039 (2010)
  [arXiv:1008.0742 [hep-ph]].



\bibitem{LHCttbar}
CMS Collaboration, note CMS PAS TOP-10-007.

\bibitem{T4thgen}
CMS Collaboration, CMS PAS EXO-11-051.

\bibitem{LHCttbarATLAS}
ATLAS Collaboration, note ATLAS-CONF-2011-087.

\bibitem{TTtoZ}
CMS Collaboration, note CMS PAS EXO-11-005.

\bibitem{Aad:2011ec}
  G.~Aad {\it et al.} [ ATLAS Collaboration ],
    [arXiv:1108.5064 [hep-ex]].

\bibitem{zb:atlas}
ATLAS Collaboration, note CERN-PH-EP-2011-133.

\bibitem{Wjets:ref}
ATLAS Collaboration, note ATLAS-CONF-2011-097.













\end{thebibliography}
